\documentclass[aps,pra,twocolumn]{revtex4-2}
\usepackage{graphicx}
\usepackage{amsmath}
\usepackage{physics}
\usepackage{hyperref}
\usepackage{mathtools}
\usepackage{amssymb}
\usepackage{orcidlink}

\hypersetup{
    colorlinks=true,
    linkcolor=blue,
    filecolor=black,
    urlcolor=blue,
    allcolors=blue,
}

\begin{document}

\title{Photonic CZ gates based on few-photon scattering from emitter--cavity systems}
\author{Mateusz Duda \orcidlink{0009-0008-3083-4686}}\email{mduda1@sheffield.ac.uk}
\author{Pieter Kok \orcidlink{0000-0002-6608-330X}}
\affiliation{School of Mathematical and Physical Sciences, University of Sheffield, Sheffield S3 7RH, United Kingdom}


\begin{abstract}
In this paper we analyze the performance of photonic CZ gates based on $N$ waveguide-coupled cavities containing two-level emitters. We derive single- and two-photon scattering matrices for one cavity using the input--output formalism of quantum optics and for ${N>1}$ cavities using the SLH formalism, allowing us to compute the CZ gate fidelity. We consider two input Gaussian wave packets that are centered on a two-photon transition of the emitter--cavity system, and find a trade-off between optimizing the single-photon and two-photon wave packet shape. While the single-photon wave packet is preserved for large emitter--cavity coupling rates, where there is little overlap with the single-photon transitions, the two-photon wave packet acquires the necessary $\pi$ phase shift for smaller coupling rates, comparable to the cavity decay rate multiplied by the number of cavities. We find that this trade-off limits the CZ gate fidelity to approximately $60\%$ when averaging over all input states.
\end{abstract}

\maketitle


\section{Introduction}\label{sec:intro}

Photons are excellent candidates for qubits in quantum technologies due to the high-speed information transfer and long coherence times they provide. In particular, photons are the natural choice for “flying qubits" that transfer information between nodes in a quantum network. Furthermore, the ease with which single-photon operations can be implemented in different encoding methods (e.g., path, time-bin, or polarization) makes them appealing for the realization of photonic quantum computing~\cite{Kok2007}. Single-qubit operations, together with an entangling two-qubit gate such as the CZ or CNOT gate, form a universal gate set that would allow us to implement any desired logical operation on a quantum computer.

Constructing a two-photon gate is a central challenge in photonic quantum computing, originating from the absence of a direct photon--photon interaction. While it has been shown that linear optics can be used to realize a universal quantum computer~\cite{Knill2001, Kok2007}, the operation of two-photon gates is probabilistic in such schemes as it requires post-selecting on successful measurement outcomes~\cite{OBrien2003, Gasparoni2004, Zhao2005, Crespi2011, Pooley2012, Pegoraro2026}. A deterministic two-photon gate would require a nonlinear medium that can mediate an effective photon--photon interaction~\cite{Chang2014}. For example, in order to implement a photonic CZ gate, a target photon in the logical state $\ket{1}_L$ needs to acquire a $\pi$ phase shift if the control photon is also in the $\ket{1}_L$ state, realizing the transformation ${\ket{11}_L \rightarrow -\ket{11}_L}$.

A significant amount of previous work has considered using cross-phase modulation induced by a $\chi^{(3)}$ Kerr nonlinearity to implement a controlled-phase gate. Here, the presence of a control pulse changes the refractive index of the medium, which imparts a phase shift on a subsequent target pulse passing through the medium. In addition to theoretical work~\cite{Schmidt1996, Harris1999, Petrosyan2004, Ottaviani2006, Zhu2010, He2012, Copetudo2026}, a range of cross-phase modulation experiments have been performed with cold atoms~\cite{Kang2003, Chen2006, Lo2010, Lo2011, Firstenberg2013, Feizpour2015, Liu2016, Tiarks2016, Tiarks2019} and atomic vapors~\cite{Matsko2003, Sagona-Stophel2020, Davis2024}, including atomic ensembles trapped in cavities~\cite{Hickman2015, Beck2016} and photonic crystal fibers~\cite{Venkataraman2013}. As the Kerr nonlinearity is proportional to the intensity of the pulses, it is generally very weak at the single-photon level~\cite{Matsuda2009}, which means that generating an appreciable phase shift per photon requires cascading many interactions~\cite{Kuriakose2022} (larger phase shifts can be achieved in the microwave domain using superconducting qubits~\cite{Rebic2009, Hoi2013}). In addition, it has been shown that even if Kerr nonlinearities could produce the conditional $\pi$ phase shift on a single photon, which is required for the CZ gate, the associated phase noise would prevent the gate fidelity from being useful for quantum computation~\cite{Shapiro2006, Gea-Banacloche2010, Dove2014}.

Another approach to controlled phase shifts involves scattering photons from emitter--cavity systems~\cite{Duan2004, Koshino2010, Johne2012, Chudzicki2013, Krastanov2022, Tian2024, Tang2024, Turchette1995, Fushman2008, Volz2014, Hacker2016}. By utilizing the strong coupling between a quantum emitter and a cavity mode, the presence of one photon in the cavity can strongly affect the state of a second incident photon, such that conditional phase shifts approaching $\pi$ can be achieved. In the experiments in Refs.~\cite{Turchette1995, Fushman2008, Volz2014, Hacker2016} the polarization degree of freedom is used, where different photon polarizations are selectively coupled to a resonator to achieve a phase shift conditioned on the polarization of a probe photon.

Photon scattering from quantum emitters is a promising route to implementing photon--photon gates~\cite{Ralph2015, Schrinski2022, Pettersson2026}, but the light--matter interaction generates spectral entanglement that distorts the photon wave packet, limiting the gate fidelity~\cite{Nysteen2017, Xu2013}. This is due to the two-photon scattering matrix, which, for a system coupled to a chiral waveguide (with photons propagating in one direction), has the general form
\begin{align}\label{eq:S2}
\begin{split}
    S_{p_1p_2k_1k_2} =&\; t(k_1)t(k_2)\left[ \delta(p_1-k_1)\delta(p_2-k_2)\right.\\
    &\left.\hspace{0.6in}+ \delta(p_1-k_2)\delta(p_2-k_1) \right]\\
    &+ iC_{p_1p_2k_1k_2} \delta(p_1+p_2-k_1-k_2),
\end{split}
\end{align}
where $k_1$ and $k_2$ denote the input photon frequencies, $p_1$ and $p_2$ denote the output photon frequencies, $t(k)$ is the single-photon transmission coefficient, and $C_{p_1p_2k_1k_2}$ is the nonlinear two-photon term responsible for frequency mixing. In Ref.~\cite{Xu2013}, Xu {\it{et al.}} showed that spectral entanglement is unavoidable in localized systems, where the delta function $\delta(p_1+p_2-k_1-k_2)$ in the nonlinear part of the scattering matrix cannot be written in terms of products of delta functions that conserve the individual photon frequencies.

Previous authors have considered different ways to overcome the wave packet distortion arising from spectral entanglement. For example, using $N$ interaction sites (e.g., an array of $N$ quantum emitters) as opposed to a single localized interaction can increase the CZ gate fidelity close to unity in the limit of large $N$~\cite{Brod2016_1, Brod2016_2, Konyk2019, Schrinski2022, Levy-Yeyati2024}. Physically, this occurs because the translational invariance of such a system enables the individual photon frequencies to be conserved, hence eliminating the spectral entanglement~\cite{Brod2016_2}. Other approaches have remained with one quantum emitter and instead utilized additional optics that aims to preserve the wave packet shape, such as sum frequency generation and a gradient echo memory that inverts the pulse shape~\cite{Ralph2015}, or temporal phase shifters~\cite{Pettersson2026}. Cavity-based approaches have considered time-dependent cavity coupling rates that provide active control over the shape of the photon wave packet extracted from the cavity~\cite{Heuck2020, Heuck2020_2, Krastanov2022, Hassan2023}. While successful, these methods introduce difficulties in the experimental realization, such as requiring waveguides with different group velocities~\cite{Levy-Yeyati2024}, counter-propagating (instead of co-propagating) photons~\cite{Brod2016_1, Schrinski2022}, time-dependent coupling rates~\cite{Krastanov2022}, complex optical components~\cite{Ralph2015, Pettersson2026}, and different qubit encoding methods such as frequency- or polarization-encoded qubits~\cite{Duan2004, Chen2021}, which are generally harder to control and manipulate in an integrated on-chip architecture compared to path- or time-bin-encoded qubits.

\begin{figure}
    \centering
    \includegraphics[width=\linewidth]{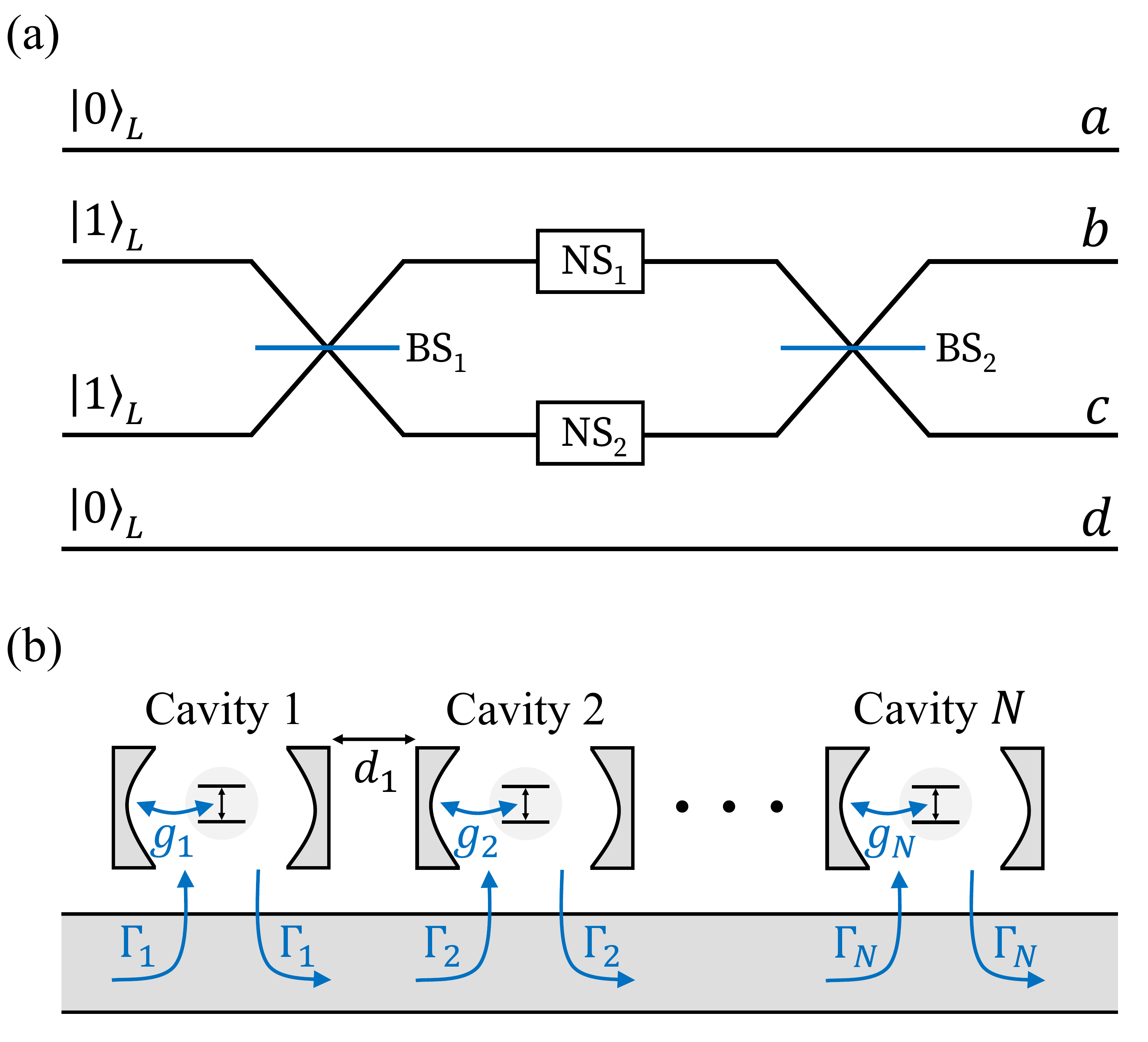}
    \caption{(a) CZ gate in the KLM scheme, implemented with two 50:50 beam splitters (BS$_1$ and BS$_2$) and two nonlinear sign gates (NS$_1$ and NS$_2$). The top (bottom) two paths correspond to the first (second) photonic qubit. (b) Our implementation of the NS gates, each consisting of $N$ waveguide-coupled cavities with two-level emitters. For cavity $j$, the emitter--cavity coupling rate is $g_j$ and the cavity decay rate is $\Gamma_j$. The separation between cavities $j$ and ${j+1}$ is $d_j$. Both NS gates are assumed to be identical.}
    \label{fig:CZ_gate_diagram}
\end{figure}

In this paper, we analyze the performance of two-level emitters in cavities as photonic CZ gates [see Fig.~\ref{fig:CZ_gate_diagram}(b)]. We consider a simple situation in which two identical photons are co-propagating through a waveguide coupled to the cavities, and there is no active control of the emitter or cavity parameters. The photonic qubits are dual-rail/path-encoded, which is favored by integrated nanophotonic platforms where the logical states $\ket{0}_L$ and $\ket{1}_L$ correspond to different waveguide channels. Similarly to Ref.~\cite{Nysteen2017}, where scattering from a two-level emitter was considered, we find a trade-off between optimizing the single-photon and two-photon components in the scattering process. While preserving the single-photon wave packets requires a large emitter--cavity coupling rate, where the photons are off-resonant with the single-photon transitions of the emitter--cavity system, the two-photon $\pi$ phase shift occurs for smaller emitter--cavity coupling rates, comparable to the cavity decay rate multiplied by the number of cavities. This trade-off limits the average gate fidelity to approximately $60\%$ for Gaussian input wave packets, and remains roughly constant for different numbers of cavities (despite spectral entanglement being reduced with more cavities, single-photon distortion prevents the fidelity from increasing significantly). This fidelity provides a baseline figure for what can be achieved with a simple passive emitter--cavity system, and future work may explore more complex setups with, e.g., multi-level emitters or time-dependent coupling rates, to try to improve on this fidelity.

The rest of this paper is organized as follows. In Section~\ref{sec:theory}, we present our CZ gate architecture and derive the single- and two-photon scattering matrices for the emitter--cavity system, which we use to calculate the CZ gate fidelity. Next, in Section~\ref{sec:results}, we present the fidelity results as a function of the input wave packet width, the emitter--cavity coupling rates, and the number of cavities. We also show the spectral profiles of the output one- and two-photon wave packets and look at different input states to demonstrate how the single- and two-photon components of the input state contribute to the fidelity. We then discuss how the fidelity is affected by photon loss and disorder in the cavity system. We conclude with a summary and outlook in Section~\ref{sec:conclusion}.


\section{Theory}\label{sec:theory}

In this section we present the fidelity equations that we use to quantify the performance of our CZ gate (Section~\ref{subsec:theory_fidelity}). The fidelity depends on single- and two-photon scattering matrices, which we derive for one emitter--cavity system in Section~\ref{subsec:theory_one_cavity} using the input--output formalism~\cite{Gardiner1985, Rephaeli2012}, and then extend the results to $N$ cavities in Section~\ref{subsec:theory_N_cavities} using the SLH formalism~\cite{Gough2009, Combes2017}.

\subsection{CZ gate fidelity}\label{subsec:theory_fidelity}

We consider the CZ gate from the KLM scheme~\cite{Knill2001}, shown in Fig.~\ref{fig:CZ_gate_diagram}(a). The setup consists of two dual-rail-encoded (path) qubits, two 50:50 beam splitters, and two nonlinear sign (NS) gates. The role of the NS gates is to flip the sign of two-photon states, while leaving zero- and one-photon states unchanged. In the ideal case, we therefore have
\begin{equation}\label{eq:NS}
    c_0\ket{0} + c_1\ket{1} + c_2\ket{2} \xrightarrow[]{\text{NS}} c_0\ket{0} + c_1\ket{1} - c_2\ket{2}
\end{equation}
in the photon number basis. Hence, when a single photon is present in one of the $\ket{1}_L$ paths in Fig.~\ref{fig:CZ_gate_diagram}, no change occurs to the state in the logical basis as the second beam splitter simply undoes the transformation of the first beam splitter. However, when both photons are in the $\ket{1}_L$ paths, they leave the first beam splitter at the same output port due to Hong-Ou-Mandel interference~\cite{Hong1987, Kok2007}. This means that the photons pass through the same NS gate, acquiring a minus sign as in Eq.~(\ref{eq:NS}). The second beam splitter then separates the photons back into their $\ket{1}_L$ paths, producing the desired transformation ${\ket{11}_L \rightarrow -\ket{11}_L}$. For a general input state
\begin{equation}\label{eq:input}
    \ket{\psi_{\text{in}}} = \alpha\ket{0,0}_L + \beta\ket{0,1}_L + \gamma\ket{1,0}_L + \delta\ket{1,1}_L
\end{equation}
(with the normalization condition $|\alpha|^2 + |\beta|^2 + |\gamma|^2 + |\delta|^2 = 1$), the ideal CZ gate output state is therefore
\begin{align}\label{eq:CZ_output_ideal}
\begin{split}
    \ket{\psi_{\text{out}}} =&\; U_{\text{CZ}}\ket{\psi_{\text{in}}} \\
    =&\; U_{\text{BS}_2}\left( U_{\text{NS}_1} \otimes U_{\text{NS}_2} \right) U_{\text{BS}_1}\ket{\psi_{\text{in}}} \\
    =&\;\alpha\ket{0,0}_L + \beta\ket{0,1}_L + \gamma\ket{1,0}_L - \delta\ket{1,1}_L,
\end{split}
\end{align}
where $U_{\text{BS}_1}$ and $U_{\text{BS}_2}$ are the unitaries corresponding to the first and second beam splitters, respectively, and $U_{\text{NS}_1}$ and $U_{\text{NS}_2}$ are the unitaries corresponding to the two NS gates that realize the transformation in Eq.~(\ref{eq:NS}) [see Fig.~\ref{fig:CZ_gate_diagram}(a)].

If we try to implement the nonlinear operation in Eq.~(\ref{eq:NS}) via a scattering process, then in general we will have
\begin{equation}
    c_0\ket{0} + c_1\ket{1} + c_2\ket{2} \rightarrow c_0\ket{0} + c_1S_1\ket{1} + c_2S_2\ket{2},
\end{equation}
where $S_1$ and $S_2$ are the single- and two-photon scattering matrices that describe the interaction. Ideally, we want an interaction where ${S_1\ket{1} = \ket{1}}$ (single-photon states are unchanged) and ${S_2\ket{2} = -\ket{2}}$ (two-photon states acquire a $\pi$ phase shift). However, processes like spontaneous emission loss into non-guided modes and spectral entanglement will prevent the NS gate operation from being realized exactly, leading to the actual output state
\begin{equation}\label{eq:CZ_output_actual}
    |\tilde{\psi}_{\text{out}}\rangle = \tilde{U}_{\text{CZ}}\ket{\psi_{\text{in}}} = U_{\text{BS}_2}\left( S_{\text{NS}_1} \otimes S_{\text{NS}_2} \right) U_{\text{BS}_1}\ket{\psi_{\text{in}}},
\end{equation}
where $S_{\text{NS}_1}$ and $S_{\text{NS}_2}$ are the scattering matrices of the physical systems that are being used to implement the NS operation. For simplicity, in what follows we will assume that the two NS gates are identical, such that ${S_{\text{NS}_2} = S_{\text{NS}_1}}$.

We can define the fidelity of a proposed CZ gate for a given input state as the modulus-squared of the overlap between the ideal output state in Eq.~(\ref{eq:CZ_output_ideal}) and the actual output state in Eq.~(\ref{eq:CZ_output_actual}):
\begin{align}\label{eq:F_CZ}
\begin{split}
    F_{\text{CZ}} =&\; |\langle \psi_{\text{out}}| \tilde{\psi}_{\text{out}} \rangle |^2\\
    =&\; \left| \bra{\psi_{\text{in}}} U_{\text{BS}_1}^{\dagger} \left(  U_{\text{NS}_1}^{\dagger} S_{\text{NS}_1} \otimes U_{\text{NS}_2}^{\dagger} S_{\text{NS}_2} \right) U_{\text{BS}_1} \ket{\psi_{\text{in}}} \right|^2.
\end{split}
\end{align}
We consider the general input state $\ket{\psi_{\text{in}}}$ in Eq.~(\ref{eq:input}) to be two identical wave packets $f(\omega)$ in a superposition of their respective paths [$a$ and $b$ for qubit 1, $c$ and $d$ for qubit 2; see Fig.~\ref{fig:CZ_gate_diagram}(a)], such that the basis states can be written in terms of mode operators as
\begin{align}\label{eq:basis}
\begin{split}
    \ket{0,0}_L =& \int {\rm d}\omega \int {\rm d}\omega' f(\omega) f(\omega') a^{\dagger}(\omega) d^{\dagger}(\omega')\ket{0,0,0,0}_{abcd},\\
    \ket{0,1}_L =& \int {\rm d}\omega \int {\rm d}\omega' f(\omega) f(\omega') a^{\dagger}(\omega) c^{\dagger}(\omega')\ket{0,0,0,0}_{abcd},\\
    \ket{1,0}_L =& \int {\rm d}\omega \int {\rm d}\omega' f(\omega) f(\omega') b^{\dagger}(\omega) d^{\dagger}(\omega')\ket{0,0,0,0}_{abcd},\\
    \ket{1,1}_L =& \int {\rm d}\omega \int {\rm d}\omega' f(\omega) f(\omega') b^{\dagger}(\omega) c^{\dagger}(\omega')\ket{0,0,0,0}_{abcd},
\end{split}
\end{align}
where $a^{\dagger}(\omega)$ is the creation operator for photons with frequency $\omega$ in path $a$ (etc. for the other paths), and $\ket{0,0,0,0}_{abcd}$ is the vacuum state, with no photons in the four paths. With the input state in Eq.~(\ref{eq:input}) and the basis states defined above, we find that the CZ gate fidelity in Eq.~(\ref{eq:F_CZ}) evaluates to
\begin{widetext}
\begin{equation}\label{eq:F_CZ_S}
    F_{\text{CZ}} = \left| |\alpha|^2 + \left( |\beta|^2 + |\gamma|^2 \right) \int {\rm d}p \int {\rm d}k\; S_{pk} f^*(p)f(k) - \frac{|\delta|^2}{2} \int {\rm d}p_1 \int {\rm d}p_2 \int {\rm d}k_1 \int {\rm d}k_2\; S_{p_1p_2k_1k_2} f^*(p_1) f^*(p_2) f(k_1) f(k_2) \right|^2,
\end{equation}
\end{widetext}
as shown in Appendix~\ref{app:fidelity}. Here $S_{pk}$ and $S_{p_1p_2k_1k_2}$ are the single-photon and two-photon scattering matrix elements for the NS gates, respectively. Substituting in ${S_{pk} = t(k)\delta(p-k)}$ and $S_{p_1p_2k_1k_2}$ from Eq.~(\ref{eq:S2})~\cite{Xu2013} allows us to write $F_{\text{CZ}}$ in the form
\begin{equation}\label{eq:F_CZ_I}
    F_{\text{CZ}} = \left| |\alpha|^2 + \left( |\beta|^2 + |\gamma|^2 \right) I_1 - |\delta|^2I_1^2 - \frac{i}{2}|\delta|^2I_2 \right|^2,
\end{equation}
where
\begin{equation}\label{eq:I_1}
    I_1 = \int {\rm d}\omega\; t(\omega) |f(\omega)|^2
\end{equation}
is the single-photon contribution, and
\begin{align}\label{eq:I_2}
\begin{split}
    I_2 =& \int {\rm d}\omega \int {\rm d}\omega' \int {\rm d}\omega''\; C(\omega, \omega', \omega'')\\
    & \times f^*(\omega) f^*(\omega') f(\omega'') f(\omega + \omega' - \omega'') 
\end{split}
\end{align}
is the two-photon contribution (see Appendix~\ref{app:fidelity}).

We also calculate the average CZ gate fidelity by integrating $F_{\text{CZ}}$ over the Haar measure ${\rm d}\mu$:
\begin{align}\label{eq:avg_F_CZ}
\begin{split}
    \bar{F}_{\text{CZ}} =& \int F_{\text{CZ}}(\alpha, \beta, \gamma, \delta) {\rm d}\mu\\
    =& \frac{1}{10}\Biggl[ \frac{3}{4} + \frac{1}{2}\left( I_1 + I_1^* \right) + 2|I_1|^2\\
    &\hspace{0.225in}+ \left| I_1^2 - I_1 + \frac{i}{2}I_2 - \frac{1}{2} \right|^2\Biggr],
\end{split}
\end{align}
as shown in Appendix~\ref{app:average_fidelity}. Note that, in the ideal case where the single-photon wave packet is unchanged by the NS gates, ${t(\omega) = 1}$ and hence ${I_1 = 1}$ due to the normalization of the wave packets. This means that if ${I_2 = 4i}$, ${F_{\text{CZ}} = 1}$ and ${\bar{F}_{\text{CZ}} = 1}$, which corresponds to a perfect CZ gate (the fidelity is one for all input states, independent of the choice of $\alpha$, $\beta$, $\gamma$, and $\delta$).

In order to calculate the fidelity $F_{\text{CZ}}$ for a specific state [Eq.~(\ref{eq:F_CZ_I})] and the average fidelity $\bar{F}_{\text{CZ}}$, we need to calculate the integrals $I_1$ in Eq.~(\ref{eq:I_1}) and $I_2$ in Eq.~(\ref{eq:I_2}). We therefore need to calculate the single-photon transmission coefficient $t(\omega)$ and the two-photon term $C(\omega,\omega',\omega'')$ for our implementation of the NS gates. In Section~\ref{subsec:theory_one_cavity}, we derive $t(\omega)$ and $C(\omega,\omega',\omega'')$ for a single emitter--cavity system using the input--output formalism~\cite{Gardiner1985, Rephaeli2012}, and then in Section~\ref{subsec:theory_N_cavities} we extend the results to the $N$-cavity system in Fig.~\ref{fig:CZ_gate_diagram}(b) using the SLH formalism~\cite{Gough2009, Combes2017}.


\subsection{Single emitter--cavity system: Input--output formalism}\label{subsec:theory_one_cavity}

\begin{figure}
    \centering
    \includegraphics[width=0.8\linewidth]{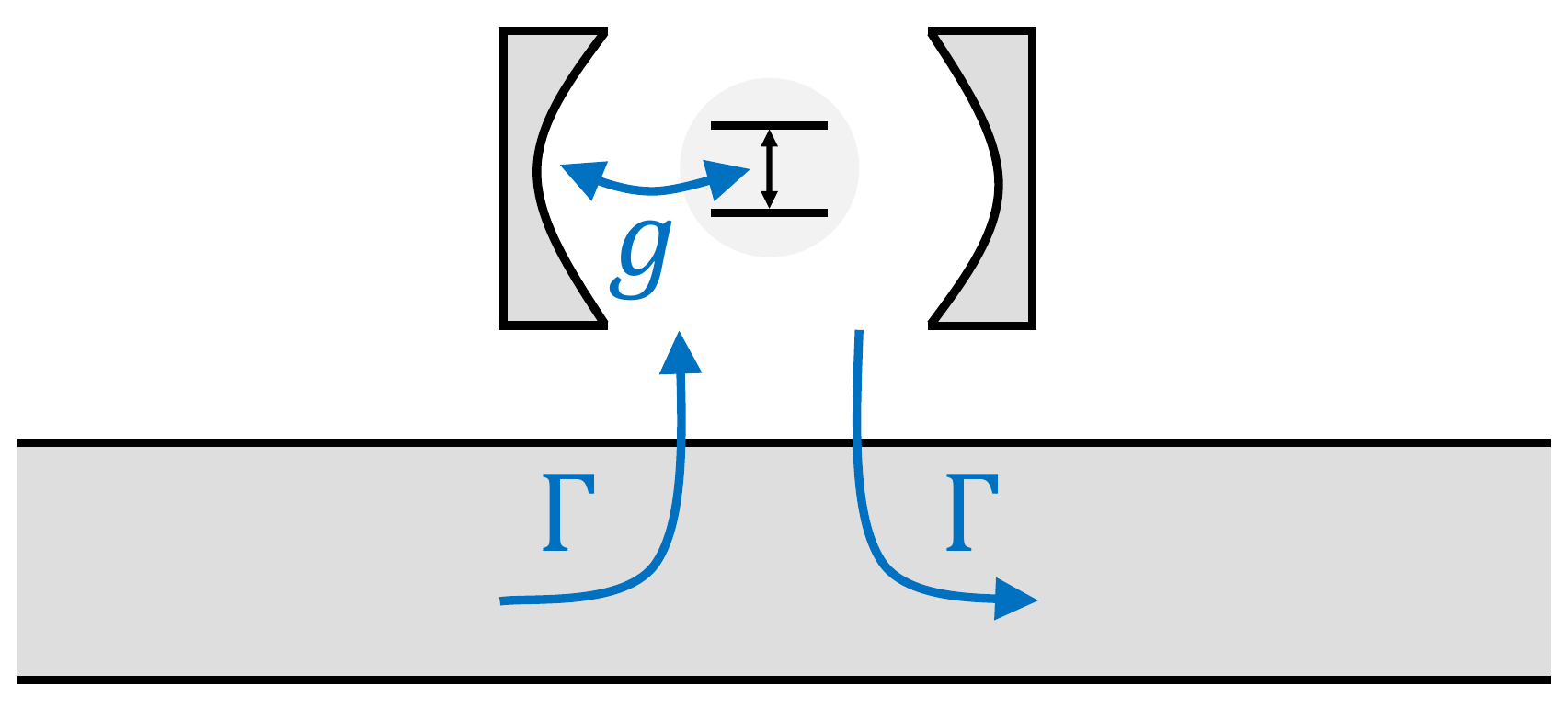}
    \caption{A single emitter--cavity system chirally coupled to a waveguide. The emitter--cavity coupling rate is $g$ and the cavity decay rate into the waveguide is $\Gamma$. We assume that $g$ and $\Gamma$ are real, and that the emitter is on resonance with the cavity.}
    \label{fig:one_cavity}
\end{figure}

In this section we derive the single- and two-photon scattering matrices for a single waveguide-coupled cavity with a two-level emitter, as shown in Fig.~\ref{fig:one_cavity}. These results were derived by Rephaeli and Fan in Ref.~\cite{Rephaeli2012}, but we include the derivation here for completeness. More details of the single- and two-photon calculations are provided in Appendices~\ref{app:one_cavity_one_photon} and \ref{app:one_cavity_two_photons}, respectively.


\subsubsection{Hamiltonian}

As shown in Fig.~\ref{fig:one_cavity}, we consider a cavity chirally coupled to a waveguide, where the cavity contains a two-level emitter. The emitter--cavity coupling rate is $g$ and the cavity decay rate into the waveguide is $\Gamma$. We assume that the emitter and cavity are in resonance, such that the transition frequency of the emitter and the resonance frequency of the cavity are both equal to $\Omega$. The Hamiltonian of the system in frequency space is given by (${\hbar = 1}$)~\cite{Duda2024}
\begin{align}\label{eq:H}
\begin{split}
    H =&\; \Omega\left( \sigma^+\sigma^- + c^\dagger c \right) + \int \omega a^\dagger(\omega)a(\omega){\rm d}\omega\\
    &+ g\left(\sigma^+c + \sigma^-c^\dagger\right) + \sqrt{\frac{\Gamma}{2\pi}} \int \left[a^\dagger(\omega)c + a(\omega)c^\dagger \right]{\rm d}\omega,
\end{split}
\end{align}
where ${\sigma^+ = \ketbra{e}{g}}$ and ${\sigma^- = \ketbra{g}{e}}$ are the raising and lowering operators for the emitter, respectively, with $\ket{g}$ being the ground state and $\ket{e}$ the excited state. In addition, $c^\dagger$ and $c$ are the creation and annihilation operators for the cavity mode, and $a^\dagger(\omega)$ and $a(\omega)$ are the waveguide mode operators for photons with frequency $\omega$, satisfying the commutation relations ${[c,c^{\dagger}] = 1}$ and ${[a(\omega), a^\dagger(\omega')] = \delta(\omega-\omega')}$, respectively. We have assumed that $g$ and $\Gamma$ are both real for simplicity, and that the cavity--waveguide coupling is independent of frequency.


\subsubsection{Single-photon scattering}\label{subsubsec:theory_one_cavity_one_photon}

The single-photon scattering matrix elements are given by
\begin{equation}\label{eq:S1}
    S_{pk} = \bra{p}S\ket{k} = \braket{p^-}{k^+} = \bra{0}a_{\text{out}}(p)a_{\text{in}}^{\dagger}(k)\ket{0},
\end{equation}
where $k$ and $p$ denote the input and output photon frequencies, respectively. The state ${\ket{k^+} = a_{\text{in}}^{\dagger}(k)\ket{0}}$ is a scattering eigenstate that evolves from the free state $\ket{k}$ in the distant past, and ${\ket{p^-} = a_{\text{out}}^{\dagger}(p)\ket{0}}$ is a scattering eigenstate that evolves into the free state $\ket{p}$ in the distant future~\cite{Fan2010}.

To calculate $S_{pk}$ for the system in Fig.~\ref{fig:one_cavity}, we need the corresponding input--output relation (see Appendix~\ref{app:one_cavity_one_photon}):
\begin{equation}\label{eq:input-output}
    a_{\text{out}}(t) = a_{\text{in}}(t) - i\sqrt{\Gamma}c(t).
\end{equation}
We perform a Fourier transform of Eq.~(\ref{eq:input-output}) from time to frequency, and substitute the result into Eq.~(\ref{eq:S1}) to obtain
\begin{align}\label{eq:S1_2}
\begin{split}
    S_{pk} =&\; \bra{0} \left[ a_{\text{in}}(p) - i\sqrt{\Gamma}c(p) \right] a_{\text{in}}^{\dagger}(k)\ket{0}\\
    =&\; \delta(p-k) - i\sqrt{\Gamma}\matrixel{0}{c(p)}{k^+},
\end{split}
\end{align}
where we used the commutator $[a_{\text{in}}(p), a_{\text{in}}^\dagger(k)] = {\delta(p-k)}$. We now calculate the matrix element ${\matrixel{0}{c(p)}{k^+}}$ by solving the Heisenberg equation for the cavity operator $c(t)$:
\begin{align}\label{eq:Heisenberg_cavity}
\begin{split}
    \frac{\rm d}{{\rm d}t}c(t) =&\; i[H, c(t)]\\
    =& -i\left( \Omega - \frac{i}{2}\Gamma \right) c(t) - ig\sigma^-(t) - i\sqrt{\Gamma}a_{\text{in}}(t),
\end{split}
\end{align}
as shown in Appendix~\ref{app:one_cavity_one_photon}. Again Fourier-transforming from time to frequency leads to
\begin{equation}
    -ipc(p) = -i\left( \Omega - \frac{i}{2}\Gamma \right) c(p) - ig\sigma^-(p) - i\sqrt{\Gamma}a_{\text{in}}(p),
\end{equation}
which can be rearranged to
\begin{equation}\label{eq:Heisenberg_c_p}
    \left[ \Delta(p) + \frac{i}{2}\Gamma \right] c(p) = g\sigma^-(p) + \sqrt{\Gamma}a_{\text{in}}(p),
\end{equation}
where ${\Delta(p) = p - \Omega}$ is the detuning between a photon of frequency $p$ and the emitter/cavity. Pre-multiplying each term in Eq.~(\ref{eq:Heisenberg_c_p}) by $\bra{0}$ and post-multiplying by $\ket{k^+}$ gives
\begin{equation}\label{eq:Heisenberg_c_p_2}
    \left[ \Delta(p) + \frac{i}{2}\Gamma \right] \matrixel{0}{c(p)}{k^+} = g\matrixel{0}{\sigma^-(p)}{k^+} + \sqrt{\Gamma}\delta(p-k),
\end{equation}
where we used ${\matrixel{0}{a_{\text{in}}(p)}{k^+} = \delta(p-k)}$. We now need to find $\matrixel{0}{\sigma^-(p)}{k^+}$ by solving the Heisenberg equation for the emitter operator $\sigma^-(t)$:
\begin{align}\label{eq:Heisenberg_emitter}
\begin{split}
    \frac{\rm d}{{\rm d}t}\sigma^-(t) =&\; i[H, \sigma^-(t)]\\
    =&\; -i\Omega\sigma^-(t) + ig\sigma_z(t)c(t),
\end{split}
\end{align}
where $\sigma_z(t) = e^{iHt}\sigma_z(0)e^{-iHt}$ with $\sigma_z(0) = \ketbra{e}{e} - \ketbra{g}{g}$. Pre-multiplying by $\bra{0}$ and post-multiplying by $\ket{k^+}$ leads to
\begin{equation}
    \frac{\rm d}{{\rm d}t}\matrixel{0}{\sigma^-(t)}{k^+} = -i\Omega\matrixel{0}{\sigma^-(t)}{k^+} - ig \matrixel{0}{c(t)}{k^+},
\end{equation}
where we used ${\bra{0}\sigma_z(t) = -\bra{0}}$. We can now Fourier transform from time to frequency,
\begin{equation}
    -ip\matrixel{0}{\sigma^-(p)}{k^+} = -i\Omega\matrixel{0}{\sigma^-(p)}{k^+} - ig \matrixel{0}{c(p)}{k^+},
\end{equation}
and obtain $\matrixel{0}{\sigma^-(p)}{k^+}$ in terms of $\matrixel{0}{c(p)}{k^+}$:
\begin{equation}\label{eq:sigma_p_matrix_el}
    \matrixel{0}{\sigma^-(p)}{k^+} = \frac{g}{\Delta(p)}\matrixel{0}{c(p)}{k^+}.
\end{equation}
Substituting this result into Eq.~(\ref{eq:Heisenberg_c_p_2}) gives us
\begin{equation}\label{eq:c_p_matrix_el}
    \matrixel{0}{c(p)}{k^+} = \frac{\Delta(p)\sqrt{\Gamma}}{\Delta^2(p) - g^2 + \frac{i}{2}\Delta(p)\Gamma} \delta(p-k),
\end{equation}
which can be substituted into Eq.~(\ref{eq:S1_2}), leading to the single-photon scattering matrix
\begin{equation}
    S_{pk} = t(p)\delta(p-k),
\end{equation}
where
\begin{equation}\label{eq:t_one_cavity}
    t(p) = \frac{\Delta^2(p) - g^2 - \frac{i}{2}\Delta(p)\Gamma}{\Delta^2(p) - g^2 + \frac{i}{2}\Delta(p)\Gamma}
\end{equation}
is the single-photon transmission coefficient, in consistency with Ref.~\cite{Rephaeli2012}.

To simplify the calculation of the two-photon scattering matrix in the next section, we write Eq.~(\ref{eq:c_p_matrix_el}) in the form
\begin{equation}\label{eq:c_p_matrix_el_3}
     \matrixel{0}{c(p)}{k^+} = s_c(p) \delta(p-k),
\end{equation}
where
\begin{equation}\label{eq:c_p_matrix_el_2}
    s_c(p) = \frac{\Delta(p)\sqrt{\Gamma}}{\Delta^2(p) - g^2 + \frac{i}{2}\Delta(p)\Gamma}
\end{equation}
is the single-photon cavity excitation amplitude. Similarly, we can write Eq.~(\ref{eq:sigma_p_matrix_el}) in the form
\begin{equation}\label{eq:sigma_p_matrix_el_2}
     \matrixel{0}{\sigma^-(p)}{k^+} = s_e(p) \delta(p-k),
\end{equation}
where
\begin{equation}\label{eq:s_e}
    s_e(p) = \frac{g\sqrt{\Gamma}}{\Delta^2(p) - g^2 + \frac{i}{2}\Delta(p)\Gamma}
\end{equation}
is the single-photon emitter excitation amplitude.


\subsubsection{Two-photon scattering}\label{subsubsec:theory_one_cavity_two_photons}

The two-photon scattering matrix elements are given by
\begin{align}\label{eq:S2_2}
\begin{split}
    S_{p_1p_2k_1k_2} =&\; \bra{p_1p_2}S\ket{k_1k_2} = \braket{p_1p_2^-}{k_1k_2^+}\\
    =&\; \bra{0}a_{\text{out}}(p_1)a_{\text{out}}(p_2)a_{\text{in}}^{\dagger}(k_1)a_{\text{in}}^{\dagger}(k_2)\ket{0}.
\end{split}
\end{align}
We will calculate these matrix elements to find the two-photon term $C_{p_1p_2k_1k_2}$ in Eq.~(\ref{eq:S2}) for the system in Fig.~\ref{fig:one_cavity}.

First, we substitute in the Fourier transform of the input--output relation in Eq.~(\ref{eq:input-output}) for $a_{\text{out}}(p_2)$ to obtain
\begin{align}\label{eq:S2_c_p2}
\begin{split}
    S_{p_1p_2k_1k_2} =&\; \bra{p_1^-}a_{\text{out}}(p_2)\ket{k_1k_2^+}\\
    =&\; \bra{p_1^-}a_{\text{in}}(p_2)\ket{k_1k_2^+} - i\sqrt{\Gamma} \bra{p_1^-}c(p_2)\ket{k_1k_2^+}.
\end{split}
\end{align}
Using the bosonic commutation relation of the input operators, we can write the first term in the last line as
\begin{equation}\label{eq:a_in_p2}
    \bra{p_1^-}a_{\text{in}}(p_2)\ket{k_1k_2^+} = S_{p_1k_2}\delta(p_2-k_1) + S_{p_1k_1}\delta(p_2-k_2),
\end{equation}
where ${S_{p_1k_2} = \braket{p_1^-}{k_2^+}}$ and ${S_{p_1k_1} = \braket{p_1^-}{k_1^+}}$ are single-photon scattering matrix elements, defined as in Eq.~(\ref{eq:S1}).

To calculate $\bra{p_1^-}c(p_2)\ket{k_1k_2^+}$, we need to use the Heisenberg equation for $c(t)$, derived in the single-photon calculation above. Returning to Eq.~(\ref{eq:Heisenberg_c_p}), relabeling $p$ with $p_2$, and pre-multiplying by $\bra{p_1^-}$ and post-multiplying by $\ket{k_1k_2^+}$ leads to
\begin{align}\label{eq:c_p2_matrix_el}
\begin{split}
    \biggl[& \Delta(p_2) + \frac{i}{2}\Gamma \biggr] \bra{p_1^-}c(p_2)\ket{k_1k_2^+}\\
    &= g\bra{p_1^-}\sigma^-(p_2)\ket{k_1k_2^+}\\
    &\hspace{0.15in}+ \sqrt{\Gamma}\left[ S_{p_1k_2}\delta(p_2-k_1) + S_{p_1k_1}\delta(p_2-k_2) \right],
\end{split}
\end{align}
where we used the result in Eq.~(\ref{eq:a_in_p2}) to obtain the last line. We now need to find $\bra{p_1^-}\sigma^-(p_2)\ket{k_1k_2^+}$ using the Heisenberg equation for $\sigma^-(t)$. Pre-multiplying Eq.~(\ref{eq:Heisenberg_emitter}) by $\bra{p_1^-}$ and post-multiplying by $\ket{k_1k_2^+}$ gives
\begin{align}\label{eq:Heisenberg_emitter_2}
\begin{split}
     \frac{\rm d}{{\rm d}t}\bra{p_1^-}\sigma^-(t)\ket{k_1k_2^+} =& -i\Omega\bra{p_1^-}\sigma^-(t)\ket{k_1k_2^+}\\
     &+ ig\bra{p_1^-}\sigma_z(t)c(t)\ket{k_1k_2^+}.
\end{split}
\end{align}
The last term in Eq.~(\ref{eq:Heisenberg_emitter_2}) is the source of the two-photon nonlinearity (in the single-photon calculation we could use ${\bra{0}\sigma_z(t) = -\bra{0}}$ to reduce this term to a single-operator matrix element, but this cannot be done in the two-photon case). Using $\sigma_z(t) = 2\sigma^+(t)\sigma^-(t) - 1$ leads to
\begin{align}\label{eq:Heisenberg_sigma_2}
\begin{split}
    \frac{\rm d}{{\rm d}t}\bra{p_1^-}\sigma^-(t)\ket{k_1k_2^+} =& -i\Omega\bra{p_1^-}\sigma^-(t)\ket{k_1k_2^+}\\
    &+ 2ig\bra{p_1^-}\sigma^+\sigma^-c(t)\ket{k_1k_2^+}\\
    &- ig \bra{p_1^-}c(t)\ket{k_1k_2^+},
\end{split}
\end{align}
where we have written $\sigma^+(t)\sigma^-(t)c(t)$ as a single time-dependent operator $\sigma^+\sigma^-c(t)$. Performing a Fourier transform from time $t$ to frequency $p_2$ gives
\begin{align}\label{eq:Heisenberg_sigma_p2}
\begin{split}
    -ip_2\bra{p_1^-}\sigma^-(p_2)\ket{k_1k_2^+} =& -i\Omega\bra{p_1^-}\sigma^-(p_2)\ket{k_1k_2^+}\\
    &+ 2ig\bra{p_1^-}\sigma^+\sigma^-c(p_2)\ket{k_1k_2^+}\\
    &- ig \bra{p_1^-}c(p_2)\ket{k_1k_2^+},
\end{split}
\end{align}
which can be rearranged for $\bra{p_1^-}\sigma^-(p_2)\ket{k_1k_2^+}$:
\begin{align}
\begin{split}
    \bra{p_1^-}\sigma^-(p_2)\ket{k_1k_2^+} =&\; \frac{g}{\Delta(p_2)} \bra{p_1^-}c(p_2)\ket{k_1k_2^+}\\
    & -\frac{2g}{\Delta(p_2)}\bra{p_1^-}\sigma^+\sigma^-c(p_2)\ket{k_1k_2^+}.
\end{split}
\end{align}
We now apply the convolution theorem to write the operator $\sigma^+\sigma^-c(p_2)$ as
\begin{equation}\label{eq:convolution_sigma}
    \sigma^+\sigma^-c(p_2) = \frac{1}{\sqrt{2\pi}} \int{\rm d}p\; \sigma^+(p) \sigma^-c(p_2+p),
\end{equation}
leading to
\begin{align}\label{eq:sigma_p2_matrix_el_conv}
\begin{split}
    &\bra{p_1^-}\sigma^-(p_2)\ket{k_1k_2^+}\\
    &= \frac{g}{\Delta(p_2)} \bra{p_1^-}c(p_2)\ket{k_1k_2^+}\\
    &\hspace{0.15in} -\frac{2}{\sqrt{2\pi}}\frac{g}{\Delta(p_2)} \int {\rm d}p \matrixel{p_1^-}{\sigma^+(p)}{0}\matrixel{0}{\sigma^-c(p_2+p)}{k_1k_2^+},
\end{split}
\end{align}
where we inserted the identity operator between $\sigma^+(p)$ and $\sigma^-c(p_2+p)$ (only the zero-excitation component of the identity contributes). Since
\begin{align}\label{eq:sigma_p_one_photon}
\begin{split}
    \matrixel{p_1^-}{\sigma^+(p)}{0} =&\; \bra{p_1^-} \left( \int {\rm d}k \ketbra{k^+}{k^+} \right)\sigma^+(p)\ket{0}\\
    =&\; \int {\rm d}k \braket{p_1^-}{k^+} \left[ \matrixel{0}{\sigma^-(p)}{k^+} \right]^\dagger\\
    =&\; \int {\rm d}k\; S_{p_1k} \left[ s_e(p) \right]^* \delta(p-k)\\
    =&\; S_{p_1p}\left[ s_e(p) \right]^*
\end{split}
\end{align}
[where we inserted an identity operator in the single-photon basis, and used the result in Eq.~(\ref{eq:sigma_p_matrix_el_2})], we have
\begin{align}\label{eq:sigma_p2_matrix_el}
\begin{split}
    &\bra{p_1^-}\sigma^-(p_2)\ket{k_1k_2^+}\\
    &= \frac{g}{\Delta(p_2)} \bra{p_1^-}c(p_2)\ket{k_1k_2^+}\\
    &\hspace{0.15in} -\frac{2}{\sqrt{2\pi}}\frac{g}{\Delta(p_2)} \int {\rm d}p\; S_{p_1p}\left[ s_e(p) \right]^*\matrixel{0}{\sigma^-c(p_2+p)}{k_1k_2^+}.
\end{split}
\end{align}
Substituting Eq.~(\ref{eq:sigma_p2_matrix_el}) into Eq.~(\ref{eq:c_p2_matrix_el}) and then the resulting equation, together with Eq.~(\ref{eq:a_in_p2}), into Eq.~(\ref{eq:S2_c_p2}), allows us to write the two-photon scattering matrix in the form
\begin{align}\label{eq:S2_linear_complete}
\begin{split}
    S_{p_1p_2k_1k_2} =&\; S_{p_1k_2}t(p_2)\delta(p_2-k_1) + S_{p_1k_1}t(p_2)\delta(p_2-k_2)\\
    &+ \frac{2i}{\sqrt{2\pi}} \frac{g^2\sqrt{\Gamma}}{\Delta^2(p_2) - g^2 + \frac{i}{2}\Delta(p_2)\Gamma}\\
    &\times\int {\rm d}p\; S_{p_1p}\left[ s_e(p) \right]^*\matrixel{0}{\sigma^-c(p_2+p)}{k_1k_2^+}\\
    =&\; S_{p_1k_2}S_{p_2k_1} + S_{p_1k_1}S_{p_2k_2}\\
    &+ \frac{2ig}{\sqrt{2\pi}} s_e(p_2) t(p_1) \left[ s_e(p_1) \right]^*\\
    &\times \matrixel{0}{\sigma^-c(p_1+p_2)}{k_1k_2^+},
\end{split}
\end{align}
where we used the result for the single-photon transmission coefficient in Eq.~(\ref{eq:t_one_cavity}), the definition of $s_e(p)$ from Eq.~(\ref{eq:s_e}), and ${S_{p_1p} = t(p_1)\delta(p_1-p)}$ to integrate over $p$. Since ${t(p_1)[s_e(p_1)]^* = s_e(p_1)}$, we have
\begin{align}\label{eq:S2_linear_complete_2}
\begin{split}
    S_{p_1p_2k_1k_2} =&\; S_{p_1k_2}S_{p_2k_1} + S_{p_1k_1}S_{p_2k_2}\\
    &+ \frac{2ig}{\sqrt{2\pi}} s_e(p_1) s_e(p_2)\matrixel{0}{\sigma^-c(p_1+p_2)}{k_1k_2^+}.
\end{split}
\end{align}
The first line in $S_{p_1p_2k_1k_2}$ above is the linear part of the scattering matrix [see Eq.~(\ref{eq:S2})], and can be written as
\begin{align}
\begin{split}
    &S_{p_1k_2}S_{p_2k_1} + S_{p_1k_1}S_{p_2k_2}\\
    &= t(k_1)t(k_2)\left[ \delta(p_1-k_1)\delta(p_2-k_2) + \delta(p_1-k_2)\delta(p_2-k_1) \right].
\end{split}
\end{align}
The last term in Eq.~(\ref{eq:S2_linear_complete_2}) is the nonlinear part that we need to calculate to find $C_{p_1p_2k_1k_2}$ in Eq.~(\ref{eq:S2}).

We now proceed to solve Heisenberg equations of operator products to find the matrix element $\matrixel{0}{\sigma^-c(p_1+p_2)}{k_1k_2^+}$ in Eq.~(\ref{eq:S2_linear_complete_2}). Additional details from the calculations are shown in Appendix~\ref{app:one_cavity_two_photons}. We start with the Heisenberg equation for $\sigma^-(t)c(t)$:
\begin{align}\label{eq:Heisenberg_sigma_c_time}
\begin{split}
    \frac{\rm d}{{\rm d}t}\sigma^-(t)c(t) =& -i\left( 2\Omega - \frac{i}{2}\Gamma \right)\sigma^-(t)c(t) + ig\sigma_z(t)c^2(t)\\
    &-i\sqrt{\Gamma}\sigma^-(t)a_{\text{in}}(t).
\end{split}
\end{align}
Pre-multiplying by $\bra{0}$, post-multiplying by $\ket{k_1k_2^+}$, and using ${\bra{0}\sigma_z(t) = -\bra{0}}$ leads to
\begin{align}
\begin{split}
    \frac{\rm d}{{\rm d}t}\matrixel{0}{\sigma^-(t)c(t)}{k_1k_2^+} =& -i\left( 2\Omega - \frac{i}{2}\Gamma \right)\matrixel{0}{\sigma^-(t)c(t)}{k_1k_2^+}\\
    &- ig\matrixel{0}{c^2(t)}{k_1k_2^+}\\
    &-i\sqrt{\Gamma}\matrixel{0}{\sigma^-(t)a_{\text{in}}(t)}{k_1k_2^+},
\end{split}
\end{align}
which becomes, after Fourier transforming to frequency,
\begin{align}\label{eq:Heisenberg_sigma_c}
\begin{split}
    -i(&p_1+p_2)\matrixel{0}{\sigma^-c(p_1+p_2)}{k_1k_2^+}\\
    =& -i\left( 2\Omega - \frac{i}{2}\Gamma \right)\matrixel{0}{\sigma^-c(p_1+p_2)}{k_1k_2^+}\\
    &- ig\matrixel{0}{c^2(p_1+p_2)}{k_1k_2^+}\\
    &-i\sqrt{\Gamma}\matrixel{0}{\sigma^-a_{\text{in}}(p_1+p_2)}{k_1k_2^+}.
\end{split}
\end{align}
The last term can be simplified using the convolution theorem:
\begin{equation}\label{eq:convolution_sigma_a_in}
    \sigma^-a_{\text{in}}(p_1+p_2) = \frac{1}{\sqrt{2\pi}} \int {\rm d}p\; \sigma^-(p)a_{\text{in}}(p_1+p_2-p),
\end{equation}
which means that
\begin{align}
\begin{split}
    &\matrixel{0}{\sigma^-a_{\text{in}}(p_1+p_2)}{k_1k_2^+}\\
    &= \frac{1}{\sqrt{2\pi}} \int {\rm d}p\; \matrixel{0}{\sigma^-(p)a_{\text{in}}(p_1+p_2-p)}{k_1k_2^+}\\
    &= \frac{1}{\sqrt{2\pi}} \left[ s_e(k_1) + s_e(k_2) \right] \delta(p_1+p_2-k_1-k_2),
\end{split}
\end{align}
where we used the commutator of the input operators and the result in Eq.~(\ref{eq:sigma_p_matrix_el_2}). Equation~(\ref{eq:Heisenberg_sigma_c}) can therefore be rearranged to give
\begin{align}\label{eq:sigma_c_matrix_el}
\begin{split}
    &\left( p_1 + p_2 - 2\Omega + \frac{i}{2}\Gamma \right) \matrixel{0}{\sigma^-c(p_1+p_2)}{k_1k_2^+}\\
    &= g \matrixel{0}{c^2(p_1+p_2)}{k_1k_2^+}\\
    &\hspace{0.15in}+ \sqrt{\frac{\Gamma}{2\pi}} \left[ s_e(k_1) + s_e(k_2) \right] \delta(p_1+p_2-k_1-k_2).
\end{split}
\end{align}
To find $\matrixel{0}{c^2(p_1+p_2)}{k_1k_2^+}$, we need to solve the Heisenberg equation for $c^2(t)$:
\begin{align}\label{eq:Heisenberg_c2_time}
\begin{split}
    \frac{\rm d}{{\rm d}t}c^2(t) =& -2i\left( \Omega - \frac{i}{2}\Gamma \right)c^2(t) - 2ig\sigma^-(t)c(t)\\
    &-2i\sqrt{\Gamma}c(t)a_{\text{in}}(t).
\end{split}
\end{align}
Again, we pre-multiply by $\bra{0}$, post-multiply by $\ket{k_1k_2^+}$, and Fourier transform from time to frequency to obtain
\begin{align}\label{eq:Heisenberg_c2}
\begin{split}
    -i(&p_1+p_2)\matrixel{0}{c^2(p_1+p_2)}{k_1k_2^+}\\
    =& -2i\left( \Omega - \frac{i}{2}\Gamma \right)\matrixel{0}{c^2(p_1+p_2)}{k_1k_2^+}\\
    &- 2ig\matrixel{0}{\sigma^-c(p_1+p_2)}{k_1k_2^+}\\
    &-2i\sqrt{\Gamma}\matrixel{0}{c\;a_{\text{in}}(p_1+p_2)}{k_1k_2^+}.
\end{split}
\end{align}
Using the convolution theorem, we have
\begin{equation}\label{eq:convolution_cavity}
    c\;a_{\text{in}}(p_1+p_2) = \frac{1}{\sqrt{2\pi}} \int {\rm d}p\; c(p)a_{\text{in}}(p_1+p_2-p),
\end{equation}
so
\begin{align}
\begin{split}
    &\matrixel{0}{c\;a_{\text{in}}(p_1+p_2)}{k_1k_2^+}\\
    &= \frac{1}{\sqrt{2\pi}} \int {\rm d}p\; \matrixel{0}{c(p)a_{\text{in}}(p_1+p_2-p)}{k_1k_2^+}\\
    &= \frac{1}{\sqrt{2\pi}} \left[ s_c(k_1) + s_c(k_2) \right] \delta(p_1+p_2-k_1-k_2),
\end{split}
\end{align}
using the commutator of the input operators and the result in Eq.~(\ref{eq:c_p_matrix_el_3}). Equation~(\ref{eq:Heisenberg_c2}) can therefore be written as
\begin{align}\label{eq:c2_matrix_el}
\begin{split}
    &\left(p_1+p_2-2\Omega+i\Gamma\right)\matrixel{0}{c^2(p_1+p_2)}{k_1k_2^+}\\
    &=2g\matrixel{0}{\sigma^-c(p_1+p_2)}{k_1k_2^+}\\
    &\hspace{0.15in}+2\sqrt{\frac{\Gamma}{2\pi}}\left[ s_c(k_1) + s_c(k_2) \right] \delta(p_1+p_2-k_1-k_2).
\end{split}
\end{align}
Equations~(\ref{eq:sigma_c_matrix_el}) and (\ref{eq:c2_matrix_el}) can be solved simultaneously for $\matrixel{0}{\sigma^-c(p_1+p_2)}{k_1k_2^+}$, and the result can be substituted into Eq.~(\ref{eq:S2_linear_complete_2}), allowing the two-photon scattering matrix to be written in the form given in Eq.~(\ref{eq:S2}), where
\begin{widetext}
\begin{equation}\label{eq:C_one_cavity}
    C_{p_1p_2k_1k_2} = \frac{g\sqrt{\Gamma}}{\pi} s_e(p_1)s_e(p_2) \left( \frac{2g\left[ s_c(k_1) + s_c(k_2)\right] + \left( k_1 + k_2 - 2\Omega + i\Gamma \right)\left[ s_e(k_1) + s_e(k_2) \right]}{(k_1+k_2-\lambda_+)(k_1+k_2-\lambda_-)} \right),
\end{equation}
\end{widetext}
with
\begin{equation}\label{eq:lambdas}
    \lambda_{\pm} = 2\Omega - \frac{3i}{4}\Gamma \pm \sqrt{2g^2 - \frac{1}{16}\Gamma^2},
\end{equation}
in consistency with Ref.~\cite{Rephaeli2012}.


\subsection{$N$ emitter--cavity systems: SLH formalism}\label{subsec:theory_N_cavities}

We now expand on the results in Section~\ref{subsec:theory_one_cavity} and derive the single-photon transmission coefficient $t(p)$ and the two-photon term $C_{p_1p_2k_1k_2}$ for $N$ emitter--cavity systems coupled to a waveguide, as shown in Fig.~\ref{fig:CZ_gate_diagram}(b).


\subsubsection{SLH parameters}

\begin{figure}[b]
    \centering
    \includegraphics[width=0.8\linewidth]{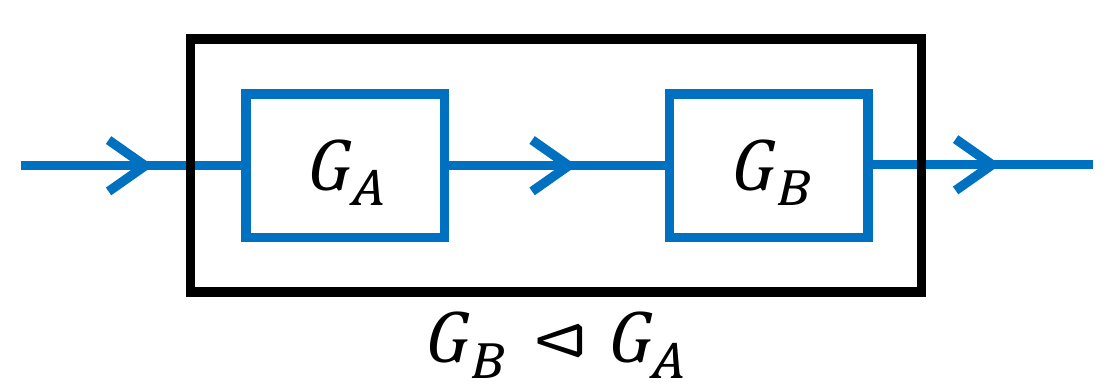}
    \caption{Illustration of the series product in Eq.~(\ref{eq:SLH_series}), where the output of system $A$ is fed into the input of system $B$.}
    \label{fig:series_product}
\end{figure}

The SLH formalism allows us to connect together multiple subsystems and derive few-photon scattering matrices for the combined system~\cite{Combes2017, Brod2016_2}. In this formalism, we describe a system with a set of three parameters:
\begin{equation}
    G = (S,L,H),
\end{equation}
where $S$ is the scattering parameter, $L$ is the coupling parameter, and $H$ is the internal system Hamiltonian. If we have two subsystems with SLH parameters $G_A = (S_A,L_A,H_A)$ and $G_B = (S_B,L_B,H_B)$, then the SLH parameters of the combined system, where the output of system $A$ is fed into the input of system $B$ (see Fig.~\ref{fig:series_product}), are given by the series product~\cite{Gough2009}:
\begin{align}\label{eq:SLH_series}
\begin{split}
   G_B \lhd G_A =&\; (S_B,L_B,H_B) \lhd (S_A,L_A,H_A)\\
   =&\; \biggl( S_BS_A, L_B + S_BL_A,\\
   &\hspace{0.15in}H_A + H_B + \frac{1}{2i}\left( L_B^\dagger S_B L_A - L_A^\dagger S_B^\dagger L_B \right) \biggr).
\end{split}
\end{align}

Using the series product, we can obtain the SLH parameters for the full $N$-cavity system in Fig.~\ref{fig:CZ_gate_diagram}(b). To do this, we need the SLH parameters for cavity $j$~\cite{Combes2017},
\begin{align}\label{eq:SLH_one_cavity}
\begin{split}
    G_{\text{cav}}^{(j)} =&\; \left(S_{\text{cav}}^{(j)}, L_{\text{cav}}^{(j)}, H_{\text{cav}}^{(j)}\right)\\
    =&\; \biggl( 1, \sqrt{\Gamma_j}c_j, \Omega_j\left(\sigma_j^+\sigma_j^- + c_j^\dagger c_j\right) + g_j\left( \sigma_j^+c_j + \sigma_j^-c_j^\dagger \right) \biggr),
\end{split}
\end{align}
where $\Gamma_j$ is the decay rate of cavity $j$ into the waveguide, $g_j$ is the emitter--cavity coupling rate in the cavity, $\Omega_j$ is the resonance frequency of emitter/cavity $j$, ${\sigma_j^+ = \ketbra{e_j}{g_j}}$ and ${\sigma_j^- = \ketbra{g_j}{e_j}}$ are the raising and lowering operators of emitter $j$, and $c_j^\dagger$ and $c_j$ are the creation and annihilation operators of cavity $j$, respectively. Note that the $L$ parameter $L_{\text{cav}}^{(j)}$ describes the coupling to the waveguide, and the $H$ parameter $H_{\text{cav}}^{(j)}$ is the Jaynes--Cummings Hamiltonian of the emitter--cavity system.

In addition, the SLH parameters of the $j$th waveguide segment (with length $d_j$) are given by
\begin{equation}\label{eq:SLH_wg}
    G_{\text{wg}}^{(j)} = \left(S_{\text{wg}}^{(j)}, L_{\text{wg}}^{(j)}, H_{\text{wg}}^{(j)}\right) = \left( e^{-i\phi_j}, 0, 0\right),
\end{equation}
which corresponds to a phase shift $\phi_j$ that photons acquire during free propagation in the waveguide (note that, for a photon of frequency $\omega$, we take this to be ${\phi_j = \omega d_j/v_g}$, where $v_g$ is the group velocity in the waveguide).

Combining the $N$ cavities and ${N-1}$ waveguide segments in series [see Fig.~\ref{fig:CZ_gate_diagram}(b)], we obtain the total SLH parameters
\begin{align}
\begin{split}
    G_{\text{tot}} =& \left( S_{\text{tot}}, L_{\text{tot}}, H_{\text{tot}} \right)\\
    =&\; G_{\text{cav}}^{(N)} \lhd G_{\text{wg}}^{(N-1)} \lhd \ldots \lhd G_{\text{cav}}^{(2)} \lhd G_{\text{wg}}^{(1)} \lhd G_{\text{cav}}^{(1)},
\end{split}
\end{align}
where
\begin{equation}\label{eq:S_tot}
    S_{\text{tot}} = e^{-i\sum_{k=1}^{N-1}\phi_k}
\end{equation}
is the $S$ parameter (which corresponds to the total phase shift in the waveguide),
\begin{equation}\label{eq:L_tot}
    L_{\text{tot}} = \sum_{j=1}^N e^{-i\sum_{k=j}^{N-1}\phi_k}\sqrt{\Gamma_j}c_j
\end{equation}
is the $L$ parameter (which describes the coupling of all the cavities to the waveguide), and
\begin{align}\label{eq:H_tot}
\begin{split}
    H_{\text{tot}} =& \sum_{j=1}^N \left[ \Omega_j\left(\sigma_j^+\sigma_j^- + c_j^\dagger c_j\right) + g_j\left( \sigma_j^+c_j + \sigma_j^-c_j^\dagger \right) \right]\\
    &+ \frac{1}{2i} \sum_{j=2}^N \sum_{k=1}^{j-1} \sqrt{\Gamma_j\Gamma_k} \left( e^{-i\sum_{l=k}^{j-1}\phi_l} c_j^\dagger c_k - e^{i\sum_{l=k}^{j-1}\phi_l} c_k^\dagger c_j \right)
\end{split}
\end{align}
is the $H$ parameter, which is a sum of Jaynes--Cummings Hamiltonians for $N$ cavities (first line) and terms that describe inter-cavity interactions mediated by the waveguide (second line). A more detailed derivation of the total SLH parameters can be found in Appendix~\ref{app:N_cavities_SLH}.

The SLH formalism allows the input--output relations and Heisenberg equations to be obtained directly from the SLH parameters of a system. In our case, where there is a single input and output channel due to chiral cavity--waveguide coupling, the input--output relation is given in terms of the total SLH parameters as~\cite{Combes2017, Brod2016_2}
\begin{equation}\label{eq:input-output_SLH}
    a_{\text{out}}(t) = S_{\text{tot}}a_{\text{in}}(t) - iL_{\text{tot}}.
\end{equation}
For a single cavity, where the SLH parameters are given in Eq.~(\ref{eq:SLH_one_cavity}), the input--output relation is consistent with Eq.~(\ref{eq:input-output}) in Section~\ref{subsubsec:theory_one_cavity_one_photon}. Furthermore, the Heisenberg equation for a system operator $A(t)$ is given in terms of the SLH parameters as
\begin{align}\label{eq:Heisenberg_SLH}
\begin{split}
    \frac{\rm d}{{\rm d}t}A(t) =&\; i\left[ H_{\text{tot}}, A(t) \right] + L_{\text{tot}}^{\dagger}A(t)L_{\text{tot}}\\
    &- \frac{1}{2}\left( L_{\text{tot}}^{\dagger} L_{\text{tot}} A(t) + A(t)L_{\text{tot}}^{\dagger} L_{\text{tot}}\right)\\
    &+ i \left[ L_{\text{tot}}^{\dagger}, A(t) \right] S_{\text{tot}}a_{\text{in}}(t) - i a_{\text{in}}^{\dagger}(t)S_{\text{tot}}^{\dagger} \left[ A(t), L_{\text{tot}} \right].
\end{split}
\end{align}

In the next subsections, we will use Eqs.~(\ref{eq:input-output_SLH}) and (\ref{eq:Heisenberg_SLH}), together with the SLH parameters in Eqs.~(\ref{eq:S_tot})-(\ref{eq:H_tot}), to derive the single-photon and two-photon scattering matrices for the $N$-cavity system.


\subsubsection{Single-photon scattering}\label{subsubsec:theory_N_cavities_one_photon}

To find the single-photon scattering matrix elements $S_{pk}$ in Eq.~(\ref{eq:S1}) for $N$ cavities, we first need the corresponding input--output relation. Substituting Eqs.~(\ref{eq:S_tot}) and (\ref{eq:L_tot}) into Eq.~(\ref{eq:input-output_SLH}) leads to
\begin{equation}\label{eq:input-output_N_cavities}
    a_{\text{out}}(t) = e^{-i\sum_{k=1}^{N-1}\phi_k} a_{\text{in}}(t) - i \sum_{j=1}^N e^{-i\sum_{k=j}^{N-1}\phi_k}\sqrt{\Gamma_j}c_j(t).
\end{equation}
Performing a Fourier transform from time to frequency and substituting the result into Eq.~(\ref{eq:S1}) gives
\begin{align}\label{eq:S1_N_cavities}
\begin{split}
    S_{pk} =& \bra{0}a_{\text{out}}(p)a_{\text{in}}^{\dagger}(k)\ket{0}\\
    =& \bra{0} \Biggl[ e^{-i\sum_{k=1}^{N-1}\phi_k(p)} a_{\text{in}}(p)\\
    &\hspace{0.2in}- i \sum_{j=1}^N e^{-i\sum_{k=j}^{N-1}\phi_k(p)}\sqrt{\Gamma_j}c_j(p) \Biggr]a_{\text{in}}^{\dagger}(k)\ket{0}\\
    =&\; e^{-i\sum_{k=1}^{N-1}\phi_k(p)} \delta(p-k)\\
    &- i \sum_{j=1}^N e^{-i\sum_{k=j}^{N-1}\phi_k(p)}\sqrt{\Gamma_j} \matrixel{0}{c_j(p)}{k^+},
\end{split}
\end{align}
where ${\phi_j(p) = pd_j/v_g}$ is the phase shift that a photon with frequency $p$ acquires between cavities $j$ and ${j+1}$.

We now need to calculate the matrix elements $\matrixel{0}{c_j(p)}{k^+}$ for ${j \in \{1,2,\ldots,N\}}$, using the Heisenberg equation for $c_j(t)$. When we use ${A(t) = c_j(t)}$ in Eq.~(\ref{eq:Heisenberg_SLH}), together with the total SLH parameters in Eqs.~(\ref{eq:S_tot})-(\ref{eq:H_tot}) and the commutation relation ${[c_j(t),c_{j'}^{\dagger}(t)] = \delta_{jj'}}$, we obtain
\begin{align}\label{eq:Heisenberg_cavity_j_time}
\begin{split}
    \frac{\rm d}{{\rm d}t}c_j(t) =& -i\left( \Omega_j - \frac{i}{2}\Gamma_j \right)c_j(t) - ig_j\sigma_j^-(t)\\
    &- i e^{-i\sum_{k=1}^{j-1}\phi_k}\sqrt{\Gamma_j} a_{\text{in}}(t)\\
    &- \sum_{l=1}^{j-1}e^{-i\sum_{k=l}^{j-1}\phi_k}\sqrt{\Gamma_j\Gamma_l}c_l(t)
\end{split}
\end{align}
(see Appendix~\ref{app:N_cavities_one_photon}). Note that, when ${j=1}$, all the summations vanish and we obtain the single-cavity Heisenberg equation in Eq.~(\ref{eq:Heisenberg_cavity}). Fourier transforming from time to frequency leads to
\begin{align}
\begin{split}
    -ipc_j(p) =& -i\left( \Omega_j - \frac{i}{2}\Gamma_j \right)c_j(p) - ig_j\sigma_j^-(p)\\
    &- i e^{-i\sum_{k=1}^{j-1}\phi_k(p)}\sqrt{\Gamma_j} a_{\text{in}}(p)\\
    &- \sum_{l=1}^{j-1}e^{-i\sum_{k=l}^{j-1}\phi_k(p)}\sqrt{\Gamma_j\Gamma_l}c_l(p),
\end{split}
\end{align}
which can be rearranged to
\begin{align}\label{eq:Heisenberg_cavity_j}
\begin{split}
    \left[ \Delta_j(p) + \frac{i}{2}\Gamma_j \right]c_j(p) =&\;  g_j\sigma_j^-(p) + e^{-i\sum_{k=1}^{j-1}\phi_k(p)}\sqrt{\Gamma_j} a_{\text{in}}(p)\\
    &-i \sum_{l=1}^{j-1}e^{-i\sum_{k=l}^{j-1}\phi_k(p)}\sqrt{\Gamma_j\Gamma_l}c_l(p),
\end{split}
\end{align}
where ${\Delta_j(p) = p - \Omega_j}$ is the detuning between a photon with frequency $p$ and emitter/cavity $j$. Pre-multiplying each term by $\bra{0}$ and post-multiplying by $\ket{k^+}$ gives
\begin{align}\label{eq:Heisenberg_c_j_p}
\begin{split}
    \biggl[ \Delta_j&(p) + \frac{i}{2}\Gamma_j \biggr]\matrixel{0}{c_j(p)}{k^+}\\
    =&\;  g_j\matrixel{0}{\sigma_j^-(p)}{k^+} + e^{-i\sum_{k=1}^{j-1}\phi_k(p)}\sqrt{\Gamma_j} \delta(p-k)\\
    &-i \sum_{l=1}^{j-1}e^{-i\sum_{k=l}^{j-1}\phi_k(p)}\sqrt{\Gamma_j\Gamma_l}\matrixel{0}{c_l(p)}{k^+},
\end{split}
\end{align}
where we used ${\matrixel{0}{a_{\text{in}}(p)}{k^+} = \delta(p-k)}$. To solve this for $\matrixel{0}{c_j(p)}{k^+}$, we need to calculate $\matrixel{0}{\sigma_j^-(p)}{k^+}$ using the Heisenberg equation for $\sigma_j^-(t)$. Using ${A(t) = \sigma_j^-(t)}$ and our SLH parameters in Eq.~(\ref{eq:Heisenberg_SLH}) leads to
\begin{equation}\label{eq:Heisenberg_emitter_j}
    \frac{\rm d}{{\rm d}t}\sigma_j^-(t) = -i\Omega_j\sigma_j^-(t) + ig_j\sigma_{z,j}(t)c_j(t),
\end{equation}
with ${\sigma_{z,j}(0) = \ketbra{e_j}{e_j} - \ketbra{g_j}{g_j}}$. This is consistent with the single-emitter Heisenberg equation in Eq.~(\ref{eq:Heisenberg_emitter}). Pre-multiplying by $\bra{0}$, post-multiplying by $\ket{k^+}$, and using ${\bra{0}\sigma_{z,j}(t) = -\bra{0}}$ gives
\begin{equation}
    \frac{\rm d}{{\rm d}t}\matrixel{0}{\sigma_j^-(t)}{k^+} = -i\Omega_j\matrixel{0}{\sigma_j^-(t)}{k^+} - ig_j \matrixel{0}{c_j(t)}{k^+}.
\end{equation}
We now Fourier transform from time to frequency,
\begin{equation}
    -ip\matrixel{0}{\sigma_j^-(p)}{k^+} = -i\Omega_j\matrixel{0}{\sigma_j^-(p)}{k^+} - ig_j \matrixel{0}{c_j(p)}{k^+},
\end{equation}
and rearrange for $\matrixel{0}{\sigma_j^-(p)}{k^+}$:
\begin{equation}\label{eq:sigma_j_p_matrix_el}
    \matrixel{0}{\sigma_j^-(p)}{k^+} = \frac{g_j}{\Delta_j(p)} \matrixel{0}{c_j(p)}{k^+}.
\end{equation}
After substituting this result into Eq.~(\ref{eq:Heisenberg_c_j_p}) and rearranging, we obtain
\begin{align}\label{eq:Heisenberg_c_j_p_2}
\begin{split}
    \biggl[ \Delta^2_j&(p) - g_j^2 + \frac{i}{2}\Delta_j(p)\Gamma_j \biggr]\matrixel{0}{c_j(p)}{k^+}\\
    =&\; e^{-i\sum_{k=1}^{j-1}\phi_k(p)}\Delta_j(p)\sqrt{\Gamma_j} \delta(p-k)\\
    &-i \Delta_j(p) \sum_{l=1}^{j-1}e^{-i\sum_{k=l}^{j-1}\phi_k(p)}\sqrt{\Gamma_j\Gamma_l}\matrixel{0}{c_l(p)}{k^+}.
\end{split}
\end{align}
Equation~(\ref{eq:Heisenberg_c_j_p_2}) is a set of $N$ coupled equations, one for each value of $j$. This set of equations can be solved iteratively because the solution for a particular value of $j$ only depends on the solutions for smaller $j$ values due to the chiral cavity--waveguide coupling (i.e., the evolution of one cavity is only affected by previous cavities in the array). For example, the equation for ${j=1}$ only contains the matrix element $\matrixel{0}{c_1(p)}{k^+}$, and can be solved straightforwardly to give
\begin{equation}\label{eq:c_1_p_matrix_el}
    \matrixel{0}{c_1(p)}{k^+} = \frac{\Delta_1(p)\sqrt{\Gamma_1}}{\Delta^2_1(p) - g_1^2 + \frac{i}{2}\Delta_1(p)\Gamma_1}\delta(p-k),
\end{equation}
in consistency with Eq.~(\ref{eq:c_p_matrix_el}). With this solution, we can find $\matrixel{0}{c_2(p)}{k^+}$ by setting ${j=2}$ in Eq.~(\ref{eq:Heisenberg_c_j_p_2}):
\begin{align}\label{eq:c_2_p_matrix_el}
\begin{split}
    \matrixel{0}{c_2(p)}{k^+} =&\; e^{-i\phi_1(p)}\frac{\Delta_2(p)\sqrt{\Gamma_2}}{\Delta^2_2(p) - g_2^2 + \frac{i}{2}\Delta_2(p)\Gamma_2}\\
    &\times\left[\frac{\Delta^2_1(p) - g_1^2 - \frac{i}{2}\Delta_1(p)\Gamma_1}{\Delta^2_1(p) - g_1^2 + \frac{i}{2}\Delta_1(p)\Gamma_1}\right]\delta(p-k).
\end{split}
\end{align}
Using Eqs.~(\ref{eq:c_1_p_matrix_el}) and (\ref{eq:c_2_p_matrix_el}), we could find $\matrixel{0}{c_3(p)}{k^+}$ by setting ${j=3}$ in Eq.~(\ref{eq:Heisenberg_c_j_p_2}). Generalizing to any $j$, we have
\begin{align}\label{eq:c_j_p_matrix_el}
\begin{split}
    \matrixel{0}{c_j(p)}{k^+} =&\; e^{-i\sum_{k=1}^{j-1}\phi_k(p)}\frac{\Delta_j(p)\sqrt{\Gamma_j}}{\Delta^2_j(p) - g_j^2 - \frac{i}{2}\Delta_j(p)\Gamma_j}\\
    &\times\prod_{l=1}^j\left[\frac{\Delta^2_l(p) - g_l^2 - \frac{i}{2}\Delta_l(p)\Gamma_l}{\Delta^2_l(p) - g_l^2 + \frac{i}{2}\Delta_l(p)\Gamma_l}\right]\delta(p-k).
\end{split}
\end{align}
We can now return to the single-photon scattering matrix in Eq.~(\ref{eq:S1_N_cavities}), and substitute in Eq.~(\ref{eq:c_j_p_matrix_el}) to obtain ${S_{pk} = t(p)\delta(p-k)}$, where
\begin{align}\label{eq:t_N_cavities}
\begin{split}
    t(p) =&\; e^{-i\sum_{k=1}^{N-1}\phi_k(p)} \Biggl( 1 - i\sum_{j=1}^N \frac{\Delta_j(p)\Gamma_j}{\Delta^2_j(p) - g_j^2 - \frac{i}{2}\Delta_j(p)\Gamma_j}\\
    &\hspace{0.9in}\times \prod_{l=1}^j\left[\frac{\Delta^2_l(p) - g_l^2 - \frac{i}{2}\Delta_l(p)\Gamma_l}{\Delta^2_l(p) - g_l^2 + \frac{i}{2}\Delta_l(p)\Gamma_l}\right] \Biggr)
\end{split}
\end{align}
is the single-photon transmission coefficient.

As in the single-cavity case (Section~\ref{subsubsec:theory_one_cavity_one_photon}), we define
\begin{equation}\label{eq:c_j_p_matrix_el_2}
    \matrixel{0}{c_j(p)}{k^+} = s_{c,j}(p)\delta(p-k),
\end{equation}
where
\begin{align}\label{eq:sc_j}
\begin{split}
    s_{c,j}(p) =&\; e^{-i\sum_{k=1}^{j-1}\phi_k(p)}\frac{\Delta_j(p)\sqrt{\Gamma_j}}{\Delta^2_j(p) - g_j^2 - \frac{i}{2}\Delta_j(p)\Gamma_j}\\
    &\times\prod_{l=1}^j\left[\frac{\Delta^2_l(p) - g_l^2 - \frac{i}{2}\Delta_l(p)\Gamma_l}{\Delta^2_l(p) - g_l^2 + \frac{i}{2}\Delta_l(p)\Gamma_l}\right]
\end{split}
\end{align}
is the single-photon excitation amplitude of cavity $j$ [see Eq.~(\ref{eq:c_j_p_matrix_el})]. In addition,
\begin{equation}\label{eq:sigma_j_p_matrix_el_2}
    \matrixel{0}{\sigma_j^-(p)}{k^+} = s_{e,j}(p)\delta(p-k),
\end{equation}
where
\begin{align}\label{eq:se_j}
\begin{split}
    s_{e,j}(p) =&\; e^{-i\sum_{k=1}^{j-1}\phi_k(p)}\frac{g_j\sqrt{\Gamma_j}}{\Delta^2_j(p) - g_j^2 - \frac{i}{2}\Delta_j(p)\Gamma_j}\\
    &\times\prod_{l=1}^j\left[\frac{\Delta^2_l(p) - g_l^2 - \frac{i}{2}\Delta_l(p)\Gamma_l}{\Delta^2_l(p) - g_l^2 + \frac{i}{2}\Delta_l(p)\Gamma_l}\right]
\end{split}
\end{align}
is the single-photon excitation amplitude of emitter $j$, which follows from Eq.~(\ref{eq:sigma_j_p_matrix_el}).


\subsubsection{Two-photon scattering}\label{subsubsec:theory_N_cavities_two_photons}

To calculate the two-photon scattering matrix elements ${S_{p_1p_2k_1k_2}}$ for the $N$-cavity system, we first substitute in the Fourier transform of the input--output relation in Eq.~(\ref{eq:input-output_N_cavities}) into Eq.~(\ref{eq:S2_2}):
\begin{align}\label{eq:S2_N_cavities}
\begin{split}
    S_{p_1p_2k_1k_2} =& \bra{0}a_{\text{out}}(p_1)a_{\text{out}}(p_2)a_{\text{in}}^{\dagger}(k_1)a_{\text{in}}^{\dagger}(k_2)\ket{0}\\
    =& \bra{p_1^-}a_{\text{out}}(p_2)\ket{k_1k_2^+}\\
    =&\; e^{-i\sum_{k=1}^{N-1}\phi_k(p_2)} \bra{p_1^-}a_{\text{in}}(p_2)\ket{k_1k_2^+}\\
    &- i \sum_{j=1}^N e^{-i\sum_{k=j}^{N-1}\phi_k(p_2)}\sqrt{\Gamma_j} \bra{p_1^-}c_j(p_2)\ket{k_1k_2^+}\\
    =&\; e^{-i\sum_{k=1}^{N-1}\phi_k(p_2)} [S_{p_1k_2}\delta(p_2-k_1) + S_{p_1k_1}\delta(p_2-k_2)]\\
    &- i \sum_{j=1}^N e^{-i\sum_{k=j}^{N-1}\phi_k(p_2)}\sqrt{\Gamma_j} \bra{p_1^-}c_j(p_2)\ket{k_1k_2^+},
\end{split}
\end{align}
where we used the result in Eq.~(\ref{eq:a_in_p2}). We now need to calculate the matrix element $\bra{p_1^-}c_j(p_2)\ket{k_1k_2^+}$, using the Heisenberg equation for $c_j(t)$ from Section~\ref{subsubsec:theory_N_cavities_one_photon} above. Returning to Eq.~(\ref{eq:Heisenberg_cavity_j}), relabeling $p$ with $p_2$, and pre-multiplying by $\bra{p_1^-}$ and post-multiplying by $\ket{k_1k_2^+}$ gives
\begin{align}\label{eq:c_j_p2_matrix_el}
\begin{split}
    &\left[ \Delta_j(p_2) + \frac{i}{2}\Gamma_j \right]\bra{p_1^-}c_j(p_2)\ket{k_1k_2^+}\\
    &=  g_j\bra{p_1^-}\sigma_j^-(p_2)\ket{k_1k_2^+}\\
    &\hspace{0.15in}+ e^{-i\sum_{k=1}^{j-1}\phi_k(p_2)}\sqrt{\Gamma_j}[S_{p_1k_2}\delta(p_2-k_1) + S_{p_1k_1}\delta(p_2-k_2)]\\
    &\hspace{0.15in}-i \sum_{l=1}^{j-1}e^{-i\sum_{k=l}^{j-1}\phi_k(p_2)}\sqrt{\Gamma_j\Gamma_l}\bra{p_1^-}c_l(p_2)\ket{k_1k_2^+},
\end{split}
\end{align}
where we again used Eq.~(\ref{eq:a_in_p2}). To solve Eq.~(\ref{eq:c_j_p2_matrix_el}) for $\bra{p_1^-}c_j(p_2)\ket{k_1k_2^+}$, we need to calculate $\bra{p_1^-}\sigma_j^-(p_2)\ket{k_1k_2^+}$ using the Heisenberg equation for $\sigma_j^-(t)$. Returning to Eq.~(\ref{eq:Heisenberg_emitter_j}), and pre-multiplying by $\bra{p_1^-}$ and post-multiplying by $\ket{k_1k_2^+}$ leads to
\begin{align}
\begin{split}
    \frac{\rm d}{{\rm d}t}\bra{p_1^-}\sigma_j^-(t)\ket{k_1k_2^+} =& -i\Omega_j\bra{p_1^-}\sigma_j^-(t)\ket{k_1k_2^+}\\
    &+ ig_j\bra{p_1^-}\sigma_{z,j}(t)c_j(t)\ket{k_1k_2^+}.
\end{split}
\end{align}
The next steps closely follow the steps below Eq.~(\ref{eq:Heisenberg_emitter_2}) from the single-cavity calculation in Section~\ref{subsubsec:theory_one_cavity_two_photons}. We use $\sigma_{z,j}(t) = 2\sigma_j^+(t)\sigma_j^-(t) - 1$ in the above to obtain
\begin{align}
\begin{split}
    \frac{\rm d}{{\rm d}t}\bra{p_1^-}\sigma_j^-(t)\ket{k_1k_2^+} =& -i\Omega_j\bra{p_1^-}\sigma_j^-(t)\ket{k_1k_2^+}\\
    &+ 2ig_j\bra{p_1^-}\sigma_j^+\sigma_j^-c_j(t)\ket{k_1k_2^+}\\
    &- ig_j\bra{p_1^-}c_j(t)\ket{k_1k_2^+}.
\end{split}
\end{align}
We then perform a Fourier transform from time $t$ to frequency $p_2$,
\begin{align}
\begin{split}
    -ip_2\bra{p_1^-}\sigma_j^-(p_2)\ket{k_1k_2^+} =& -i\Omega_j\bra{p_1^-}\sigma_j^-(p_2)\ket{k_1k_2^+}\\
    &+ 2ig_j\bra{p_1^-}\sigma_j^+\sigma_j^-c_j(p_2)\ket{k_1k_2^+}\\
    &- ig_j\bra{p_1^-}c_j(p_2)\ket{k_1k_2^+},
\end{split}
\end{align}
and solve for $\bra{p_1^-}\sigma_j^-(p_2)\ket{k_1k_2^+}$:
\begin{align}
\begin{split}
    \bra{p_1^-}\sigma_j^-(p_2)\ket{k_1k_2^+} =&\; \frac{g_j}{\Delta_j(p_2)}\bra{p_1^-}c_j(p_2)\ket{k_1k_2^+}\\
    &- \frac{2g_j}{\Delta_j(p_2)}\bra{p_1^-}\sigma_j^+\sigma_j^-c_j(p_2)\ket{k_1k_2^+}.
\end{split}
\end{align}
Applying the convolution theorem as in Eq.~(\ref{eq:convolution_sigma}) means that
\begin{align}\label{eq:sigma_j_matrix_el}
\begin{split}
    &\bra{p_1^-}\sigma_j^-(p_2)\ket{k_1k_2^+}\\
    &= \frac{g_j}{\Delta_j(p_2)} \bra{p_1^-}c_j(p_2)\ket{k_1k_2^+}\\
    &\hspace{0.15in}-\frac{2}{\sqrt{2\pi}}\frac{g_j}{\Delta_j(p_2)} \int {\rm d}p \matrixel{p_1^-}{\sigma_j^+(p)}{0}\matrixel{0}{\sigma_j^-c_j(p_2+p)}{k_1k_2^+},
\end{split}
\end{align}
where we also inserted the identity operator between $\sigma_j^+(p)$ and $\sigma_j^-c_j(p_2+p)$ as in Eq.~(\ref{eq:sigma_p2_matrix_el_conv}). Similarly to Eq.~(\ref{eq:sigma_p_one_photon}), we have
\begin{align}
\begin{split}
    \matrixel{p_1^-}{\sigma_j^+(p)}{0} =&\; \bra{p_1^-} \left( \int {\rm d}k \ketbra{k^+}{k^+} \right)\sigma_j^+(p)\ket{0}\\
    =&\; \int {\rm d}k \braket{p_1^-}{k^+} \left[ \matrixel{0}{\sigma_j^-(p)}{k^+} \right]^\dagger\\
    =&\; \int {\rm d}k\; S_{p_1k} \left[ s_{e,j}(p) \right]^* \delta(p-k)\\
    =&\; S_{p_1p}\left[ s_{e,j}(p) \right]^*.
\end{split}
\end{align}
Substituting this into Eq.~(\ref{eq:sigma_j_matrix_el}) and using ${S_{p_1p} = t(p_1)\delta(p_1-p)}$ leads to
\begin{align}
\begin{split}
    \bra{p_1^-}&\sigma_j^-(p_2)\ket{k_1k_2^+}\\
    =&\; \frac{g_j}{\Delta_j(p_2)} \bra{p_1^-}c_j(p_2)\ket{k_1k_2^+}\\
    & -\frac{2}{\sqrt{2\pi}}\frac{g_j}{\Delta_j(p_2)} t(p_1)\left[ s_{e,j}(p_1) \right]^* \matrixel{0}{\sigma_j^-c_j(p_1+p_2)}{k_1k_2^+}.
\end{split}
\end{align}
We can now use this result in Eq.~(\ref{eq:c_j_p2_matrix_el}) to obtain
\begin{align}\label{eq:c_j_p2_matrix_el_2}
\begin{split}
    &\left[ \Delta^2_j(p_2) - g^2_j + \frac{i}{2}\Delta_j(p_2)\Gamma_j \right]\bra{p_1^-}c_j(p_2)\ket{k_1k_2^+}\\
    &= -\frac{2g^2_j}{\sqrt{2\pi}}t(p_1)\left[ s_{e,j}(p_1) \right]^* \matrixel{0}{\sigma_j^-c_j(p_1+p_2)}{k_1k_2^+}\\
    &\hspace{0.15in}+ e^{-i\sum_{k=1}^{j-1}\phi_k(p_2)}\Delta_j(p_2)\sqrt{\Gamma_j}[S_{p_1k_2}\delta(p_2-k_1) + S_{p_1k_1}\delta(p_2-k_2)]\\
    &\hspace{0.15in}-i\Delta_j(p_2) \sum_{l=1}^{j-1}e^{-i\sum_{k=l}^{j-1}\phi_k(p_2)}\sqrt{\Gamma_j\Gamma_l}\bra{p_1^-}c_l(p_2)\ket{k_1k_2^+}.
\end{split}
\end{align}
To make further progress in calculating $\bra{p_1^-}c_j(p_2)\ket{k_1k_2^+}$, we need to find $\matrixel{0}{\sigma_j^-c_j(p_1+p_2)}{k_1k_2^+}$, which requires calculating Heisenberg equations of operator products (see Appendix~\ref{app:N_cavities_two_photons} for more details). Using ${A(t) = \sigma_j^-(t)c_l(t)}$ together with our total SLH parameters in Eq.~(\ref{eq:Heisenberg_SLH}) gives the Heisenberg equation
\begin{align}\label{eq:Heisenberg_sigma_j_c_l}
\begin{split}
    \frac{\rm d}{{\rm d}t} \sigma_j^-(t)c_l(t) =& -i\left( \Omega_j + \Omega_l - \frac{i}{2}\Gamma_l \right)\sigma_j^-(t)c_l(t)\\
    &+ ig_j \sigma_{z,j}(t)c_j(t)c_l(t)\\
    &- ie^{-i\sum_{k=1}^{l-1}\phi_{k}}\sqrt{\Gamma_l}\sigma_j^-(t)a_{\text{in}}(t)\\
    &- \sum_{l'=1}^{l-1}e^{-i\sum_{k=l'}^{l-1}\phi_k}\sqrt{\Gamma_l\Gamma_{l'}}\sigma_j^-(t)c_{l'}(t).
\end{split}
\end{align}
Pre-multiplying by $\bra{0}$, post-multiplying by $\ket{k_1k_2^+}$, and using ${\bra{0}\sigma_{z,j}(t) = -\bra{0}}$ leads to
\begin{align}
\begin{split}
    \frac{\rm d}{{\rm d}t}& \matrixel{0}{\sigma_j^-(t)c_l(t)}{k_1k_2^+}\\
    =& -i\left( \Omega_j + \Omega_l - \frac{i}{2}\Gamma_l \right)\matrixel{0}{\sigma_j^-(t)c_l(t)}{k_1k_2^+}\\
    &- ig_j \matrixel{0}{c_j(t)c_l(t)}{k_1k_2^+}\\
    &- ie^{-i\sum_{k=1}^{l-1}\phi_{k}}\sqrt{\Gamma_l}\matrixel{0}{\sigma_j^-(t)a_{\text{in}}(t)}{k_1k_2^+}\\
    &- \sum_{l'=1}^{l-1}e^{-i\sum_{k=l'}^{l-1}\phi_k}\sqrt{\Gamma_l\Gamma_{l'}}\matrixel{0}{\sigma_j^-(t)c_{l'}(t)}{k_1k_2^+},
\end{split}
\end{align}
which becomes, after Fourier transforming from time to frequency,
\begin{align}\label{eq:sigma_c_j_matrix_el}
\begin{split}
    -i(&p_1+p_2) \matrixel{0}{\sigma_j^-c_l(p_1+p_2)}{k_1k_2^+}\\
    =& -i\left( \Omega_j + \Omega_l - \frac{i}{2}\Gamma_l \right)\matrixel{0}{\sigma_j^-c_l(p_1+p_2)}{k_1k_2^+}\\
    &- ig_j \matrixel{0}{c_jc_l(p_1+p_2)}{k_1k_2^+}\\
    &- ie^{-i\sum_{k=1}^{l-1}\phi_{k}(p_1+p_2)}\sqrt{\Gamma_l}\matrixel{0}{\sigma_j^-a_{\text{in}}(p_1+p_2)}{k_1k_2^+}\\
    &- \sum_{l'=1}^{l-1}e^{-i\sum_{k=l'}^{l-1}\phi_k(p_1+p_2)}\sqrt{\Gamma_l\Gamma_{l'}}\matrixel{0}{\sigma_j^-c_{l'}(p_1+p_2)}{k_1k_2^+}.
\end{split}
\end{align}
Using the convolution theorem as in Eq.~(\ref{eq:convolution_sigma_a_in}) gives
\begin{align}
\begin{split}
    &\matrixel{0}{\sigma_j^-a_{\text{in}}(p_1+p_2)}{k_1k_2^+}\\
    &= \frac{1}{\sqrt{2\pi}} \int {\rm d}p\; \matrixel{0}{\sigma_j^-(p)a_{\text{in}}(p_1+p_2-p)}{k_1k_2^+}\\
    &= \frac{1}{\sqrt{2\pi}} \left[ s_{e,j}(k_1) + s_{e,j}(k_2) \right] \delta(p_1+p_2-k_1-k_2),
\end{split}
\end{align}
where we made use of Eq.~(\ref{eq:sigma_j_p_matrix_el_2}) and the commutation relation of the input operators to obtain the last line. Equation~(\ref{eq:sigma_c_j_matrix_el}) can therefore be rearranged to
\begin{widetext}
\begin{align}\label{eq:N2_equations_1}
\begin{split}
    &\left(p_1+p_2 - \Omega_j - \Omega_l + \frac{i}{2}\Gamma_l\right) \matrixel{0}{\sigma_j^-c_l(p_1+p_2)}{k_1k_2^+}\\
    &= g_j \matrixel{0}{c_jc_l(p_1+p_2)}{k_1k_2^+} + e^{-i\sum_{k=1}^{l-1}\phi_{k}(p_1+p_2)}\sqrt{\frac{\Gamma_l}{2\pi}} \left[ s_{e,j}(k_1) + s_{e,j}(k_2) \right] \delta(p_1+p_2-k_1-k_2)\\
    &\hspace{0.15in}-i \sum_{l'=1}^{l-1}e^{-i\sum_{k=l'}^{l-1}\phi_k(p_1+p_2)}\sqrt{\Gamma_l\Gamma_{l'}}\matrixel{0}{\sigma_j^-c_{l'}(p_1+p_2)}{k_1k_2^+}.
\end{split}
\end{align}
This equation contains the matrix element $\matrixel{0}{c_jc_l(p_1+p_2)}{k_1k_2^+}$, so we need to solve the Heisenberg equation for $c_j(t)c_l(t)$. Using ${A(t) = c_j(t)c_l(t)}$ and the total SLH parameters in Eq.~(\ref{eq:Heisenberg_SLH}) gives
\begin{align}\label{eq:Heisenberg_c2_time_2}
\begin{split}
    \frac{\rm d}{{\rm d}t}c_j(t)c_l(t) =& -i\left( \Omega_j + \Omega_l - \frac{i}{2}\Gamma_j - \frac{i}{2}\Gamma_l \right)c_j(t)c_l(t) - ig_l\sigma_l^-(t)c_j(t) - ig_j\sigma_j^-(t)c_l(t)\\
    &- ie^{-i\sum_{k=1}^{l-1}\phi_{k}}\sqrt{\Gamma_l}c_j(t)a_{\text{in}}(t) - ie^{-i\sum_{k=1}^{j-1}\phi_{k}}\sqrt{\Gamma_j}c_l(t)a_{\text{in}}(t)\\
    &- \sum_{l'=1}^{l-1} e^{-i\sum_{k=l'}^{l-1}\phi_k} \sqrt{\Gamma_l\Gamma_{l'}} c_j(t)c_{l'}(t) - \sum_{l'=1}^{j-1} e^{-i\sum_{k=l'}^{j-1}\phi_k} \sqrt{\Gamma_j\Gamma_{l'}} c_l(t)c_{l'}(t).
\end{split}
\end{align}
We now pre-multiply by $\bra{0}$, post-multiply by $\ket{k_1k_2^+}$, and perform a Fourier transform to obtain
\begin{align}\label{eq:c_j_c_l_matrix_el}
\begin{split}
    -i(p_1+p_2)\matrixel{0}{c_jc_l(p_1+p_2)}{k_1k_2^+} =& -i\left( \Omega_j + \Omega_l - \frac{i}{2}\Gamma_j - \frac{i}{2}\Gamma_l \right)\matrixel{0}{c_jc_l(p_1+p_2)}{k_1k_2^+}\\
    &- ig_l\matrixel{0}{\sigma_l^-c_j(p_1+p_2)}{k_1k_2^+} - ig_j\matrixel{0}{\sigma_j^-c_l(p_1+p_2)}{k_1k_2^+}\\
    &- ie^{-i\sum_{k=1}^{l-1}\phi_{k}(p_1+p_2)}\sqrt{\Gamma_l}\matrixel{0}{c_ja_{\text{in}}(p_1+p_2)}{k_1k_2^+}\\
    &- ie^{-i\sum_{k=1}^{j-1}\phi_{k}(p_1+p_2)}\sqrt{\Gamma_j}\matrixel{0}{c_la_{\text{in}}(p_1+p_2)}{k_1k_2^+}\\
    &- \sum_{l'=1}^{l-1} e^{-i\sum_{k=l'}^{l-1}\phi_k(p_1+p_2)} \sqrt{\Gamma_l\Gamma_{l'}} \matrixel{0}{c_jc_{l'}(p_1+p_2)}{k_1k_2^+}\\
    &- \sum_{l'=1}^{j-1} e^{-i\sum_{k=l'}^{j-1}\phi_k(p_1+p_2)} \sqrt{\Gamma_j\Gamma_{l'}} \matrixel{0}{c_lc_{l'}(p_1+p_2)}{k_1k_2^+}.
\end{split}
\end{align}
Using the convolution theorem as in Eq.~(\ref{eq:convolution_cavity}) means that
\begin{align}
\begin{split}
    \matrixel{0}{c_ja_{\text{in}}(p_1+p_2)}{k_1k_2^+} =&\; \frac{1}{\sqrt{2\pi}} \int {\rm d}p\; \matrixel{0}{c_j(p)a_{\text{in}}(p_1+p_2-p)}{k_1k_2^+}\\
    =&\; \frac{1}{\sqrt{2\pi}} \left[ s_{c,j}(k_1) + s_{c,j}(k_2) \right] \delta(p_1+p_2-k_1-k_2),
\end{split}
\end{align}
where we also used the result in Eq.~(\ref{eq:c_j_p_matrix_el_2}). Substituting this into Eq.~(\ref{eq:c_j_c_l_matrix_el}) and rearranging leads to
\begin{align}\label{eq:N2_equations_2}
\begin{split}
    &\left(p_1+p_2 - \Omega_j - \Omega_l + \frac{i}{2}\Gamma_j + \frac{i}{2}\Gamma_l\right)\matrixel{0}{c_jc_l(p_1+p_2)}{k_1k_2^+}\\
    &= g_l\matrixel{0}{\sigma_l^-c_j(p_1+p_2)}{k_1k_2^+} + g_j\matrixel{0}{\sigma_j^-c_l(p_1+p_2)}{k_1k_2^+}\\
    &\hspace{0.15in}+ e^{-i\sum_{k=1}^{l-1}\phi_{k}(p_1+p_2)}\sqrt{\frac{\Gamma_l}{2\pi}} \left[ s_{c,j}(k_1) + s_{c,j}(k_2) \right] \delta(p_1+p_2-k_1-k_2)\\
    &\hspace{0.15in}+ e^{-i\sum_{k=1}^{j-1}\phi_{k}(p_1+p_2)}\sqrt{\frac{\Gamma_j}{2\pi}} \left[ s_{c,l}(k_1) + s_{c,l}(k_2) \right] \delta(p_1+p_2-k_1-k_2)\\
    &\hspace{0.15in}-i \sum_{l'=1}^{l-1} e^{-i\sum_{k=l'}^{l-1}\phi_k(p_1+p_2)} \sqrt{\Gamma_l\Gamma_{l'}} \matrixel{0}{c_jc_{l'}(p_1+p_2)}{k_1k_2^+}\\
    &\hspace{0.15in}-i \sum_{l'=1}^{j-1} e^{-i\sum_{k=l'}^{j-1}\phi_k(p_1+p_2)} \sqrt{\Gamma_j\Gamma_{l'}} \matrixel{0}{c_lc_{l'}(p_1+p_2)}{k_1k_2^+}.
\end{split}
\end{align}

Equations~(\ref{eq:N2_equations_1}) and (\ref{eq:N2_equations_2}) are a closed set of equations that can be solved to obtain the matrix elements $\matrixel{0}{\sigma_j^-c_j(p_1+p_2)}{k_1k_2^+}$ (both equations are a set of $N^2$ equations, as both indices $j$ and $l$ take values from $1$ to $N$). We do this by collecting the $2N^2$ equations into a matrix equation, and performing matrix inversion numerically. This provides solutions of the form
\begin{equation}
    \matrixel{0}{\sigma_j^-c_j(p_1+p_2)}{k_1k_2^+} = \alpha^{(j)}_{p_1p_2k_1k_2}\delta(p_1+p_2-k_1-k_2).
\end{equation}
This can be substituted into Eq.~(\ref{eq:c_j_p2_matrix_el_2}) to obtain a closed set of equations for the matrix elements $\bra{p_1^-}c_j(p_2)\ket{k_1k_2^+}$:
\begin{align}\label{eq:c_j_p2_matrix_el_3}
\begin{split}
    \left[ \Delta^2_j(p_2) - g^2_j + \frac{i}{2}\Delta_j(p_2)\Gamma_j \right]\bra{p_1^-}c_j(p_2)\ket{k_1k_2^+} =& -\frac{2g^2_j}{\sqrt{2\pi}}t(p_1)\left[ s_{e,j}(p_1) \right]^* \alpha^{(j)}_{p_1p_2k_1k_2}\delta(p_1+p_2-k_1-k_2)\\
    &+ e^{-i\sum_{k=1}^{j-1}\phi_k(p_2)}\Delta_j(p_2)\sqrt{\Gamma_j}[S_{p_1k_2}\delta(p_2-k_1) + S_{p_1k_1}\delta(p_2-k_2)]\\
    &-i\Delta_j(p_2) \sum_{l=1}^{j-1}e^{-i\sum_{k=l}^{j-1}\phi_k(p_2)}\sqrt{\Gamma_j\Gamma_l}\bra{p_1^-}c_l(p_2)\ket{k_1k_2^+},
\end{split}
\end{align}
which can be solved iteratively, similarly to the single-photon case (see Appendix~\ref{app:N_cavities_two_photons}). Finally, we can substitute the result for $\bra{p_1^-}c_j(p_2)\ket{k_1k_2^+}$ into the two-photon scattering matrix in Eq.~(\ref{eq:S2_N_cavities}), giving a result of the form in Eq.~(\ref{eq:S2}), where
\begin{align}\label{eq:C_N_cavities}
\begin{split}
    C_{p_1p_2k_1k_2} =&\; \frac{2}{\sqrt{2\pi}}t(p_1) \Biggl[ \sum_{j=1}^N \left( \frac{g_j^2\left[s_{e,j}(p_1)\right]^*\alpha^{(j)}_{p_1p_2k_1k_2}\sqrt{\Gamma_j} e^{-i\sum_{k=j}^{N-1}\phi_k(p_2)}}{\Delta_j^2(p_2) - g_j^2 + \frac{i}{2}\Delta_j(p_2)\Gamma_j} \right)\\
    &\hspace{0.7in}-i \sum_{j=2}^N \frac{\Delta_j(p_2)\Gamma_j}{\Delta_j^2(p_2) - g_j^2 - \frac{i}{2}\Delta_j(p_2)\Gamma_j} \sum_{l=1}^{j-1}\left( \frac{g_l^2\left[s_{e,l}(p_1)\right]^*\alpha^{(l)}_{p_1p_2k_1k_2}\sqrt{\Gamma_l}e^{-i\sum_{k=l}^{N-1}\phi_k(p_2)}}{\Delta_l^2(p_2) - g_l^2 - \frac{i}{2}\Delta_l(p_2)\Gamma_l} \right)\\
    &\hspace{3in}\times \prod_{m=l}^j\left[ \frac{\Delta_m^2(p_2) - g_m^2 - \frac{i}{2}\Delta_m(p_2)\Gamma_m}{\Delta_m^2(p_2) - g_m^2 + \frac{i}{2}\Delta_m(p_2)\Gamma_m} \right]   \Biggr]
\end{split}
\end{align}
is the nonlinear two-photon term. In the next section, we use the results for $t(p)$ and $C_{p_1p_2k_1k_2}$ to calculate the CZ gate fidelity for different numbers of cavities $N$.
\end{widetext}


\section{Results}\label{sec:results}


\subsection{Input wave packets}

\begin{figure}[b]
    \centering
    \includegraphics[width=0.8\linewidth]{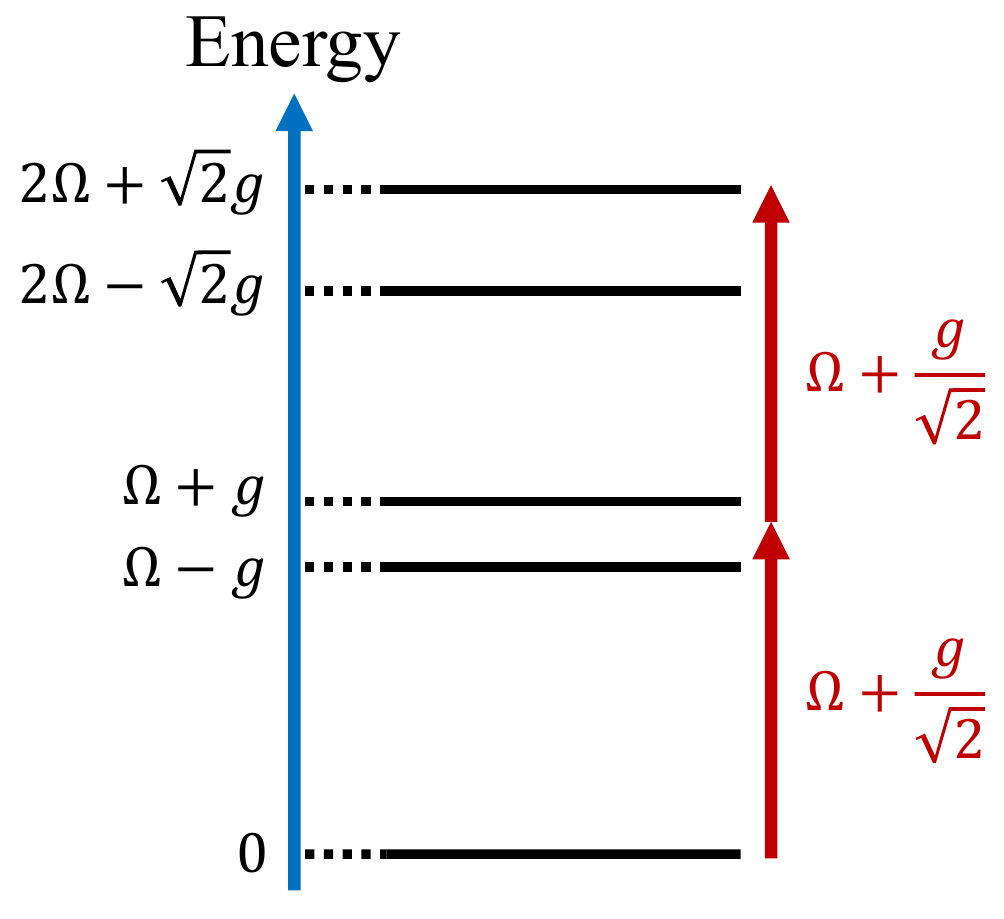}
    \caption{The first few energy levels of the emitter--cavity system in Fig.~\ref{fig:one_cavity}. The central frequencies of the input wave packets are $\Omega+g/\sqrt{2}$, as illustrated by the red arrows. The two photons are centered on the upper two-photon transition, with energy $2\Omega + \sqrt{2}g$. However, a single photon is detuned from the closest single-photon transition, with energy ${\Omega+g}$, by ${(1-1/\sqrt{2})g}$.}
    \label{fig:energy_levels}
\end{figure}

To evaluate the CZ gate fidelity for the $N$-cavity system in Fig.~\ref{fig:CZ_gate_diagram}, we consider two identical Gaussian input wave packets given by
\begin{equation}\label{eq:input_Gaussian}
    f(\omega) = \left[ \frac{4\text{ln}(2)}{\pi \sigma^2} \right]^{1/4} e^{-2\text{ln}(2)(\omega-\omega_0)^2/\sigma^2}
\end{equation}
(which satisfy the normalization condition $\int{\rm d}\omega|f(\omega)|^2=1$), where $\omega_0$ is the central frequency and $\sigma$ is the full width at half-maximum. These wave packets enter the general input state $\ket{\psi_{\text{in}}}$ in Eq.~(\ref{eq:input}) through the basis states defined in Eq.~(\ref{eq:basis}). Consider first the situation where each NS gate consists of one emitter--cavity system, as in Fig.~\ref{fig:one_cavity}. The ideal output state $\ket{\psi_{\text{out}}}$ in Eq.~(\ref{eq:CZ_output_ideal}) would be obtained if the photon--cavity interaction produces a $\pi$ phase shift on the two-photon wave packet, while producing no change in the single-photon wave packet [see Eq.~(\ref{eq:NS})]. Therefore, we choose the central frequency ${\omega_0 = \Omega + g/\sqrt{2}}$, such that the two photons are centered on a two-photon transition of the emitter--cavity system, while being off resonance with the single-photon transitions with energy $\Omega \pm g$ (see Fig.~\ref{fig:energy_levels}). Substituting this choice of $\omega_0$ into Eq.~(\ref{eq:input_Gaussian}) leads to
\begin{equation}\label{eq:Gaussian_detuning}
    f(\Delta) = \left[ \frac{4\text{ln}(2)}{\pi \sigma^2} \right]^{1/4} e^{-2\text{ln}(2)\left(\Delta-g/\sqrt{2}\right)^2/\sigma^2},
\end{equation}
in terms of the detuning ${\Delta = \omega - \Omega}$. The single- and two-photon input wave packets are shown in Fig.~\ref{fig:wave_packets}. In the appendices, we also express the single-photon transmission coefficient $t(p)$ and the two-photon term $C_{p_1p_2k_1k_2}$ in terms of detunings from the emitters/cavities, allowing us to write the integrals $I_1$ and $I_2$ in Eqs.~(\ref{eq:I_1}) and (\ref{eq:I_2}) in terms of detunings, relative to an arbitrary emitter/cavity frequency.

To prevent any distortion of the single-photon wave packet, the photons should not have any overlap with the closest single-photon transition (the one with energy ${\Omega+g}$; see Fig.~\ref{fig:energy_levels}). Due to the finite cavity linewidth $\Gamma$ and wave packet width $\sigma$, the energy difference between a single photon centered at ${\Omega+g/\sqrt{2}}$ and this transition should therefore exceed the sum of $\Gamma$ and $\sigma$. Since the energy difference between the peak of the wave packet and the transition is $\Omega + g - (\Omega + g/\sqrt{2}) = (2-\sqrt{2})g/2$, we have
\begin{equation}
    \left(\frac{2-\sqrt{2}}{2}\right)g > \Gamma + \sigma,
\end{equation}
or ${g \gtrsim 3.4(\Gamma + \sigma)}$. If this condition is satisfied, the change to the single-photon component of the input state should be negligible, while the two-photon component will be modified by the interaction with the cavity system.

For a chain of $N$ identical chirally-coupled cavities, the total transmission coefficient becomes a product of $N$ single-cavity transmission coefficients (ignoring the propagation phases, which only amount to a global phase under chiral coupling). This means that ${t(p) = e^{i\theta(p)} \rightarrow [t(p)]^N = e^{iN\theta(p)}}$, where ${\theta(p)}$ is the frequency-dependent phase that a single photon acquires when scattering from one cavity. As the total phase shift scales with $N$, the above condition needs to be modified to
\begin{equation}\label{eq:inequality}
    g \gtrsim 3.4(N\Gamma + \sigma)
\end{equation}
to ensure that the photon does not acquire a phase shift when passing through the $N$ cavities.

In Section~\ref{subsec:results_identical_cavities}, we present results for the CZ gate fidelity for the ideal case of $N$ identical cavities, and in Section~\ref{subsec:results_disorder} we demonstrate the effect of photon loss and disorder on the fidelity.


\subsection{CZ gate fidelity---$N$ identical cavities}\label{subsec:results_identical_cavities}

\begin{figure}[t]
    \centering
    \includegraphics[width=\linewidth]{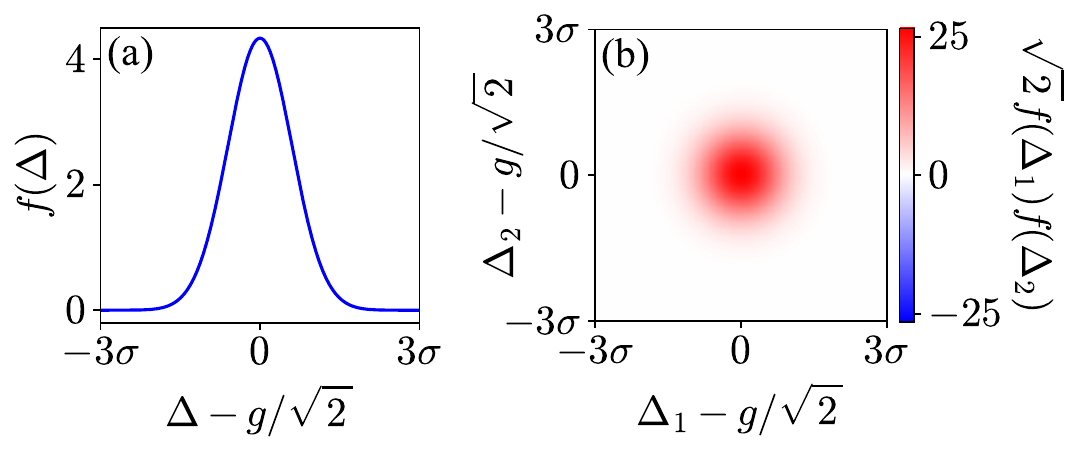}
    \caption{(a) Single-photon Gaussian wave packet from Eq.~(\ref{eq:Gaussian_detuning}). (b) Two-photon input wave packet, given by a product of two identical Gaussian functions. This can be calculated by projecting the two-photon state in Eq.~(\ref{eq:input_states_app}) onto detunings $\Delta_1$ and $\Delta_2$.}
    \label{fig:wave_packets}
\end{figure}

\begin{figure*}
    \centering
    \includegraphics[width=0.9\linewidth]{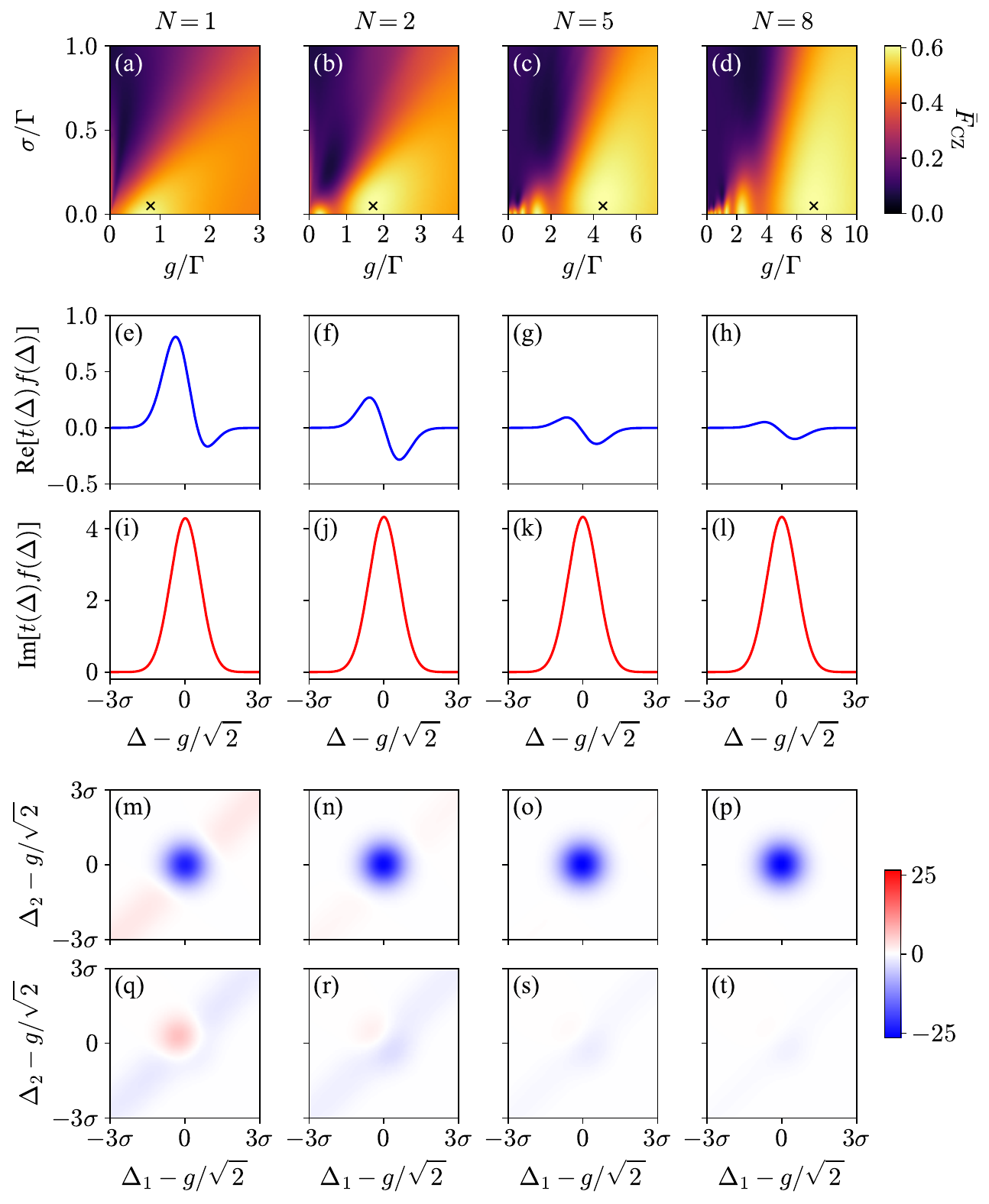}
    \caption{Average fidelity results and output wave packets for the case where each NS gate consists of $N$ identical waveguide-coupled cavities. The columns show results for ${N=1}$, $2$, $5$, and $8$ (from left to right). (a)-(d) Average fidelity $\bar{F}_{\text{CZ}}$ [Eq.~(\ref{eq:avg_F_CZ})] as a function of the wave packet width $\sigma$ and the emitter--cavity coupling rate $g$. The black cross in each of these figures indicates the maximum fidelity for ${\sigma/\Gamma = 0.05}$, where ${g/\Gamma = 0.81}$, $1.71$, $4.44$, and $7.14$ for ${N=1}$, $2$, $5$, and $8$, respectively. (e)-(h) Real part and (i)-(l) imaginary part of the output single-photon wave packet $t(\Delta)f(\Delta)$, with ${\sigma/\Gamma = 0.05}$ and $g/\Gamma$ corresponding to the black crosses in (a)-(d). (m)-(p) Real part and (q)-(t) imaginary part of the output two-photon wave packet [Eq.~(\ref{eq:two-photon_output})], with the same parameters being used as for the single-photon wave packets.}
    \label{fig:fidelity_identical_cavities}
\end{figure*}

Figure~\ref{fig:fidelity_identical_cavities} shows results for the average CZ gate fidelity $\bar{F}_{\text{CZ}}$ from Eq.~(\ref{eq:avg_F_CZ}), as well as examples of single- and two-photon output wave packets, for different numbers of identical cavities $N$. The single-photon output wave packet after each NS gate is simply given by $t(\Delta)f(\Delta)$, where $t(\Delta)$ is given in Eq.~(\ref{eq:t_one_cavity_app}) for one cavity and in Eq.~(\ref{eq:tN_app}) for the more general $N$-cavity case. The two-photon output wave packet after the NS gates is calculated by taking the two-photon state in Eq.~(\ref{eq:S_NS_2}) and projecting it onto the detunings $\Delta_1$ and $\Delta_2$, giving, for example,
\begin{align}\label{eq:two-photon_output}
\begin{split}
    &\bra{\Delta_1,\Delta_2}S_{\text{NS}_1} \ket{2}_b\\
    &= \bra{0}b(\Delta_1)b(\Delta_2)S_{\text{NS}_1} \ket{2}_b\\
    &= \sqrt{2}\;t(\Delta_1)t(\Delta_2)f(\Delta_1)f(\Delta_2)\\
    &\hspace{0.15in}+ \frac{i}{\sqrt{2}}\int {\rm d}\Delta_3\; C(\Delta_1,\Delta_2,\Delta_3) f(\Delta_3) f(\Delta_1+\Delta_2-\Delta_3)
\end{split}
\end{align}
(this would be identical for the second NS gate, since we assumed them to be identical). The expression for $C(\Delta_1,\Delta_2,\Delta_3)$ is given in Eq.~(\ref{eq:C_one_cavity_app_2}) for one cavity and in Eq.~(\ref{eq:C_N_cavities_app}) for $N$ cavities.

To obtain the results for identical cavities, we set ${\Omega_j = \Omega}$, ${g_j = g}$, and ${\Gamma_j = \Gamma}$ for all ${j\in\{1,\ldots,N\}}$. We also set the propagation phases ${\phi_j = 0}$ for all $j$ because the distances between the cavities have no effect due to chiral coupling (there is no interference as there are no counter-propagating photons).

The first row of Fig.~\ref{fig:fidelity_identical_cavities} shows the average fidelity $\bar{F}_{\text{CZ}}$ as a function of the wave packet width $\sigma$ and the emitter--cavity coupling rate $g$ for ${N=1}$, $2$, $5$, and $8$ cavities (from left to right). The maximum value of $\bar{F}_{\text{CZ}}$ is approximately $60\%$ and remains roughly constant when $N$ is increased. As expected, the highest fidelities occur for narrow wave packets (${\sigma/\Gamma \lesssim 0.1}$), since a narrower wave packet leads to a better overlap with the two-photon transition (see Fig.~\ref{fig:energy_levels}), ensuring the two-photon phase shift is applied across the entire wave packet. As an example, the optimal fidelity for an input wave packet width ${\sigma = 0.05\Gamma}$ is indicated with the black crosses in Fig.~\ref{fig:fidelity_identical_cavities}(a)-(d). For larger $N$, larger fidelities become achievable for wider wave packets, since increasing $N$ amplifies the total phase shift across the frequency range of the wave packet.

We can understand why the average fidelity is limited to about $60\%$ by observing the output wave packets. In the remaining rows of Fig.~\ref{fig:fidelity_identical_cavities}, we show the real and imaginary parts of the single- and two-photon output wave packets. The parameters used to obtain the wave packets in each column correspond to the black cross in the first figure of the respective column, where $\sigma/\Gamma = 0.05$ and ${g/\Gamma = 0.81}$, $1.71$, $4.44$, and $7.14$ for ${N=1}$, $2$, $5$, and $8$, respectively. We can see that as $N$ is increased, the real part of the single-photon wave packet vanishes [second row, Fig.~\ref{fig:fidelity_identical_cavities}(e)-(h)], and the input Gaussian wave packet becomes purely imaginary [third row, Fig.~\ref{fig:fidelity_identical_cavities}(i)-(l)], i.e., the single-photon wave packet acquires a $\pi/2$ phase shift. This is because, with the parameters listed above, we do not satisfy the condition in Eq.~(\ref{eq:inequality}), so the wave packet is not sufficiently detuned from the single-photon transitions of the emitter--cavity system to prevent a single-photon phase shift from occurring. Nevertheless, these parameters maximize the average fidelity because the two-photon output wave packet is optimized, as can be seen in the last two rows of the figure [Fig.~\ref{fig:fidelity_identical_cavities}(m)-(t)]. When $N$ is increased, the imaginary part of the two-photon wave packet vanishes and the real part becomes the same as the input wave packet in Fig.~\ref{fig:wave_packets}(b), with an additional $\pi$ phase shift as required for the CZ gate. We can observe in the real part that increasing $N$ reduces the amplitude of frequency components at larger detunings from the wave packet center, which is a consequence of spectral entanglement being reduced for larger $N$~\cite{Brod2016_1, Brod2016_2, Konyk2019, Schrinski2022, Levy-Yeyati2024}.

The results in Fig.~\ref{fig:fidelity_identical_cavities} demonstrate that there is a trade-off between optimizing the single- and two-photon components of the output state: while the single-photon component favors strong emitter--cavity coupling, as indicated by Eq.~(\ref{eq:inequality}), the two-photon component is optimized when ${g \lesssim N\Gamma}$ for the chosen input wave packet [see the black crosses in Fig.~\ref{fig:fidelity_identical_cavities}(a)-(d)]. If the emitter--cavity coupling rate $g$ is too large compared to the cavity decay rate $\Gamma$ into the waveguide, the phase shift on the two-photon wave packet vanishes (coupling from the waveguide to the cavities becomes weak relative to $g$, so the nonlinearity essentially disappears). The fact that the single- and two-photon wave packets are optimized at different values of $g$ is the reason behind the relatively low average fidelity.

\begin{figure}
    \centering
    \includegraphics[width=\linewidth]{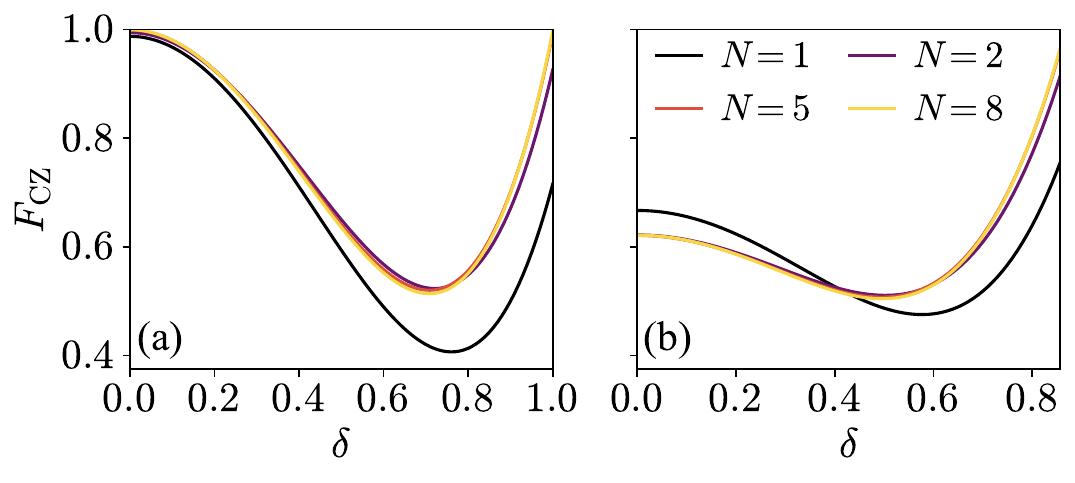}
    \caption{CZ gate fidelity as a function of the amplitude $\delta$ of the state $\ket{1,1}_L$ in the input state $\ket{\psi_{\text{in}}}$ [Eq.~(\ref{eq:input})], for ${N=1}$, $2$, $5$, and $8$ identical cavities. The remaining amplitudes are ${\beta = \gamma}$ and (a) ${\alpha=0}$ and (b) ${\alpha = 0.5}$.}
    \label{fig:fidelity_vs_delta}
\end{figure}

To further investigate how the single-photon and two-photon wave packets contribute to the CZ gate fidelity, we calculate $F_{\text{CZ}}$ in Eq.~(\ref{eq:F_CZ_I}) for a range of different input states by sweeping the amplitude $\delta$ of the $\ket{1,1}_L$ state [see Eq.~(\ref{eq:input}) for the general input state]. We use the same wave packet and cavity parameters as those indicated by the black crosses in Fig.~\ref{fig:fidelity_identical_cavities}(a)-(d), which give the optimal average fidelity for the wave packet width ${\sigma = 0.05\Gamma}$. The results for $F_{\text{CZ}}$ as a function of $\delta$ are shown in Fig.~\ref{fig:fidelity_vs_delta}, where we fix the amplitudes of the $\ket{0,1}_L$ and $\ket{1,0}_L$ states to be equal (${\beta=\gamma}$), and we set the amplitude of the $\ket{0,0}_L$ state to be ${\alpha = 0}$ in (a) and ${\alpha = 0.5}$ in (b). The normalization condition $|\alpha|^2 + |\beta|^2 + |\gamma|^2 + |\delta|^2 = 1$ with all amplitudes real and ${\beta=\gamma}$ becomes ${2\gamma^2+\delta^2 = 1}$ when ${\alpha = 0}$ and ${2\gamma^2+\delta^2 = 0.75}$ when ${\alpha = 0.5}$. By sweeping $\delta$, we change the fraction of the one- and two-photon components in the input state, since the $\ket{1,1}_L$ state corresponds to both photons passing through each NS gate (due to Hong-Ou-Mandel interference at the first beam splitter), while the $\ket{0,1}_L$ and $\ket{1,0}_L$ states correspond to one photon passing through the NS gates [see Fig.~\ref{fig:CZ_gate_diagram}(a)].

In Fig.~\ref{fig:fidelity_vs_delta}(a), where ${\alpha=0}$, the input state is given by $\ket{\psi_{\text{in}}} = \gamma(\ket{0,1}_L + \ket{1,0}_L) + \delta\ket{1,1}_L$. When ${\delta=0}$, only the single-photon output wave packet contributes to the fidelity. In this case, the $\pi/2$ phase that we observe in Fig.~\ref{fig:fidelity_identical_cavities}(i)-(l) becomes a global phase, resulting in a CZ gate fidelity close to one for all cavity numbers $N$ (the wave packet shape is approximately preserved, up to the global phase). On the opposite extreme, when ${\delta = 1}$, only the two-photon output wave packet contributes, and we see that $F_{\text{CZ}}$ increases to one as $N$ is increased, in agreement with Fig.~\ref{fig:fidelity_identical_cavities}(m)-(p), where the wave packet shape approaches that of the input wave packet (with a global phase of $\pi$). In the intermediate regime, where ${0<\delta<1}$, the fidelity drops significantly for all $N$ because the input state becomes a superposition of the $\ket{0,1}_L$, $\ket{1,0}_L$, and $\ket{1,1}_L$ states, and the phase acquired by the single-photon wave packet becomes an undesired relative phase.

In Fig.~\ref{fig:fidelity_vs_delta}(b) we set ${\alpha = 0.5}$, such that $\ket{\psi_{\text{in}}} = 0.5\ket{0,0}_L + \gamma(\ket{0,1}_L + \ket{1,0}_L) + \delta\ket{1,1}_L$. When ${\delta = 0}$, the fidelity is significantly lower than in Fig.~\ref{fig:fidelity_vs_delta}(a) because the phase on the single-photon wave packet is a relative phase due to the presence of the $\ket{0,0}_L$ state. This shows that the phase acquired by the single-photon wave packet from the interaction with the cavities is the main reason for the reduction in fidelity below unity. When $\delta$ reaches its maximum value of $\sqrt{3}/2$ for this value of $\alpha$ (corresponding to ${\gamma=0}$), the fidelity becomes close to one as $N$ is increased, showing that the two-photon wave packet is close to optimal.


\subsection{CZ gate fidelity with photon loss and disorder}\label{subsec:results_disorder}

\begin{figure}
    \centering
    \includegraphics[width=\linewidth]{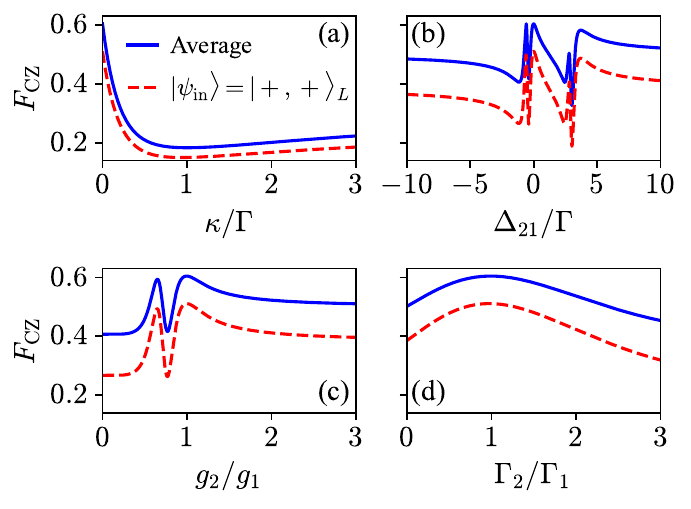}
    \caption{CZ gate fidelity for ${N=2}$ cavities as a function of (a) the photon loss rate of the emitters/cavities, (b) the detuning between the cavities, (c) the ratio of the emitter--cavity coupling rates $g_1$ and $g_2$, and (d) the ratio of the cavity decay rates $\Gamma_1$ and $\Gamma_2$. The solid blue lines show the average fidelity $\bar{F}_{\text{CZ}}$ [Eq.~(\ref{eq:avg_F_CZ})] and the dashed red lines show the fidelity for the input state $\ket{+,+}_L$ [Eq.~(\ref{eq:F_CZ_I}) with ${\alpha=\beta=\gamma=\delta=1/2}$].}
    \label{fig:fidelity_loss_disorder}
\end{figure}

Figure~\ref{fig:fidelity_loss_disorder} presents CZ gate fidelity results where we introduce imperfections into the system with ${N=2}$ cavities, for input Gaussian wave packets with width ${\sigma = 0.05\Gamma}$. In particular, Fig.~\ref{fig:fidelity_loss_disorder}(a) shows the fidelity as a function of the photon loss rate $\kappa$ of the cavities and emitters (assumed to be identical), which we introduce using the substitution ${\Omega \rightarrow \Omega - i\kappa/2}$ in the scattering matrix results~\cite{Rephaeli2013}. We show both the average fidelity $\bar{F}_{\text{CZ}}$ [Eq.~(\ref{eq:avg_F_CZ})] and the fidelity for the input state ${\ket{\psi_{\text{in}}} = \ket{+,+}_L}$ [where $\ket{+}_L = (\ket{0}_L + \ket{1}_L)/\sqrt{2}$], corresponding to Eq.~(\ref{eq:F_CZ_I}) with the input state coefficients ${\alpha=\beta=\gamma=\delta=1/2}$. As expected, the fidelity decreases when the loss rate $\kappa$ is increased. For large loss rates that exceed the cavity decay rate into the waveguide (${\kappa>\Gamma}$), the fidelity gradually increases again because coupling to the cavity becomes weak relative to $\kappa$ (the photons effectively do not enter the cavities and the transmission coefficient tends to one: ${I_1 \rightarrow 1}$ and ${I_2 \rightarrow 0}$ in the limit of large $\kappa$, so ${\bar{F}_{\text{CZ}} \rightarrow 0.4}$).

In the remaining parts of Fig.~\ref{fig:fidelity_loss_disorder}, instead of photon loss we introduce disorder into the two-cavity system. In Fig.~\ref{fig:fidelity_loss_disorder}(b), we show the fidelity as a function of the detuning ${\Delta_{21} = \Omega_2 - \Omega_1}$ between the emitters/cavities. In Fig.~\ref{fig:fidelity_loss_disorder}(c) and (d), we show the fidelity as a function of the ratio of the emitter--cavity coupling rates $g_1$ and $g_2$ and the cavity decay rates $\Gamma_1$ and $\Gamma_2$, respectively. In all cases, we observe that the highest fidelities occur for identical cavities. When ${\Delta_{21} = 0}$, ${g_2 = g_1}$, and ${\Gamma_2 = \Gamma_1}$, we obtain a peak average fidelity of approximately $60\%$, corresponding to the black cross in Fig.~\ref{fig:fidelity_identical_cavities}(b). We also find that the CZ gate fidelity corresponding to the input state $\ket{+,+}_L$ is below average, with a peak value of about $51\%$.


\section{Conclusion}\label{sec:conclusion}

In conclusion, we have analyzed the performance of an array of $N$ cavities with two-level emitters in a photonic CZ gate. We found that increasing the number of cavities reduces spectral entanglement arising from two-photon scattering~\cite{Brod2016_1, Brod2016_2, Konyk2019, Schrinski2022, Levy-Yeyati2024}, resulting in a close-to-optimal two-photon wave packet with a $\pi$ phase shift already with only a few cavities [Fig.~\ref{fig:fidelity_identical_cavities}(n)-(p)]. However, for the parameters where the two-photon wave packet is optimized, the single-photon wave packet is distorted due to a significant overlap with the single-photon transitions of the emitter--cavity system. This overlap can be reduced by increasing the emitter--cavity coupling rates to satisfy the condition in Eq.~(\ref{eq:inequality}), but for large emitter--cavity coupling rates the two-photon nonlinearity vanishes. The trade-off between optimizing the single-photon and two-photon components of the input state limits the fidelity of the CZ gate to approximately $60\%$ for Gaussian input wave packets, when averaged over all possible logical input states.

In this paper, we considered a simple passive system with no active control of the emitter or cavity parameters. Future work may look at more complex setups to try to improve the fidelity results presented here, for example, using multi-level emitters or time-dependent coupling rates that can provide greater control of the output wave packet shape~\cite{Heuck2020, Heuck2020_2, Krastanov2022, Hassan2023}.


\section*{Acknowledgments}

The authors thank Matias Bundgaard-Nielsen for helpful discussions. MD was supported by the Engineering and Physical Sciences Research Council (Grant No. EP/W524360/1). PK acknowledges financial support from the UK Engineering and Physical Sciences Research Council and the IQN Quantum Hub (Grant No. EP/Z533208/1).


\appendix
\begin{widetext}


\section{CZ gate fidelity derivation}\label{app:fidelity}

In this appendix, we derive the CZ gate fidelity in Eq.~(\ref{eq:F_CZ_S}), which then leads to the expression in Eq.~(\ref{eq:F_CZ_I}). We begin by substituting the basis states in Eq.~(\ref{eq:basis}) into Eq.~(\ref{eq:CZ_output_ideal}) to obtain the ideal CZ gate output state
\begin{equation}\label{eq:out_ideal}
    \ket{\psi_{\text{out}}} = \int {\rm d}\omega \int {\rm d}\omega' f(\omega) f(\omega')\Bigl[ \alpha a^{\dagger}(\omega) d^{\dagger}(\omega') + \beta a^{\dagger}(\omega) c^{\dagger}(\omega') + \gamma b^{\dagger}(\omega) d^{\dagger}(\omega') - \delta b^{\dagger}(\omega) c^{\dagger}(\omega')  \Bigr]\ket{0,0,0,0}_{abcd}.
\end{equation}
To calculate the fidelity ${F_{\text{CZ}} = |\langle \psi_{\text{out}}| \tilde{\psi}_{\text{out}} \rangle |^2}$ [Eq.~(\ref{eq:F_CZ})], we need to find the actual output state for our gate, $| \tilde{\psi}_{\text{out}} \rangle$ [Eq.~(\ref{eq:CZ_output_actual})], and calculate the inner product with the ideal state above. Substituting the basis states in Eq.~(\ref{eq:basis}) into Eq.~(\ref{eq:input}) gives the input state
\begin{equation}
    \ket{\psi_{\text{in}}} = \int {\rm d}\omega \int {\rm d}\omega' f(\omega) f(\omega')\Bigl[ \alpha a^{\dagger}(\omega) d^{\dagger}(\omega') + \beta a^{\dagger}(\omega) c^{\dagger}(\omega') + \gamma b^{\dagger}(\omega) d^{\dagger}(\omega') + \delta b^{\dagger}(\omega) c^{\dagger}(\omega')  \Bigr]\ket{0,0,0,0}_{abcd},
\end{equation}
which means that Eq.~(\ref{eq:CZ_output_actual}) becomes
\begin{align}\label{eq:out_1}
\begin{split}
    |\tilde{\psi}_{\text{out}}\rangle =&\; U_{\text{BS}_2}\left( S_{\text{NS}_1} \otimes S_{\text{NS}_2} \right) U_{\text{BS}_1}\int {\rm d}\omega \int {\rm d}\omega' f(\omega) f(\omega') \Bigl[ \alpha a^{\dagger}(\omega) d^{\dagger}(\omega') + \beta a^{\dagger}(\omega) c^{\dagger}(\omega')\\
    &\hspace{2.9in}+ \gamma b^{\dagger}(\omega) d^{\dagger}(\omega') + \delta b^{\dagger}(\omega) c^{\dagger}(\omega') \Bigr] \ket{0,0,0,0}_{abcd}.
\end{split}
\end{align}
The unitary operator for the first beam splitter (BS$_1$) transforms modes $b$ and $c$ according to
\begin{equation}
    b^{\dagger}(\omega) \xrightarrow[]{\text{BS$_1$}} \frac{1}{\sqrt{2}}\left[ b^{\dagger}(\omega) - c^{\dagger}(\omega) \right] \quad \text{and} \quad c^{\dagger}(\omega) \xrightarrow[]{\text{BS$_1$}} \frac{1}{\sqrt{2}}\left[ b^{\dagger}(\omega) + c^{\dagger}(\omega) \right],
\end{equation}
such that Eq.~(\ref{eq:out_1}) becomes
\begin{align}\label{eq:out_2}
\begin{split}
    |\tilde{\psi}_{\text{out}}\rangle =&\; U_{\text{BS}_2}\left( S_{\text{NS}_1} \otimes S_{\text{NS}_2} \right) \int {\rm d}\omega \int {\rm d}\omega' f(\omega) f(\omega') \biggl( \alpha a^{\dagger}(\omega) d^{\dagger}(\omega') + \frac{\beta}{\sqrt{2}}\left[ a^{\dagger}(\omega) b^{\dagger}(\omega') + a^{\dagger}(\omega) c^{\dagger}(\omega') \right]\\
    &\hspace{2.6in}+ \frac{\gamma}{\sqrt{2}} \left[ b^{\dagger}(\omega) d^{\dagger}(\omega') - c^{\dagger}(\omega) d^{\dagger}(\omega') \right]\\
    &\hspace{2.6in}+ \frac{\delta}{2} \left[b^{\dagger}(\omega) b^{\dagger}(\omega') - c^{\dagger}(\omega) c^{\dagger}(\omega')\right] \biggr) \ket{0,0,0,0}_{abcd}
\end{split}
\end{align}
after applying $U_{\text{BS}_1}$. The last line in Eq.~(\ref{eq:out_2}) corresponds to the situation where both photons approach the first beam splitter and exit through the same port due to Hong-Ou-Mandel interference. Using
\begin{align}\label{eq:input_states_app}
\begin{gathered}
    \ket{1}_a = \int {\rm d}\omega f(\omega)a^{\dagger}(\omega)\ket{0}_a,\\
    \ket{2}_b = \frac{1}{\sqrt{2}} \int {\rm d}\omega \int {\rm d}\omega' f(\omega) f(\omega') b^{\dagger}(\omega) b^{\dagger}(\omega') \ket{0}_b,
\end{gathered}
\end{align}
and similarly for the other modes, we can write Eq.~(\ref{eq:out_2}) as
\begin{align}\label{eq:out_3}
\begin{split}
    |\tilde{\psi}_{\text{out}}\rangle =&\; U_{\text{BS}_2}\left( S_{\text{NS}_1} \otimes S_{\text{NS}_2} \right) \Bigl[ \alpha \ket{1,0,0,1}_{abcd} + \frac{\beta}{\sqrt{2}}\left( \ket{1,1,0,0}_{abcd} + \ket{1,0,1,0}_{abcd} \right)\\
    &+ \frac{\gamma}{\sqrt{2}} \left( \ket{0,1,0,1}_{abcd} - \ket{0,0,1,1}_{abcd} \right) + \frac{\delta}{\sqrt{2}} \left(\ket{0,2,0,0}_{abcd} - \ket{0,0,2,0}_{abcd}\right) \Bigr].
\end{split}
\end{align}
We now apply the scattering matrix $S_{\text{NS}_1}$ to mode $b$, and $S_{\text{NS}_2}$ to mode $c$ [see Fig.~\ref{fig:CZ_gate_diagram}(a)]. From Eq.~(\ref{eq:out_3}), it follows that these modes can contain zero, one, or two photons. In the trivial case of zero photons, the states do not change:
\begin{equation}\label{eq:S_NS_0}
    S_{\text{NS}_1} \ket{0}_b = \ket{0}_b, \quad S_{\text{NS}_2} \ket{0}_c = \ket{0}_c.
\end{equation}
In the case of one photon, the resulting states are given in terms of single-photon scattering matrix elements $S_{pk}$:
\begin{align}\label{eq:S_NS_1}
\begin{split}
    S_{\text{NS}_1} \ket{1}_b =& \int {\rm d}p \int {\rm d}k\; S_{pk} f(k) b^{\dagger}(p)\ket{0}_b,\\
    S_{\text{NS}_2} \ket{1}_c =& \int {\rm d}p \int {\rm d}k\; S_{pk} f(k) c^{\dagger}(p)\ket{0}_c,
\end{split}
\end{align}
where we have assumed that $S_{\text{NS}_1}$ and $S_{\text{NS}_2}$ have the same matrix elements $S_{pk}$, as mentioned in the main text. For two photons, the outputs of the NS gates are given in terms of two-photon scattering matrix elements $S_{p_1p_2k_1k_2}$:
\begin{align}\label{eq:S_NS_2}
\begin{split}
    S_{\text{NS}_1} \ket{2}_b =&\; \frac{1}{2\sqrt{2}} \int {\rm d}p_1 \int {\rm d}p_2 \int {\rm d}k_1 \int {\rm d}k_2\; S_{p_1p_2k_1k_2} f(k_1) f(k_2) b^{\dagger}(p_1) b^{\dagger}(p_2) \ket{0}_b,\\
    S_{\text{NS}_2} \ket{2}_c =&\; \frac{1}{2\sqrt{2}} \int {\rm d}p_1 \int {\rm d}p_2 \int {\rm d}k_1 \int {\rm d}k_2\; S_{p_1p_2k_1k_2} f(k_1) f(k_2) c^{\dagger}(p_1) c^{\dagger}(p_2) \ket{0}_c,
\end{split}
\end{align}
where we have also assumed the same matrix elements in the two-photon case. Using Eqs.~(\ref{eq:S_NS_0})-(\ref{eq:S_NS_2}) in Eq.~(\ref{eq:out_3}) leads to
\begin{align}\label{eq:out_4}
\begin{split}
    |\tilde{\psi}_{\text{out}}\rangle = U_{\text{BS}_2}&\biggl(\alpha \int {\rm d}\omega \int {\rm d}\omega' f(\omega) f(\omega') a^{\dagger}(\omega) d^{\dagger}(\omega')\\
    &+ \frac{1}{\sqrt{2}} \int {\rm d}\omega \int {\rm d}p \int {\rm d}k\; S_{pk}f(\omega)f(k) \left[ \beta a^{\dagger}(\omega) b^{\dagger}(p) + \beta a^{\dagger}(\omega) c^{\dagger}(p) + \gamma b^{\dagger}(p) d^{\dagger}(\omega) - \gamma c^{\dagger}(p) d^{\dagger}(\omega) \right]\\
    &+ \frac{\delta}{4} \int {\rm d}p_1 \int {\rm d}p_2 \int {\rm d}k_1 \int {\rm d}k_2\; S_{p_1p_2k_1k_2} f(k_1) f(k_2) \left[b^{\dagger}(p_1) b^{\dagger}(p_2) - c^{\dagger}(p_1) c^{\dagger}(p_2)\right] \biggr) \ket{0,0,0,0}_{abcd}.
\end{split}
\end{align}
We now need to apply the second beam splitter to modes $b$ and $c$, which performs the following mode transformations:
\begin{equation}
    b^{\dagger}(\omega) \xrightarrow[]{\text{BS$_2$}} \frac{1}{\sqrt{2}}\left[ b^{\dagger}(\omega) + c^{\dagger}(\omega) \right] \quad \text{and} \quad c^{\dagger}(\omega) \xrightarrow[]{\text{BS$_2$}} \frac{1}{\sqrt{2}}\left[ c^{\dagger}(\omega) - b^{\dagger}(\omega) \right].
\end{equation}
Equation~(\ref{eq:out_4}) then becomes
\begin{align}\label{eq:out_5}
\begin{split}
    |\tilde{\psi}_{\text{out}}\rangle = &\biggl(\alpha \int {\rm d}\omega \int {\rm d}\omega' f(\omega) f(\omega') a^{\dagger}(\omega) d^{\dagger}(\omega') + \int {\rm d}\omega \int {\rm d}p \int {\rm d}k\; S_{pk}f(\omega)f(k) \left[ \beta a^{\dagger}(\omega) c^{\dagger}(p) + \gamma b^{\dagger}(p) d^{\dagger}(\omega) \right]\\
    &+ \frac{\delta}{2} \int {\rm d}p_1 \int {\rm d}p_2 \int {\rm d}k_1 \int {\rm d}k_2\; S_{p_1p_2k_1k_2} f(k_1) f(k_2) b^{\dagger}(p_1) c^{\dagger}(p_2) \biggr) \ket{0,0,0,0}_{abcd}.
\end{split}
\end{align}
The CZ gate fidelity $F_{\text{CZ}}$ in Eq.~(\ref{eq:F_CZ_S}) is then obtained by calculating the inner product between the ideal CZ gate output state in Eq.~(\ref{eq:out_ideal}) and the actual output state in Eq.~(\ref{eq:out_5}), and then taking the modulus squared. We calculate the inner product using the bosonic commutation relations of the mode operators (${[a(\omega), a^{\dagger}(\omega')] = \delta(\omega-\omega')}$ etc.). To get from Eq.~(\ref{eq:F_CZ_S}) to Eq.~(\ref{eq:F_CZ_I}), we simply substitute ${S_{pk} = t(k)\delta(p-k)}$ and $S_{p_1p_2k_1k_2} = t(k_1)t(k_2)[\delta({p_1-k_1})\delta({p_2-k_2}) + \delta({p_1-k_2})\delta({p_2-k_1})] + iC_{p_1p_2k_1k_2}\delta({p_1+p_2-k_1-k_2})$ into Eq.~(\ref{eq:F_CZ_S}). After integrating over the delta functions, we find
\begin{align}\label{eq:F_CZ_int}
\begin{split}
    F_{\text{CZ}} =&\; \biggl| |\alpha|^2 + \left( |\beta|^2 + |\gamma|^2 \right) \int {\rm d}p\; t(p) |f(p)|^2 - |\delta|^2 \int {\rm d}p_1 \int {\rm d}p_2\; t(p_1) t(p_2) |f(p_1)|^2 |f(p_2)|^2\\
    &- \frac{i}{2}|\delta|^2 \int {\rm d}p_1 \int {\rm d}p_2 \int {\rm d}k_1\; C_{p_1,p_2,k_1,p_1+p_2-k_1} f^*(p_1) f^*(p_2) f(k_1) f(p_1+p_2-k_1) \biggr|^2\\
    =&\; \biggl| |\alpha|^2 + \left( |\beta|^2 + |\gamma|^2 \right) \int {\rm d}\omega\; t(\omega) |f(\omega)|^2 - |\delta|^2 \left( \int {\rm d}\omega\; t(\omega) |f(\omega)|^2 \right)^2\\
    &- \frac{i}{2}|\delta|^2 \int {\rm d}\omega \int {\rm d}\omega' \int {\rm d}\omega''\; C(\omega, \omega', \omega'') f^*(\omega) f^*(\omega') f(\omega'') f(\omega+\omega'-\omega'') \biggr|^2,
\end{split}
\end{align}
where, to get from the first line to the second line, we relabeled the frequency variables and wrote $C$ as a function of three variables, since the fourth frequency variable was integrated over using the delta function $\delta({p_1+p_2-k_1-k_2})$ (where $k_2$ was set to ${p_1+p_2-k_1}$). Equation~(\ref{eq:F_CZ_int}) is equivalent to Eq.~(\ref{eq:F_CZ_I}) in the main text [where the integrals $I_1$ and $I_2$ are given in Eqs.~(\ref{eq:I_1}) and (\ref{eq:I_2}), respectively].


\section{Average CZ gate fidelity}\label{app:average_fidelity}

To obtain the average CZ gate fidelity $\bar{F}_{\text{CZ}}$ in Eq.~(\ref{eq:avg_F_CZ}), we integrate over all possible input states using the Haar measure, following the supplemental material of Ref.~\cite{Brod2016_1}. We first parameterize the complex coefficients $\alpha$, $\beta$, $\gamma$, and $\delta$ in the input state $\ket{\psi_{\text{in}}}$ [Eq.~(\ref{eq:input})]. A complex number can be represented by two real parameters (its real and imaginary parts), so we would need eight parameters for our four coefficients. However, the constraint $|\alpha|^2 + |\beta|^2 + |\gamma|^2 + |\delta|^2 = 1$ allows one of the variables to be fixed, so seven real parameters are required in total:
\begin{align}\label{eq:parameterization}
\begin{gathered}
    \alpha = e^{i\chi_0}\text{cos}\theta_0,\\
    \beta = e^{i\chi_1}\text{sin}\theta_0\;\text{cos}\theta_1,\\
    \gamma = e^{i\chi_2}\text{sin}\theta_0\;\text{sin}\theta_1\;\text{cos}\theta_2,\\
    \delta = e^{i\chi_3}\text{sin}\theta_0\;\text{sin}\theta_1\;\text{sin}\theta_2,
\end{gathered}
\end{align}
where ${0 \leq \chi_i < 2\pi}$ and ${0 \leq \theta_i < \pi/2}$. The Haar measure for this parameter space is given by
\begin{equation}
    {\rm d}\mu = \frac{48}{(2\pi)^4} \text{sin}^5\theta_0\;\text{cos}\theta_0\;\text{sin}^3\theta_1\;\text{cos}\theta_1\;\text{sin}\theta_2\;\text{cos}\theta_2\; {\rm d}\chi_0 {\rm d}\chi_1 {\rm d}\chi_2 {\rm d}\chi_3 {\rm d}\theta_0 {\rm d}\theta_1 {\rm d}\theta_2,
\end{equation}
which satisfies $\int {\rm d}\mu = 1$ when integrated over the full range of parameters. The fidelity averaged over all input states is then given by Eq.~(\ref{eq:F_CZ_I}) with the parameterization in Eq.~(\ref{eq:parameterization}), integrated over all the parameters with respect to the Haar measure:
\begin{align}
\begin{split}
    \bar{F}_{\text{CZ}} =&\; 48 \int_0^{\pi/2} {\rm d}\theta_0 \int_0^{\pi/2} {\rm d}\theta_1 \int_0^{\pi/2} {\rm d}\theta_2\; \text{sin}^5\theta_0\;\text{cos}\theta_0\;\text{sin}^3\theta_1\;\text{cos}\theta_1\;\text{sin}\theta_2\;\text{cos}\theta_2\\
    &\times \left| \text{cos}^2\theta_0 + \text{sin}^2\theta_0\left( \text{cos}^2\theta_1 + \text{sin}^2\theta_1\;\text{cos}^2\theta_2 \right)I_1 - \text{sin}^2\theta_0\;\text{sin}^2\theta_1\;\text{sin}^2\theta_2 \left( I_1^2 + \frac{i}{2}I_2 \right)  \right|^2,
\end{split}
\end{align}
where we have already integrated over the parameters $\chi_i$, since $F_{\text{CZ}}$ in Eq.~(\ref{eq:F_CZ_I}) does not depend on them. Multiplying out the modulus squared and collecting terms with the same trigonometric functions gives
\begin{align}
\begin{split}
    \bar{F}_{\text{CZ}} =&\; 48 \int_0^{\pi/2} {\rm d}\theta_0 \int_0^{\pi/2} {\rm d}\theta_1 \int_0^{\pi/2} {\rm d}\theta_2\; \text{sin}^5\theta_0\;\text{cos}\theta_0\;\text{sin}^3\theta_1\;\text{cos}\theta_1\;\text{sin}\theta_2\;\text{cos}\theta_2\\
    &\times \Biggl( \text{cos}^4\theta_0 + \left|I_1\right|^2\text{sin}^4\theta_0\;\text{cos}^4\theta_1 + \left(I_1 + I_1^*\right)\text{sin}^2\theta_0\;\text{cos}^2\theta_0\;\text{cos}^2\theta_1 + \left| I_1 \right|^2 \text{sin}^4\theta_0\;\text{sin}^4\theta_1\;\text{cos}^4\theta_2\\
    &\hspace{0.3in}+ \left| I_1^2 + \frac{i}{2}I_2 \right|^2 \text{sin}^4\theta_0\;\text{sin}^4\theta_1\;\text{sin}^4\theta_2 + \left(I_1 + I_1^*\right)\text{sin}^2\theta_0\;\text{cos}^2\theta_0\;\text{sin}^2\theta_1\;\text{cos}^2\theta_2\\
    &\hspace{0.3in}- \left[ I_1^2 + \frac{i}{2}I_2 + \left(I_1^*\right)^2 - \frac{i}{2}I_2^* \right]\text{sin}^2\theta_0\;\text{cos}^2\theta_0\;\text{sin}^2\theta_1\;\text{sin}^2\theta_2 + 2|I_1|^2\text{sin}^4\theta_0\;\text{sin}^2\theta_1\;\text{cos}^2\theta_1\;\text{cos}^2\theta_2\\
    &\hspace{0.3in}- \left( I_1^*\left(I_1^2 + \frac{i}{2}I_2\right) + I_1\left[ \left(I_1^*\right)^2 - \frac{i}{2}I_2^* \right] \right)\text{sin}^4\theta_0\;\text{sin}^2\theta_1\;\text{cos}^2\theta_1\;\text{sin}^2\theta_2\\
    &\hspace{0.3in}- \left( I_1^*\left(I_1^2 + \frac{i}{2}I_2\right) + I_1\left[ \left(I_1^*\right)^2 - \frac{i}{2}I_2^* \right] \right)\text{sin}^4\theta_0\;\text{sin}^4\theta_1\;\text{sin}^2\theta_2\;\text{cos}^2\theta_2\Biggr).
\end{split}
\end{align}
All the integrals over $\theta_0$, $\theta_1$, and $\theta_2$ can be evaluated analytically by substitution, and simplifying the terms eventually leads to the final expression in Eq.~(\ref{eq:avg_F_CZ}).


\section{One cavity---single-photon scattering}\label{app:one_cavity_one_photon}

The input--output relation in Eq.~(\ref{eq:input-output}) for a single emitter--cavity system can be derived from the Heisenberg equation for the waveguide operator $a(\omega,t)$:
\begin{equation}
    \frac{\rm d}{{\rm d}t}a(\omega,t) = i\left[H, a(\omega,t)\right] = -i\omega a(\omega,t) - i\sqrt{\frac{\Gamma}{2\pi}} c(t),
\end{equation}
where we used the Hamiltonian in Eq.~(\ref{eq:H}) and the bosonic commutation relations of the mode operators. Multiplying by $e^{i\omega t}$ and rearranging leads to
\begin{equation}\label{eq:Heisenberg_wg}
    \frac{\rm d}{{\rm d}t}\left[ a(\omega,t) e^{i\omega t} \right] = -i\sqrt{\frac{\Gamma}{2\pi}} c(t) e^{i\omega t}.
\end{equation}
Relabeling $t$ with $t'$, and integrating from an “input time" $t_0$ to some time $t$ gives
\begin{equation}
    a(\omega,t)e^{i\omega t} - a(\omega,t_0)e^{i\omega t_0} = -i\sqrt{\frac{\Gamma}{2\pi}} \int_{t_0}^t {\rm d}t' c(t') e^{i\omega t'}.
\end{equation}
We now multiply by $e^{-i\omega t}$, divide by $\sqrt{2\pi}$, and integrate over all $\omega$:
\begin{align}\label{eq:Heisenberg_wg_in}
\begin{split}
    \frac{1}{\sqrt{2\pi}} \int a(\omega,t){\rm d}\omega - a_{\text{in}}(t) =& -i\sqrt{\Gamma} \int_{t_0}^t {\rm d}t' c(t') \left( \frac{1}{2\pi} \int {\rm d}\omega\; e^{-i\omega(t-t')} \right)\\
    =& -i\sqrt{\Gamma}\int_{t_0}^t {\rm d}t' c(t') \delta(t-t')\\
    =& -\frac{i}{2} \sqrt{\Gamma} c(t),
\end{split}
\end{align}
where we acquired a factor of $1/2$ because the delta function $\delta(t-t')$ is centered at one of the integration limits, and we used the definition
\begin{equation}
    a_{\text{in}}(t) = \frac{1}{\sqrt{2\pi}} \int a(\omega,t_0) e^{-i\omega(t-t_0)} {\rm d}\omega
\end{equation}
of an input operator in the input--output formalism~\cite{Gardiner1985}. Returning to Eq.~(\ref{eq:Heisenberg_wg}), relabeling $t$ with $t'$, integrating from some time $t$ to an “output time" $t_1$, and repeating the remaining steps outlined above gives
\begin{equation}\label{eq:Heisenberg_wg_out}
    a_{\text{out}}(t) - \frac{1}{\sqrt{2\pi}} \int a(\omega,t){\rm d}\omega = -\frac{i}{2} \sqrt{\Gamma} c(t),
\end{equation}
where
\begin{equation}
    a_{\text{out}}(t) = \frac{1}{\sqrt{2\pi}} \int a(\omega,t_1) e^{-i\omega(t-t_1)} {\rm d}\omega
\end{equation}
is the definition of an output operator in the input--output formalism~\cite{Gardiner1985}. Rearranging Eqs.~(\ref{eq:Heisenberg_wg_in}) and (\ref{eq:Heisenberg_wg_out}), we obtain
\begin{subequations}
\begin{equation}\label{eq:a_in}
    a_{\text{in}}(t) = \frac{1}{\sqrt{2\pi}} \int a(\omega,t){\rm d}\omega + \frac{i}{2} \sqrt{\Gamma} c(t),
\end{equation}
\begin{equation}
    a_{\text{out}}(t) = \frac{1}{\sqrt{2\pi}} \int a(\omega,t){\rm d}\omega - \frac{i}{2} \sqrt{\Gamma} c(t),
\end{equation}
\end{subequations}
which immediately leads to the input--output relation in Eq.~(\ref{eq:input-output}).

To obtain the Heisenberg equation for the cavity operator [Eq.~(\ref{eq:Heisenberg_cavity})], we first use the Hamiltonian in Eq.~(\ref{eq:H}) to find
\begin{equation}
    \frac{\rm d}{{\rm d}t}c(t) = i\left[H, c(t)\right] = -i\Omega c(t) - ig \sigma^-(t) - i \sqrt{\frac{\Gamma}{2\pi}} \int a(\omega,t) {\rm d }\omega.
\end{equation}
We then use the result for the input operator $a_{\text{in}}(t)$ from Eq.~(\ref{eq:a_in}) to eliminate the integral:
\begin{align}
\begin{split}
    \frac{\rm d}{{\rm d}t}c(t) =& -i\Omega c(t) - ig \sigma^-(t) - i \sqrt{\Gamma} \left[ a_{\text{in}}(t) - \frac{i}{2}\sqrt{\Gamma}c(t) \right]\\
    =& -i\left( \Omega - \frac{i}{2}\Gamma \right) c(t) - ig\sigma^-(t) - i\sqrt{\Gamma}a_{\text{in}}(t),
\end{split}
\end{align}
as in Eq.~(\ref{eq:Heisenberg_cavity}).

To Fourier transform the input--output relation and Heisenberg equations from time $t$ to frequency $p$, we use
\begin{align}\label{eq:Fourier}
\begin{split}
    a_{\text{in}}(t) =&\; \frac{1}{\sqrt{2\pi}} \int a_{\text{in}}(p) e^{-ipt} {\rm d}p,\\
    a_{\text{out}}(t) =&\; \frac{1}{\sqrt{2\pi}} \int a_{\text{out}}(p) e^{-ipt} {\rm d}p,\\
    c(t) =&\; \frac{1}{\sqrt{2\pi}} \int c(p) e^{-ipt} {\rm d}p,\\
    \sigma^-(t) =&\; \frac{1}{\sqrt{2\pi}} \int \sigma^-(p) e^{-ipt} {\rm d}p.
\end{split}
\end{align}
The simple relationship in Eq.~(\ref{eq:Fourier}) between $a_{\text{in/out}}(t)$ from the input--output formalism and $a_{\text{in/out}}(p)$ in the scattering matrix holds provided that we take the input time $t_0$ to be the distant past (${t_0 \rightarrow -\infty}$) and the output time $t_1$ to be the distant future (${t_1 \rightarrow \infty}$), long before/after any interactions take place~\cite{Fan2010}.

When we calculate the CZ gate fidelity, we integrate over detunings as opposed to absolute frequencies. We therefore need to express the single-photon transmission coefficient $t(p)$ in Eq.~(\ref{eq:t_one_cavity}) as a function of the detuning $\Delta$ between the scattered photon and the emitter/cavity. This is straightforward in the case of $t(p)$, as it is already expressed in terms of the detuning ${\Delta(p) = p - \Omega}$. Defining the detuning variable ${\Delta = \Delta(p)}$ means that
\begin{equation}\label{eq:t_one_cavity_app}
    t(\Delta) = \frac{\Delta^2 - g^2 - \frac{i}{2}\Gamma\Delta}{\Delta^2 - g^2 + \frac{i}{2}\Gamma\Delta}.
\end{equation}


\section{One cavity---two-photon scattering}\label{app:one_cavity_two_photons}

When we Fourier transform the Heisenberg equations derived in Section~\ref{subsubsec:theory_one_cavity_two_photons}, we treat products of operators as a single operator. For example, to get from Eq.~(\ref{eq:Heisenberg_sigma_2}) to Eq.~(\ref{eq:Heisenberg_sigma_p2}), we use
\begin{equation}\label{eq:Fourier_product}
    \sigma^+\sigma^-c(t) = \frac{1}{\sqrt{2\pi}}\int \sigma^+\sigma^-c(p) e^{-ipt} {\rm d}p,
\end{equation}
and then apply the convolution theorem in Eq.~(\ref{eq:convolution_sigma}) to separate the operators.

To find the Heisenberg equation for $\sigma^-(t)c(t)$ [Eq.~(\ref{eq:Heisenberg_sigma_c_time})], we calculate the commutator with the Hamiltonian in Eq.~(\ref{eq:H}):
\begin{align}
\begin{split}
    \frac{\rm d}{{\rm d}t}\sigma^-(t)c(t) =&\; i\left[H, \sigma^-(t)c(t)\right]\\
    =& -2i\Omega \sigma^-(t)c(t) + ig\sigma_z(t)c^2(t) - i\sqrt{\frac{\Gamma}{2\pi}}\sigma^-(t) \int a(\omega,t){\rm d}\omega.
\end{split}
\end{align}
We then use Eq.~(\ref{eq:a_in}) to eliminate the integral in the last term:
\begin{align}
\begin{split}
    \frac{\rm d}{{\rm d}t}\sigma^-(t)c(t) =& -2i\Omega \sigma^-(t)c(t) + ig\sigma_z(t)c^2(t) - i\sqrt{\Gamma}\sigma^-(t) \left[a_{\text{in}}(t) - \frac{i}{2}\sqrt{\Gamma}c(t)\right]\\
    =& -i\left( 2\Omega - \frac{i}{2}\Gamma \right)\sigma^-(t)c(t) + ig\sigma_z(t)c^2(t) -i\sqrt{\Gamma}\sigma^-(t)a_{\text{in}}(t),
\end{split}
\end{align}
as in Eq.~(\ref{eq:Heisenberg_sigma_c_time}). Later, when we Fourier transform from time $t$ to frequency $p$, we treat each operator product as a single operator [like in Eq.~(\ref{eq:Fourier_product})], and then replace $p$ with ${p_1+p_2}$ to find $\matrixel{0}{\sigma^-c(p_1+p_2)}{k_1k_2^+}$ [see Eq.~(\ref{eq:Heisenberg_sigma_c})].

Similarly, to find the Heisenberg equation for $c^2(t)$ [Eq.~(\ref{eq:Heisenberg_c2_time})], we solve
\begin{align}
\begin{split}
    \frac{\rm d}{{\rm d}t}c^2(t) =&\; i\left[H, c^2(t)\right]\\
    =& - 2i\Omega c^2(t) - 2ig\sigma^-(t)c(t) - 2i\sqrt{\frac{\Gamma}{2\pi}}c(t) \int a(\omega,t) {\rm d}\omega,
\end{split}
\end{align}
and again use Eq.~(\ref{eq:a_in}) to eliminate the integral:
\begin{align}
\begin{split}
    \frac{\rm d}{{\rm d}t}c^2(t) =& - 2i\Omega c^2(t) - 2ig\sigma^-(t)c(t) - 2i\sqrt{\Gamma}c(t) \left[a_{\text{in}}(t) - \frac{i}{2}\sqrt{\Gamma}c(t)\right]\\
    =& -2i\left( \Omega - \frac{i}{2}\Gamma \right)c^2(t) - 2ig\sigma^-(t)c(t) -2i\sqrt{\Gamma}c(t)a_{\text{in}}(t),
\end{split}
\end{align}
as in Eq.~(\ref{eq:Heisenberg_c2_time}). We then Fourier transform from time $t$ to frequency $p$, and replace $p$ with ${p_1+p_2}$ to find $\matrixel{0}{c^2(p_1+p_2)}{k_1k_2^+}$ [see Eq.~(\ref{eq:Heisenberg_c2})].

Once we obtain simultaneous equations involving $\matrixel{0}{\sigma^-c(p_1+p_2)}{k_1k_2^+}$ and $\matrixel{0}{c^2(p_1+p_2)}{k_1k_2^+}$ from the Heisenberg equations for $\sigma^-(t)c(t)$ and $c^2(t)$, we solve for $\matrixel{0}{\sigma^-c(p_1+p_2)}{k_1k_2^+}$ and substitute the result into Eq.~(\ref{eq:S2_linear_complete_2}) to find the two-photon scattering matrix. Rearranging Eq.~(\ref{eq:c2_matrix_el}) for $\matrixel{0}{c^2(p_1+p_2)}{k_1k_2^+}$ and substituting the result into Eq.~(\ref{eq:sigma_c_matrix_el}) leads to
\begin{equation}
    \matrixel{0}{\sigma^-c(p_1+p_2)}{k_1k_2^+} = \sqrt{\frac{\Gamma}{2\pi}} \left( \frac{2g\left[ s_c(k_1) + s_c(k_2) \right] + \left(p_1+p_2-2\Omega + i\Gamma\right)\left[ s_e(k_1) + s_e(k_2) \right]}{\left(p_1+p_2-2\Omega+\frac{i}{2}\Gamma\right)\left(p_1+p_2-2\Omega+i\Gamma\right) - 2g^2} \right)\delta(p_1+p_2-k_1-k_2).
\end{equation}
This means that Eq.~(\ref{eq:S2_linear_complete_2}) becomes
\begin{align}\label{eq:C_one_cavity_app}
\begin{split}
    S_{p_1p_2k_1k_2} =&\; S_{p_1k_2}S_{p_2k_1} + S_{p_1k_1}S_{p_2k_2}\\
    &+ i\frac{g\sqrt{\Gamma}}{\pi}s_e(p_1)s_e(p_2) \left( \frac{2g\left[ s_c(k_1) + s_c(k_2) \right] + \left(k_1+k_2-2\Omega + i\Gamma\right)\left[ s_e(k_1) + s_e(k_2) \right]}{\left(k_1+k_2-2\Omega+\frac{i}{2}\Gamma\right)\left(k_1+k_2-2\Omega+i\Gamma\right) - 2g^2} \right)\delta(p_1+p_2-k_1-k_2)\\
    =&\; S_{p_1k_2}S_{p_2k_1} + S_{p_1k_1}S_{p_2k_2} + iC_{p_1p_2k_1k_2}\delta(p_1+p_2-k_1-k_2),
\end{split}
\end{align}
where we replaced ${p_1+p_2}$ with ${k_1+k_2}$ using the delta function $\delta(p_1+p_2-k_1-k_2)$. To get to Eq.~(\ref{eq:C_one_cavity}), we factorized the denominator into the form $(k_1+k_2-\lambda_+)(k_1+k_2-\lambda_-)$, where $\lambda_+$ and $\lambda_-$ are given in Eq.~(\ref{eq:lambdas}).

We now express $C_{p_1p_2k_1k_2}$ in terms of detunings from the emitter/cavity. In particular, we first set $k_2 = p_1 + p_2 - k_1$ using the delta function, and define the detuning variables ${\Delta = p_1-\Omega}$, ${\Delta' = p_2-\Omega}$, and ${\Delta'' = k_1-\Omega}$. The result in Eq.~(\ref{eq:C_one_cavity_app}) then implies that
\begin{equation}\label{eq:C_one_cavity_app_2}
    C(\Delta, \Delta', \Delta'') = \frac{g\sqrt{\Gamma}}{\pi} s_e(\Delta) s_e(\Delta') \left( \frac{2g\left[ s_c(\Delta'') + s_c(\Delta+\Delta'-\Delta'') \right] + \left(\Delta + \Delta' + i\Gamma\right)\left[ s_e(\Delta'') + s_e(\Delta+\Delta'-\Delta'') \right]}{\left(\Delta+\Delta'+\frac{i}{2}\Gamma\right)\left(\Delta+\Delta'+i\Gamma\right) - 2g^2} \right),
\end{equation}
where
\begin{equation}
    s_c(\Delta) =  \frac{\Delta\sqrt{\Gamma}}{\Delta^2 - g^2 + \frac{i}{2}\Gamma\Delta} \quad \text{and} \quad s_e(\Delta) = \frac{g\sqrt{\Gamma}}{\Delta^2 - g^2 + \frac{i}{2}\Gamma\Delta}
\end{equation}
[see Eqs.~(\ref{eq:c_p_matrix_el_2}) and (\ref{eq:s_e})].


\section{N cavities---SLH parameters}\label{app:N_cavities_SLH}

Here we show how we calculate the total SLH parameters for the $N$-cavity system in Fig.~\ref{fig:CZ_gate_diagram}(b) [Eq.~(\ref{eq:S_tot})-(\ref{eq:H_tot})]. First, we obtain the SLH parameters for a unit cell, consisting of cavity $j$ and the adjacent waveguide segment of length $d_j$ (see Fig.~\ref{fig:unit_cell}). This involves calculating the series product between the SLH parameters for the cavity [Eq.~(\ref{eq:SLH_one_cavity})] and the waveguide segment [Eq.~(\ref{eq:SLH_wg})]:
\begin{equation}\label{eq:SLH_unit_cell}
    G_j = G_{\text{wg}}^{(j)} \lhd G_{\text{cav}}^{(j)} = \biggl( e^{-i\phi_j},\; e^{-i\phi_j} \sqrt{\Gamma_j} c_j, \Omega_j\left(\sigma_j^+\sigma_j^- + c_j^\dagger c_j\right) + g_j\left( \sigma_j^+c_j + \sigma_j^-c_j^\dagger \right) \biggr).
\end{equation}
We can then combine the unit cells to find the total SLH parameters. For example, when we combine the first two unit cells (${j=1}$ and ${j=2}$), we obtain
\begin{align}
\begin{split}
    G_2 \lhd G_1 =&\; \biggl( e^{-i(\phi_1+\phi_2)},\; e^{-i(\phi_1+\phi_2)}\sqrt{\Gamma_1}c_1 + e^{-i\phi_2}\sqrt{\Gamma_2}c_2,\\
    &\hspace{0.05in} \sum_{j=1}^2 \left[ \Omega_j\left(\sigma_j^+\sigma_j^- + c_j^\dagger c_j\right) + g_j\left( \sigma_j^+c_j + \sigma_j^-c_j^\dagger \right) \right] + \frac{1}{2i} \sqrt{\Gamma_2\Gamma_1} \left( e^{-i\phi_1}c_2^{\dagger}c_1 - e^{i\phi_1}c_1^{\dagger}c_2 \right) \biggr).
\end{split}
\end{align}
Adding the third unit cell gives
\begin{align}
\begin{split}
    G_3 \lhd G_2 \lhd G_1 =&\; \biggl( e^{-i(\phi_1+\phi_2+\phi_3)},\; e^{-i(\phi_1+\phi_2+\phi_3)}\sqrt{\Gamma_1}c_1 + e^{-i(\phi_2+\phi_3)}\sqrt{\Gamma_2}c_2 + e^{-i\phi_3}\sqrt{\Gamma_3}c_3,\\
    &\hspace{0.05in} \sum_{j=1}^3 \left[ \Omega_j\left(\sigma_j^+\sigma_j^- + c_j^\dagger c_j\right) + g_j\left( \sigma_j^+c_j + \sigma_j^-c_j^\dagger \right) \right] + \frac{1}{2i} \sqrt{\Gamma_2\Gamma_1} \left( e^{-i\phi_1}c_2^{\dagger}c_1 - e^{i\phi_1}c_1^{\dagger}c_2 \right)\\
    &\hspace{0.1in} + \frac{1}{2i} \sqrt{\Gamma_3\Gamma_1} \left( e^{-i(\phi_1+\phi_2)}c_3^{\dagger}c_1 - e^{i(\phi_1+\phi_2)}c_1^{\dagger}c_3 \right) + \frac{1}{2i} \sqrt{\Gamma_3\Gamma_2} \left( e^{-i\phi_2}c_3^{\dagger}c_2 - e^{i\phi_2}c_2^{\dagger}c_3 \right) \biggr).
\end{split}
\end{align}
Generalizing to $N$ unit cells leads to
\begin{align}
\begin{split}
    G_N \lhd \ldots \lhd G_3 \lhd G_2 \lhd G_1 =&\; \biggl( e^{-i\sum_{k=1}^N\phi_k},\; \sum_{j=1}^N e^{-i\sum_{k=j}^N\phi_k}\sqrt{\Gamma_j}c_j,\; \sum_{j=1}^N \left[ \Omega_j\left(\sigma_j^+\sigma_j^- + c_j^\dagger c_j\right) + g_j\left( \sigma_j^+c_j + \sigma_j^-c_j^\dagger \right) \right]\\
    &\hspace{0.1in}  + \frac{1}{2i} \sum_{j=2}^N \sum_{k=1}^{j-1} \sqrt{\Gamma_j\Gamma_k} \left( e^{-i\sum_{l=k}^{j-1}\phi_l} c_j^\dagger c_k - e^{i\sum_{l=k}^{j-1}\phi_l} c_k^\dagger c_j \right) \biggr).
\end{split}
\end{align}
The final unit cell contains the phase $\phi_N$ from $G_{\text{wg}}^{(N)}$. This is a global phase that appears after the last cavity, so we can remove it by setting ${\phi_N = 0}$. This leads us to the total SLH parameters in Eqs.~(\ref{eq:S_tot})-(\ref{eq:H_tot}).

\begin{figure}
    \centering
    \includegraphics[width=0.35\linewidth]{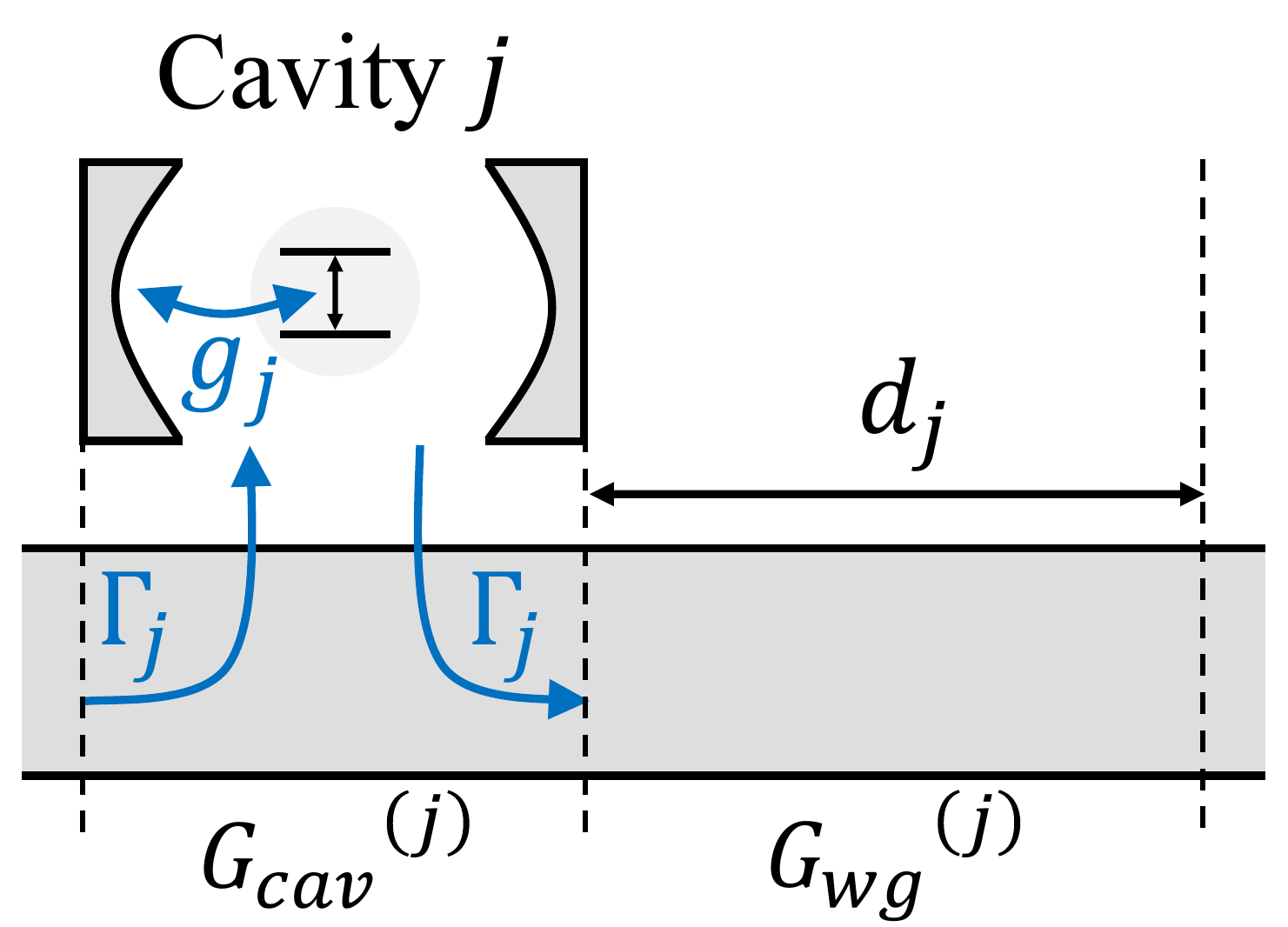}
    \caption{A unit cell of the $N$-cavity system in Fig.~\ref{fig:CZ_gate_diagram}(b), consisting of cavity $j$ and the adjacent waveguide segment of length $d_j$. The SLH parameters for this unit cell are given by the series product of the SLH parameters for the cavity and the waveguide segment [Eq.~(\ref{eq:SLH_unit_cell})].}
    \label{fig:unit_cell}
\end{figure}


\section{N cavities---single-photon scattering}\label{app:N_cavities_one_photon}

In this appendix, we show additional details of the calculation of the single-photon scattering matrix for $N$ cavities (Section~\ref{subsubsec:theory_N_cavities_one_photon}). To derive the Heisenberg equation for $c_j(t)$ in Eq.~(\ref{eq:Heisenberg_cavity_j_time}), we first use Eq.~(\ref{eq:Heisenberg_SLH}) and the SLH parameters in Eqs.~(\ref{eq:S_tot})-(\ref{eq:H_tot}) to obtain
\begin{align}\label{eq:Heisenberg_c_j_app}
\begin{split}
    \frac{\rm d}{{\rm d}t}c_j(t) =&\; i\left[ H_{\text{tot}}, c_j(t) \right] + L_{\text{tot}}^{\dagger}c_j(t)L_{\text{tot}} - \frac{1}{2}\left( L_{\text{tot}}^{\dagger} L_{\text{tot}} c_j(t) + c_j(t)L_{\text{tot}}^{\dagger} L_{\text{tot}}\right)\\
    &+ i \left[ L_{\text{tot}}^{\dagger}, c_j(t) \right] S_{\text{tot}}a_{\text{in}}(t) - i a_{\text{in}}^{\dagger}(t)S_{\text{tot}}^{\dagger} \left[ c_j(t), L_{\text{tot}} \right]\\[0.05in]
    =& -i\Omega_jc_j(t) - ig_j\sigma_j^-(t) - i e^{-i\sum_{k=1}^{j-1}\phi_k} \sqrt{\Gamma_j} a_{\text{in}}(t)\\
    &-\frac{1}{2} \sum_{j'=2}^N \sum_{l=1}^{j'-1} \sqrt{\Gamma_{j'}\Gamma_l} \left( \delta_{jj'} e^{-i\sum_{k=l}^{j'-1}\phi_k}c_l(t) - \delta_{jl} e^{i\sum_{k=l}^{j'-1}\phi_k} c_{j'}(t) \right) - \frac{1}{2} \sum_{j'=1}^N \sqrt{\Gamma_j\Gamma_{j'}}e^{i\left( \sum_{k=j}^{N-1}\phi_k - \sum_{k=j'}^{N-1}\phi_k \right)} c_{j'}(t).
\end{split}
\end{align}
To simplify this expression, we look at different values of $j$ individually, write out the summations explicitly, and cancel all the terms that are the same but have opposite signs. For example, when ${j=1}$, the terms containing $\delta_{jj'}$ are all zero because ${j' \geq 2}$. The remaining terms can be summed together, leading to
\begin{equation}
    \frac{\rm d}{{\rm d}t}c_1(t) = -i\left(\Omega_1-\frac{i}{2}\Gamma_1\right) c_1(t) - ig_1\sigma_1^-(t) - i\sqrt{\Gamma_1}a_{\text{in}}(t).
\end{equation}
Similar simplifications can be made for higher values of $j$ (except more and more terms will be nonzero), eventually producing the Heisenberg equation for $c_j(t)$ in Eq.~(\ref{eq:Heisenberg_cavity_j_time}). To transform the Heisenberg equations from time to frequency later in the derivation, we use Fourier transforms of the same form as in Eq.~(\ref{eq:Fourier}).

As in the single-cavity case, we express the transmission coefficient $t(p)$ in Eq.~(\ref{eq:t_N_cavities}) as a function of detuning. In this case, we have a detuning
${\Delta_j(p) = p - \Omega_j}$ that will generally by different for each emitter--cavity system, since each emitter/cavity can have a different resonance frequency $\Omega_j$. Nevertheless, we can define the detuning variable ${\Delta = \Delta_1(p)}$ relative to the first cavity, and write the detuning from cavity $j$ in terms of $\Delta$ as
\begin{equation}
    \Delta_j(p) = p - \Omega_j = \left(p-\Omega_1\right) - \left(\Omega_j-\Omega_1\right) = \Delta - \Delta_{j1},
\end{equation}
where $\Delta_{j1}$ is the detuning between cavity $j$ and the first cavity. We can therefore write the transmission coefficient in Eq.~(\ref{eq:t_N_cavities}) as
\begin{align}\label{eq:tN_app}
\begin{split}
    t(\Delta) =&\; e^{-i\sum_{k=1}^{N-1}\phi_k(\Delta)} \Biggl( 1 - i\sum_{j=1}^N \frac{\left(\Delta - \Delta_{j1}\right)\Gamma_j}{\left(\Delta - \Delta_{j1}\right)^2 - g_j^2 - \frac{i}{2}\left(\Delta - \Delta_{j1}\right)\Gamma_j} \prod_{l=1}^j\left[\frac{\left(\Delta - \Delta_{l1}\right)^2 - g_l^2 - \frac{i}{2}\left(\Delta - \Delta_{l1}\right)\Gamma_l}{\left(\Delta - \Delta_{l1}\right)^2 - g_l^2 + \frac{i}{2}\left(\Delta - \Delta_{l1}\right)\Gamma_l}\right] \Biggr),
\end{split}
\end{align}
where ${\phi_k(\Delta) = d_k\Delta/v_g + \phi_{d_k}}$, and ${\phi_{d_k} = \Omega_1d_k/v_g}$ is a constant phase.


\section{N cavities---two-photon scattering}\label{app:N_cavities_two_photons}

In this final appendix we show details of the two-photon scattering matrix calculation for $N$ cavities (Section~\ref{subsubsec:theory_N_cavities_two_photons}). To derive the Heisenberg equation for $\sigma_j^-(t)c_l(t)$ in Eq.~(\ref{eq:Heisenberg_sigma_j_c_l}), we use Eq.~(\ref{eq:Heisenberg_SLH}) with the SLH parameters in Eqs.~(\ref{eq:S_tot})-(\ref{eq:H_tot}):
\begin{align}
\begin{split}
    \frac{\rm d}{{\rm d}t}\sigma_j^-(t)c_l(t) =&\; i\left[ H_{\text{tot}}, \sigma_j^-(t)c_l(t) \right] + L_{\text{tot}}^{\dagger}\sigma_j^-(t)c_l(t)L_{\text{tot}} - \frac{1}{2}\left( L_{\text{tot}}^{\dagger} L_{\text{tot}} \sigma_j^-(t)c_l(t) + \sigma_j^-(t)c_l(t)L_{\text{tot}}^{\dagger} L_{\text{tot}}\right)\\
    &+ i \left[ L_{\text{tot}}^{\dagger}, \sigma_j^-(t)c_l(t) \right] S_{\text{tot}}a_{\text{in}}(t) - i a_{\text{in}}^{\dagger}(t)S_{\text{tot}}^{\dagger} \left[ \sigma_j^-(t)c_l(t), L_{\text{tot}} \right]\\[0.05in]
    =& -i\left(\Omega_j+\Omega_l\right)\sigma_j^-(t)c_l(t) + ig_j\sigma_{z,j}(t)c_j(t)c_l(t) - i e^{-i\sum_{k=1}^{l-1}\phi_k} \sqrt{\Gamma_l} \sigma_j^-(t) a_{\text{in}}(t)\\
    &-\frac{1}{2} \sum_{j'=2}^N \sum_{l'=1}^{j'-1} \sqrt{\Gamma_{j'}\Gamma_{l'}} \left( \delta_{lj'} e^{-i\sum_{k=l'}^{j'-1}\phi_k}\sigma_j^-(t)c_{l'}(t) - \delta_{ll'} e^{i\sum_{k=l'}^{j'-1}\phi_k} \sigma_j^-(t)c_{j'}(t) \right)\\
    &- \frac{1}{2} \sum_{j'=1}^N \sqrt{\Gamma_l\Gamma_{j'}}e^{i\left( \sum_{k=l}^{N-1}\phi_k - \sum_{k=j'}^{N-1}\phi_k \right)} \sigma_j^-(t)c_{j'}(t).
\end{split}
\end{align}
This is similar to Eq.~(\ref{eq:Heisenberg_c_j_app}), and can be simplified in a similar way by looking at different values of $j$ and $l$ separately, expanding the summations, and canceling terms that are the same but have an opposite sign. This allows us to write the above Heisenberg equation in the form given in Eq.~(\ref{eq:Heisenberg_sigma_j_c_l}). To get to Eq.~(\ref{eq:sigma_c_j_matrix_el}), we then pre-multiply by $\bra{0}$, post-multiply by $\ket{k_1k_2^+}$, Fourier transform from time $t$ to frequency $p$, and replace $p$ with ${p_1+p_2}$.

We follow a similar procedure with the Heisenberg equation for $c_j(t)c_l(t)$. Again using Eq.~(\ref{eq:Heisenberg_SLH}) with the SLH parameters in Eqs.~(\ref{eq:S_tot})-(\ref{eq:H_tot}) gives
\begin{align}
\begin{split}
    \frac{\rm d}{{\rm d}t}c_j(t)c_l(t) =&\; i\left[ H_{\text{tot}}, c_j(t)c_l(t) \right] + L_{\text{tot}}^{\dagger}c_j(t)c_l(t)L_{\text{tot}} - \frac{1}{2}\left( L_{\text{tot}}^{\dagger} L_{\text{tot}} c_j(t)c_l(t) + c_j(t)c_l(t)L_{\text{tot}}^{\dagger} L_{\text{tot}}\right)\\
    &+ i \left[ L_{\text{tot}}^{\dagger}, c_j(t)c_l(t) \right] S_{\text{tot}}a_{\text{in}}(t) - i a_{\text{in}}^{\dagger}(t)S_{\text{tot}}^{\dagger} \left[ c_j(t)c_l(t), L_{\text{tot}} \right]\\[0.05in]
    =& -i\left(\Omega_j+\Omega_l\right)c_j(t)c_l(t) - ig_l\sigma_l^-(t)c_j(t) - ig_j\sigma_j^-(t)c_l(t)\\
    &- i e^{-i\sum_{k=1}^{l-1}\phi_k} \sqrt{\Gamma_l} c_j(t) a_{\text{in}}(t) - i e^{-i\sum_{k=1}^{j-1}\phi_k} \sqrt{\Gamma_j} c_l(t) a_{\text{in}}(t)\\
    &-\frac{1}{2} \sum_{j'=2}^N \sum_{l'=1}^{j'-1} \sqrt{\Gamma_{j'}\Gamma_{l'}} \Bigl( \delta_{lj'} e^{-i\sum_{k=l'}^{j'-1}\phi_k}c_j(t)c_{l'}(t) + \delta_{jj'} e^{-i\sum_{k=l'}^{j'-1}\phi_k}c_l(t)c_{l'}(t)\\
    &\hspace{1.35in}- \delta_{ll'} e^{i\sum_{k=l'}^{j'-1}\phi_k} c_j(t)c_{j'}(t) - \delta_{jl'} e^{i\sum_{k=l'}^{j'-1}\phi_k} c_l(t)c_{j'}(t) \Bigr)\\
    &- \frac{1}{2} \sum_{j'=1}^N \sqrt{\Gamma_l\Gamma_{j'}}e^{i\left( \sum_{k=l}^{N-1}\phi_k - \sum_{k=j'}^{N-1}\phi_k \right)} c_j(t)c_{j'}(t)\\
    &- \frac{1}{2} \sum_{j'=1}^N \sqrt{\Gamma_j\Gamma_{j'}}e^{i\left( \sum_{k=j}^{N-1}\phi_k - \sum_{k=j'}^{N-1}\phi_k \right)} c_l(t)c_{j'}(t).
\end{split}
\end{align}
We then look at different $(j,l)$ pairs separately, and collect together terms that do not cancel to write this equation in the form given in Eq.~(\ref{eq:Heisenberg_c2_time_2}). To get to Eq.~(\ref{eq:c_j_c_l_matrix_el}), we perform a Fourier transform as outlined above.

After obtaining the Heisenberg equations of the operator products and solving for $\matrixel{0}{\sigma_j^-c_j(p_1+p_2)}{k_1k_2^+}$ as outlined in the main text, we finally solve for $\bra{p_1^-}c_j(p_2)\ket{k_1k_2^+}$ using Eq.~(\ref{eq:c_j_p2_matrix_el_3}). We do this iteratively, in a similar way to how we solved Eq.~(\ref{eq:Heisenberg_c_j_p_2}) for $\matrixel{0}{c_j(p)}{k^+}$ in the single-photon case (Section~\ref{subsubsec:theory_N_cavities_one_photon}). When ${j=1}$, Eq.~(\ref{eq:c_j_p2_matrix_el_3}) depends only on $\bra{p_1^-}c_1(p_2)\ket{k_1k_2^+}$, and can be solved straightforwardly to give
\begin{align}
\begin{split}
    \bra{p_1^-}c_1(p_2)\ket{k_1k_2^+} =&\; s_{c,1}(p_2) \left[ S_{p_1k_2}\delta(p_2-k_1) + S_{p_1k_1}\delta(p_2-k_2) \right]\\
    &- \frac{2}{\sqrt{2\pi}}t(p_1) \frac{g_1^2\left[s_{e,1}(p_1)\right]^*\alpha^{(1)}_{p_1p_2k_1k_2}}{\Delta_1^2(p_2) - g_1^2 + \frac{i}{2}\Delta_1(p_2)\Gamma_1}\delta(p_1+p_2-k_1-k_2),
\end{split}
\end{align}
where we used the result in Eq.~(\ref{eq:sc_j}) with ${j=1}$. With this solution, we can find $\bra{p_1^-}c_2(p_2)\ket{k_1k_2^+}$ by setting ${j=2}$ in Eq.~(\ref{eq:c_j_p2_matrix_el_3}):
\begin{align}
\begin{split}
    \bra{p_1^-}c_2(p_2)\ket{k_1k_2^+} =&\; s_{c,2}(p_2) \left[ S_{p_1k_2}\delta(p_2-k_1) + S_{p_1k_1}\delta(p_2-k_2) \right]\\
    &+ \frac{2}{\sqrt{2\pi}}t(p_1) \Biggl( \frac{i\Delta_2(p_2)\sqrt{\Gamma_2\Gamma_1}g_1^2\left[s_{e,1}(p_1)\right]^*\alpha^{(1)}_{p_1p_2k_1k_2}e^{-i\phi_1(p_2)}}{\prod_{m=1,2}\left[ \Delta_m^2(p_2) - g_m^2 + \frac{i}{2}\Delta_m(p_2)\Gamma_m \right]}\\
    &\hspace{0.8in}- \frac{g_2^2\left[s_{e,2}(p_1)\right]^*\alpha^{(2)}_{p_1p_2k_1k_2}}{\Delta_2^2(p_2) - g_2^2 + \frac{i}{2}\Delta_2(p_2)\Gamma_2} \Biggr) \delta(p_1+p_2-k_1-k_2).
\end{split}
\end{align}
In general, for any $j$, we obtain
\begin{align}
\begin{split}
    \bra{p_1^-}c_j(p_2)\ket{k_1k_2^+} =&\; s_{c,j}(p_2) \left[ S_{p_1k_2}\delta(p_2-k_1) + S_{p_1k_1}\delta(p_2-k_2) \right]\\
    &+ \frac{2}{\sqrt{2\pi}}t(p_1) \Biggl( \frac{i\Delta_j(p_2)\sqrt{\Gamma_j}}{\Delta_j^2(p_2) - g_j^2 - \frac{i}{2}\Delta_j(p_2)\Gamma_j} \sum_{l=1}^{j-1} \frac{g_l^2 \left[s_{e,l}(p_1)\right]^* \alpha^{(l)}_{p_1p_2k_1k_2}\sqrt{\Gamma_l}e^{-i\sum_{k=l}^{j-1}\phi_k(p_2)}}{\Delta_l^2(p_2) - g_l^2 - \frac{i}{2}\Delta_l(p_2)\Gamma_l}\\
    &\hspace{2.65in}\times \prod_{m=l}^j\left[ \frac{\Delta_m^2(p_2) - g_m^2 - \frac{i}{2}\Delta_m(p_2)\Gamma_m}{\Delta_m^2(p_2) - g_m^2 + \frac{i}{2}\Delta_m(p_2)\Gamma_m} \right]\\
    &\hspace{0.8in}- \frac{g_j^2\left[s_{e,j}(p_1)\right]^*\alpha^{(j)}_{p_1p_2k_1k_2}}{\Delta_j^2(p_2) - g_j^2 + \frac{i}{2}\Delta_j(p_2)\Gamma_j} \Biggr) \delta(p_1+p_2-k_1-k_2).
\end{split}
\end{align}
Substituting this result into Eq.~(\ref{eq:S2_N_cavities}) and using
\begin{equation}
    e^{-i\sum_{k=1}^{N-1}\phi_k(p_2)} - i\sum_{j=1}^N e^{-i\sum_{k=j}^{N-1}\phi_k(p_2)} \sqrt{\Gamma_j}s_{c,j}(p_2) = t(p_2)
\end{equation}
allows us to write the two-photon scattering matrix in the form given in Eq.~(\ref{eq:S2}), where $C_{p_1p_2k_1k_2}$ is given in Eq.~(\ref{eq:C_N_cavities}).

Finally, we express $C_{p_1p_2k_1k_2}$ in terms of detunings. We set ${k_2 = p_1+p_2-k_1}$ in Eq.~(\ref{eq:C_N_cavities}) using the delta function $\delta(p_1+p_2-k_1-k_2)$ in the scattering matrix, as in Appendix~\ref{app:one_cavity_two_photons}. Then, similarly to Appendix~\ref{app:N_cavities_one_photon}, we define the detuning variables ${\Delta = p_1 - \Omega_1}$, ${\Delta' = p_2 - \Omega_1}$, and ${\Delta'' = k_1 - \Omega_1}$ (relative to the first emitter/cavity), and write the detuning from cavity $j$ in terms of these variables, e.g., ${\Delta_j(p_2) = p_2 - \Omega_j = \Delta' - \Delta_{j1}}$, where ${\Delta_{j1} = \Omega_j - \Omega_1}$ is the detuning between cavity $j$ and the first cavity, as defined in Appendix~\ref{app:N_cavities_one_photon}. We therefore have
\begin{align}\label{eq:C_N_cavities_app}
\begin{split}
    C(\Delta, \Delta', \Delta'') =&\; \frac{2}{\sqrt{2\pi}}t(\Delta) \Biggl( \sum_{j=1}^N \left( \frac{g_j^2\left[s_{e,j}(\Delta)\right]^*\alpha^{(j)}(\Delta,\Delta',\Delta'')\sqrt{\Gamma_j} e^{-i\sum_{k=j}^{N-1}\phi_k(\Delta')}}{(\Delta' - \Delta_{j1})^2 - g_j^2 + \frac{i}{2}(\Delta' - \Delta_{j1})\Gamma_j} \right)\\
    &\hspace{0.7in}-i \sum_{j=2}^N \frac{(\Delta' - \Delta_{j1})\Gamma_j}{(\Delta' - \Delta_{j1})^2 - g_j^2 - \frac{i}{2}(\Delta' - \Delta_{j1})\Gamma_j} \sum_{l=1}^{j-1}\left( \frac{g_l^2\left[s_{e,l}(\Delta)\right]^*\alpha^{(l)}(\Delta,\Delta',\Delta'')\sqrt{\Gamma_l}e^{-i\sum_{k=l}^{N-1}\phi_k(\Delta')}}{(\Delta' - \Delta_{l1})^2 - g_l^2 - \frac{i}{2}(\Delta' - \Delta_{l1})\Gamma_l} \right)\\
    &\hspace{3.5in}\times \prod_{m=l}^j\left[ \frac{(\Delta' - \Delta_{m1})^2 - g_m^2 - \frac{i}{2}(\Delta' - \Delta_{m1})\Gamma_m}{(\Delta' - \Delta_{m1})^2 - g_m^2 + \frac{i}{2}(\Delta' - \Delta_{m1})\Gamma_m} \right] \Biggr),
\end{split}
\end{align}
where $t(\Delta)$ is given in Eq.~(\ref{eq:tN_app}),
\begin{subequations}
\begin{equation}
    s_{c,j}(\Delta) = e^{-i\sum_{k=1}^{j-1}\phi_k(\Delta)}\frac{(\Delta - \Delta_{j1})\sqrt{\Gamma_j}}{(\Delta - \Delta_{j1})^2 - g_j^2 - \frac{i}{2}(\Delta - \Delta_{j1})\Gamma_j} \prod_{l=1}^j\left[\frac{(\Delta - \Delta_{l1})^2 - g_l^2 - \frac{i}{2}(\Delta - \Delta_{l1})\Gamma_l}{(\Delta - \Delta_{l1})^2 - g_l^2 + \frac{i}{2}(\Delta - \Delta_{l1})\Gamma_l}\right],
\end{equation}
\begin{equation}
    s_{e,j}(\Delta) = e^{-i\sum_{k=1}^{j-1}\phi_k(\Delta)}\frac{g_j\sqrt{\Gamma_j}}{(\Delta - \Delta_{j1})^2 - g_j^2 - \frac{i}{2}(\Delta - \Delta_{j1})\Gamma_j} \prod_{l=1}^j\left[\frac{(\Delta - \Delta_{l1})^2 - g_l^2 - \frac{i}{2}(\Delta - \Delta_{l1})\Gamma_l}{(\Delta - \Delta_{l1})^2 - g_l^2 + \frac{i}{2}(\Delta - \Delta_{l1})\Gamma_l}\right]
\end{equation}
\end{subequations}
[see Eqs.~(\ref{eq:sc_j}) and (\ref{eq:se_j})], ${\phi_k(\Delta) = d_k\Delta/v_g + \phi_{d_k}}$ (as in Appendix~\ref{app:N_cavities_one_photon}), and $\alpha^{(j)}(\Delta,\Delta',\Delta'')$ is calculated in the same way as $\alpha^{(j)}_{p_1p_2k_1k_2}$, except that we set ${k_2 = p_1 + p_2 - k_1}$ and substitute in the detuning variables before we solve the relevant matrix equation numerically.
\end{widetext}


\begin{thebibliography}{70}%
\makeatletter
\providecommand \@ifxundefined [1]{%
 \@ifx{#1\undefined}
}%
\providecommand \@ifnum [1]{%
 \ifnum #1\expandafter \@firstoftwo
 \else \expandafter \@secondoftwo
 \fi
}%
\providecommand \@ifx [1]{%
 \ifx #1\expandafter \@firstoftwo
 \else \expandafter \@secondoftwo
 \fi
}%
\providecommand \natexlab [1]{#1}%
\providecommand \enquote  [1]{``#1''}%
\providecommand \bibnamefont  [1]{#1}%
\providecommand \bibfnamefont [1]{#1}%
\providecommand \citenamefont [1]{#1}%
\providecommand \href@noop [0]{\@secondoftwo}%
\providecommand \href [0]{\begingroup \@sanitize@url \@href}%
\providecommand \@href[1]{\@@startlink{#1}\@@href}%
\providecommand \@@href[1]{\endgroup#1\@@endlink}%
\providecommand \@sanitize@url [0]{\catcode `\\12\catcode `\$12\catcode `\&12\catcode `\#12\catcode `\^12\catcode `\_12\catcode `\%12\relax}%
\providecommand \@@startlink[1]{}%
\providecommand \@@endlink[0]{}%
\providecommand \url  [0]{\begingroup\@sanitize@url \@url }%
\providecommand \@url [1]{\endgroup\@href {#1}{\urlprefix }}%
\providecommand \urlprefix  [0]{URL }%
\providecommand \Eprint [0]{\href }%
\providecommand \doibase [0]{https://doi.org/}%
\providecommand \selectlanguage [0]{\@gobble}%
\providecommand \bibinfo  [0]{\@secondoftwo}%
\providecommand \bibfield  [0]{\@secondoftwo}%
\providecommand \translation [1]{[#1]}%
\providecommand \BibitemOpen [0]{}%
\providecommand \bibitemStop [0]{}%
\providecommand \bibitemNoStop [0]{.\EOS\space}%
\providecommand \EOS [0]{\spacefactor3000\relax}%
\providecommand \BibitemShut  [1]{\csname bibitem#1\endcsname}%
\let\auto@bib@innerbib\@empty
\bibitem [{\citenamefont {Kok}\ \emph {et~al.}(2007)\citenamefont {Kok}, \citenamefont {Munro}, \citenamefont {Nemoto}, \citenamefont {Ralph}, \citenamefont {Dowling},\ and\ \citenamefont {Milburn}}]{Kok2007}%
  \BibitemOpen
  \bibfield  {author} {\bibinfo {author} {\bibfnamefont {P.}~\bibnamefont {Kok}}, \bibinfo {author} {\bibfnamefont {W.~J.}\ \bibnamefont {Munro}}, \bibinfo {author} {\bibfnamefont {K.}~\bibnamefont {Nemoto}}, \bibinfo {author} {\bibfnamefont {T.~C.}\ \bibnamefont {Ralph}}, \bibinfo {author} {\bibfnamefont {J.~P.}\ \bibnamefont {Dowling}},\ and\ \bibinfo {author} {\bibfnamefont {G.~J.}\ \bibnamefont {Milburn}},\ }\bibfield  {title} {\bibinfo {title} {Linear optical quantum computing with photonic qubits},\ }\href {https://doi.org/10.1103/RevModPhys.79.135} {\bibfield  {journal} {\bibinfo  {journal} {Rev. Mod. Phys.}\ }\textbf {\bibinfo {volume} {79}},\ \bibinfo {pages} {135} (\bibinfo {year} {2007})}\BibitemShut {NoStop}%
\bibitem [{\citenamefont {Knill}\ \emph {et~al.}(2001)\citenamefont {Knill}, \citenamefont {Laflamme},\ and\ \citenamefont {Milburn}}]{Knill2001}%
  \BibitemOpen
  \bibfield  {author} {\bibinfo {author} {\bibfnamefont {E.}~\bibnamefont {Knill}}, \bibinfo {author} {\bibfnamefont {R.}~\bibnamefont {Laflamme}},\ and\ \bibinfo {author} {\bibfnamefont {G.~J.}\ \bibnamefont {Milburn}},\ }\bibfield  {title} {\bibinfo {title} {A scheme for efficient quantum computation with linear optics},\ }\href {https://doi.org/10.1038/35051009} {\bibfield  {journal} {\bibinfo  {journal} {Nature}\ }\textbf {\bibinfo {volume} {409}},\ \bibinfo {pages} {46} (\bibinfo {year} {2001})}\BibitemShut {NoStop}%
\bibitem [{\citenamefont {O'Brien}\ \emph {et~al.}(2003)\citenamefont {O'Brien}, \citenamefont {Pryde}, \citenamefont {White}, \citenamefont {Ralph},\ and\ \citenamefont {Branning}}]{OBrien2003}%
  \BibitemOpen
  \bibfield  {author} {\bibinfo {author} {\bibfnamefont {J.~L.}\ \bibnamefont {O'Brien}}, \bibinfo {author} {\bibfnamefont {G.~J.}\ \bibnamefont {Pryde}}, \bibinfo {author} {\bibfnamefont {A.~G.}\ \bibnamefont {White}}, \bibinfo {author} {\bibfnamefont {T.~C.}\ \bibnamefont {Ralph}},\ and\ \bibinfo {author} {\bibfnamefont {D.}~\bibnamefont {Branning}},\ }\bibfield  {title} {\bibinfo {title} {{Demonstration of an all-optical quantum controlled-NOT gate}},\ }\href {https://doi.org/10.1038/nature02054} {\bibfield  {journal} {\bibinfo  {journal} {Nature}\ }\textbf {\bibinfo {volume} {426}},\ \bibinfo {pages} {264} (\bibinfo {year} {2003})}\BibitemShut {NoStop}%
\bibitem [{\citenamefont {Gasparoni}\ \emph {et~al.}(2004)\citenamefont {Gasparoni}, \citenamefont {Pan}, \citenamefont {Walther}, \citenamefont {Rudolph},\ and\ \citenamefont {Zeilinger}}]{Gasparoni2004}%
  \BibitemOpen
  \bibfield  {author} {\bibinfo {author} {\bibfnamefont {S.}~\bibnamefont {Gasparoni}}, \bibinfo {author} {\bibfnamefont {J.-W.}\ \bibnamefont {Pan}}, \bibinfo {author} {\bibfnamefont {P.}~\bibnamefont {Walther}}, \bibinfo {author} {\bibfnamefont {T.}~\bibnamefont {Rudolph}},\ and\ \bibinfo {author} {\bibfnamefont {A.}~\bibnamefont {Zeilinger}},\ }\bibfield  {title} {\bibinfo {title} {{Realization of a Photonic Controlled-NOT Gate Sufficient for Quantum Computation}},\ }\href {https://doi.org/10.1103/PhysRevLett.93.020504} {\bibfield  {journal} {\bibinfo  {journal} {Phys. Rev. Lett.}\ }\textbf {\bibinfo {volume} {93}},\ \bibinfo {pages} {020504} (\bibinfo {year} {2004})}\BibitemShut {NoStop}%
\bibitem [{\citenamefont {Zhao}\ \emph {et~al.}(2005)\citenamefont {Zhao}, \citenamefont {Zhang}, \citenamefont {Chen}, \citenamefont {Zhang}, \citenamefont {Du}, \citenamefont {Yang},\ and\ \citenamefont {Pan}}]{Zhao2005}%
  \BibitemOpen
  \bibfield  {author} {\bibinfo {author} {\bibfnamefont {Z.}~\bibnamefont {Zhao}}, \bibinfo {author} {\bibfnamefont {A.-N.}\ \bibnamefont {Zhang}}, \bibinfo {author} {\bibfnamefont {Y.-A.}\ \bibnamefont {Chen}}, \bibinfo {author} {\bibfnamefont {H.}~\bibnamefont {Zhang}}, \bibinfo {author} {\bibfnamefont {J.-F.}\ \bibnamefont {Du}}, \bibinfo {author} {\bibfnamefont {T.}~\bibnamefont {Yang}},\ and\ \bibinfo {author} {\bibfnamefont {J.-W.}\ \bibnamefont {Pan}},\ }\bibfield  {title} {\bibinfo {title} {{Experimental Demonstration of a Nondestructive Controlled-NOT Quantum Gate for Two Independent Photon Qubits}},\ }\href {https://doi.org/10.1103/PhysRevLett.94.030501} {\bibfield  {journal} {\bibinfo  {journal} {Phys. Rev. Lett.}\ }\textbf {\bibinfo {volume} {94}},\ \bibinfo {pages} {030501} (\bibinfo {year} {2005})}\BibitemShut {NoStop}%
\bibitem [{\citenamefont {Crespi}\ \emph {et~al.}(2011)\citenamefont {Crespi}, \citenamefont {Ramponi}, \citenamefont {Osellame}, \citenamefont {Sansoni}, \citenamefont {Bongioanni}, \citenamefont {Sciarrino}, \citenamefont {Vallone},\ and\ \citenamefont {Mataloni}}]{Crespi2011}%
  \BibitemOpen
  \bibfield  {author} {\bibinfo {author} {\bibfnamefont {A.}~\bibnamefont {Crespi}}, \bibinfo {author} {\bibfnamefont {R.}~\bibnamefont {Ramponi}}, \bibinfo {author} {\bibfnamefont {R.}~\bibnamefont {Osellame}}, \bibinfo {author} {\bibfnamefont {L.}~\bibnamefont {Sansoni}}, \bibinfo {author} {\bibfnamefont {I.}~\bibnamefont {Bongioanni}}, \bibinfo {author} {\bibfnamefont {F.}~\bibnamefont {Sciarrino}}, \bibinfo {author} {\bibfnamefont {G.}~\bibnamefont {Vallone}},\ and\ \bibinfo {author} {\bibfnamefont {P.}~\bibnamefont {Mataloni}},\ }\bibfield  {title} {\bibinfo {title} {Integrated photonic quantum gates for polarization qubits},\ }\href {https://doi.org/10.1038/ncomms1570} {\bibfield  {journal} {\bibinfo  {journal} {Nat. Commun.}\ }\textbf {\bibinfo {volume} {2}},\ \bibinfo {pages} {566} (\bibinfo {year} {2011})}\BibitemShut {NoStop}%
\bibitem [{\citenamefont {Pooley}\ \emph {et~al.}(2012)\citenamefont {Pooley}, \citenamefont {Ellis}, \citenamefont {Patel}, \citenamefont {Bennett}, \citenamefont {Chan}, \citenamefont {Farrer}, \citenamefont {Ritchie},\ and\ \citenamefont {Shields}}]{Pooley2012}%
  \BibitemOpen
  \bibfield  {author} {\bibinfo {author} {\bibfnamefont {M.~A.}\ \bibnamefont {Pooley}}, \bibinfo {author} {\bibfnamefont {D.~J.~P.}\ \bibnamefont {Ellis}}, \bibinfo {author} {\bibfnamefont {R.~B.}\ \bibnamefont {Patel}}, \bibinfo {author} {\bibfnamefont {A.~J.}\ \bibnamefont {Bennett}}, \bibinfo {author} {\bibfnamefont {K.~H.~A.}\ \bibnamefont {Chan}}, \bibinfo {author} {\bibfnamefont {I.}~\bibnamefont {Farrer}}, \bibinfo {author} {\bibfnamefont {D.~A.}\ \bibnamefont {Ritchie}},\ and\ \bibinfo {author} {\bibfnamefont {A.~J.}\ \bibnamefont {Shields}},\ }\bibfield  {title} {\bibinfo {title} {{Controlled-NOT gate operating with single photons}},\ }\href {https://doi.org/10.1063/1.4719077} {\bibfield  {journal} {\bibinfo  {journal} {Appl. Phys. Lett.}\ }\textbf {\bibinfo {volume} {100}},\ \bibinfo {pages} {211103} (\bibinfo {year} {2012})}\BibitemShut {NoStop}%
\bibitem [{\citenamefont {Pegoraro}\ \emph {et~al.}(2026)\citenamefont {Pegoraro}, \citenamefont {Held}, \citenamefont {Lammers}, \citenamefont {Brecht},\ and\ \citenamefont {Silberhorn}}]{Pegoraro2026}%
  \BibitemOpen
  \bibfield  {author} {\bibinfo {author} {\bibfnamefont {F.}~\bibnamefont {Pegoraro}}, \bibinfo {author} {\bibfnamefont {P.}~\bibnamefont {Held}}, \bibinfo {author} {\bibfnamefont {J.}~\bibnamefont {Lammers}}, \bibinfo {author} {\bibfnamefont {B.}~\bibnamefont {Brecht}},\ and\ \bibinfo {author} {\bibfnamefont {C.}~\bibnamefont {Silberhorn}},\ }\href@noop {} {\bibinfo {title} {{Demonstration of a quantum C-NOT Gate in a Time-Multiplexed fully reconfigurable photonic processor}}} (\bibinfo {year} {2026}),\ \Eprint {https://arxiv.org/abs/2412.02478} {arXiv:2412.02478 [quant-ph]} \BibitemShut {NoStop}%
\bibitem [{\citenamefont {Chang}\ \emph {et~al.}(2014)\citenamefont {Chang}, \citenamefont {Vuletić},\ and\ \citenamefont {Lukin}}]{Chang2014}%
  \BibitemOpen
  \bibfield  {author} {\bibinfo {author} {\bibfnamefont {D.~E.}\ \bibnamefont {Chang}}, \bibinfo {author} {\bibfnamefont {V.}~\bibnamefont {Vuletić}},\ and\ \bibinfo {author} {\bibfnamefont {M.~D.}\ \bibnamefont {Lukin}},\ }\bibfield  {title} {\bibinfo {title} {Quantum nonlinear optics — photon by photon},\ }\href {https://doi.org/10.1038/nphoton.2014.192} {\bibfield  {journal} {\bibinfo  {journal} {Nat. Photon.}\ }\textbf {\bibinfo {volume} {8}},\ \bibinfo {pages} {685} (\bibinfo {year} {2014})}\BibitemShut {NoStop}%
\bibitem [{\citenamefont {Schmidt}\ and\ \citenamefont {Imamoglu}(1996)}]{Schmidt1996}%
  \BibitemOpen
  \bibfield  {author} {\bibinfo {author} {\bibfnamefont {H.}~\bibnamefont {Schmidt}}\ and\ \bibinfo {author} {\bibfnamefont {A.}~\bibnamefont {Imamoglu}},\ }\bibfield  {title} {\bibinfo {title} {{Giant Kerr nonlinearities obtained by electromagnetically induced transparency}},\ }\href {https://doi.org/10.1364/OL.21.001936} {\bibfield  {journal} {\bibinfo  {journal} {Opt. Lett.}\ }\textbf {\bibinfo {volume} {21}},\ \bibinfo {pages} {1936} (\bibinfo {year} {1996})}\BibitemShut {NoStop}%
\bibitem [{\citenamefont {Harris}\ and\ \citenamefont {Hau}(1999)}]{Harris1999}%
  \BibitemOpen
  \bibfield  {author} {\bibinfo {author} {\bibfnamefont {S.~E.}\ \bibnamefont {Harris}}\ and\ \bibinfo {author} {\bibfnamefont {L.~V.}\ \bibnamefont {Hau}},\ }\bibfield  {title} {\bibinfo {title} {{Nonlinear Optics at Low Light Levels}},\ }\href {https://doi.org/10.1103/PhysRevLett.82.4611} {\bibfield  {journal} {\bibinfo  {journal} {Phys. Rev. Lett.}\ }\textbf {\bibinfo {volume} {82}},\ \bibinfo {pages} {4611} (\bibinfo {year} {1999})}\BibitemShut {NoStop}%
\bibitem [{\citenamefont {Petrosyan}\ and\ \citenamefont {Malakyan}(2004)}]{Petrosyan2004}%
  \BibitemOpen
  \bibfield  {author} {\bibinfo {author} {\bibfnamefont {D.}~\bibnamefont {Petrosyan}}\ and\ \bibinfo {author} {\bibfnamefont {Y.~P.}\ \bibnamefont {Malakyan}},\ }\bibfield  {title} {\bibinfo {title} {Magneto-optical rotation and cross-phase modulation via coherently driven four-level atoms in a tripod configuration},\ }\href {https://doi.org/10.1103/PhysRevA.70.023822} {\bibfield  {journal} {\bibinfo  {journal} {Phys. Rev. A}\ }\textbf {\bibinfo {volume} {70}},\ \bibinfo {pages} {023822} (\bibinfo {year} {2004})}\BibitemShut {NoStop}%
\bibitem [{\citenamefont {Ottaviani}\ \emph {et~al.}(2006)\citenamefont {Ottaviani}, \citenamefont {Rebi\ifmmode~\acute{c}\else \'{c}\fi{}}, \citenamefont {Vitali},\ and\ \citenamefont {Tombesi}}]{Ottaviani2006}%
  \BibitemOpen
  \bibfield  {author} {\bibinfo {author} {\bibfnamefont {C.}~\bibnamefont {Ottaviani}}, \bibinfo {author} {\bibfnamefont {S.}~\bibnamefont {Rebi\ifmmode~\acute{c}\else \'{c}\fi{}}}, \bibinfo {author} {\bibfnamefont {D.}~\bibnamefont {Vitali}},\ and\ \bibinfo {author} {\bibfnamefont {P.}~\bibnamefont {Tombesi}},\ }\bibfield  {title} {\bibinfo {title} {{Quantum phase-gate operation based on nonlinear optics: Full quantum analysis}},\ }\href {https://doi.org/10.1103/PhysRevA.73.010301} {\bibfield  {journal} {\bibinfo  {journal} {Phys. Rev. A}\ }\textbf {\bibinfo {volume} {73}},\ \bibinfo {pages} {010301(R)} (\bibinfo {year} {2006})}\BibitemShut {NoStop}%
\bibitem [{\citenamefont {Zhu}(2010)}]{Zhu2010}%
  \BibitemOpen
  \bibfield  {author} {\bibinfo {author} {\bibfnamefont {Y.}~\bibnamefont {Zhu}},\ }\bibfield  {title} {\bibinfo {title} {{Large Kerr nonlinearities on cavity-atom polaritons}},\ }\href {https://doi.org/10.1364/OL.35.000303} {\bibfield  {journal} {\bibinfo  {journal} {Opt. Lett.}\ }\textbf {\bibinfo {volume} {35}},\ \bibinfo {pages} {303} (\bibinfo {year} {2010})}\BibitemShut {NoStop}%
\bibitem [{\citenamefont {He}\ and\ \citenamefont {Scherer}(2012)}]{He2012}%
  \BibitemOpen
  \bibfield  {author} {\bibinfo {author} {\bibfnamefont {B.}~\bibnamefont {He}}\ and\ \bibinfo {author} {\bibfnamefont {A.}~\bibnamefont {Scherer}},\ }\bibfield  {title} {\bibinfo {title} {Continuous-mode effects and photon-photon phase gate performance},\ }\href {https://doi.org/10.1103/PhysRevA.85.033814} {\bibfield  {journal} {\bibinfo  {journal} {Phys. Rev. A}\ }\textbf {\bibinfo {volume} {85}},\ \bibinfo {pages} {033814} (\bibinfo {year} {2012})}\BibitemShut {NoStop}%
\bibitem [{\citenamefont {Copetudo}\ \emph {et~al.}(2026)\citenamefont {Copetudo}, \citenamefont {Kasper}, \citenamefont {Krisnanda}, \citenamefont {Veyrac}, \citenamefont {Qin}, \citenamefont {Ng},\ and\ \citenamefont {Gao}}]{Copetudo2026}%
  \BibitemOpen
  \bibfield  {author} {\bibinfo {author} {\bibfnamefont {A.}~\bibnamefont {Copetudo}}, \bibinfo {author} {\bibfnamefont {A.~M.}\ \bibnamefont {Kasper}}, \bibinfo {author} {\bibfnamefont {T.}~\bibnamefont {Krisnanda}}, \bibinfo {author} {\bibfnamefont {G.}~\bibnamefont {Veyrac}}, \bibinfo {author} {\bibfnamefont {S.}~\bibnamefont {Qin}}, \bibinfo {author} {\bibfnamefont {H.~K.}\ \bibnamefont {Ng}},\ and\ \bibinfo {author} {\bibfnamefont {Y.~Y.}\ \bibnamefont {Gao}},\ }\href@noop {} {\bibinfo {title} {A direct controlled-phase gate between microwave photons}} (\bibinfo {year} {2026}),\ \Eprint {https://arxiv.org/abs/2603.15587} {arXiv:2603.15587 [quant-ph]} \BibitemShut {NoStop}%
\bibitem [{\citenamefont {Kang}\ and\ \citenamefont {Zhu}(2003)}]{Kang2003}%
  \BibitemOpen
  \bibfield  {author} {\bibinfo {author} {\bibfnamefont {H.}~\bibnamefont {Kang}}\ and\ \bibinfo {author} {\bibfnamefont {Y.}~\bibnamefont {Zhu}},\ }\bibfield  {title} {\bibinfo {title} {{Observation of Large Kerr Nonlinearity at Low Light Intensities}},\ }\href {https://doi.org/10.1103/PhysRevLett.91.093601} {\bibfield  {journal} {\bibinfo  {journal} {Phys. Rev. Lett.}\ }\textbf {\bibinfo {volume} {91}},\ \bibinfo {pages} {093601} (\bibinfo {year} {2003})}\BibitemShut {NoStop}%
\bibitem [{\citenamefont {Chen}\ \emph {et~al.}(2006)\citenamefont {Chen}, \citenamefont {Wang}, \citenamefont {Wang},\ and\ \citenamefont {Yu}}]{Chen2006}%
  \BibitemOpen
  \bibfield  {author} {\bibinfo {author} {\bibfnamefont {Y.-F.}\ \bibnamefont {Chen}}, \bibinfo {author} {\bibfnamefont {C.-Y.}\ \bibnamefont {Wang}}, \bibinfo {author} {\bibfnamefont {S.-H.}\ \bibnamefont {Wang}},\ and\ \bibinfo {author} {\bibfnamefont {I.~A.}\ \bibnamefont {Yu}},\ }\bibfield  {title} {\bibinfo {title} {{Low-Light-Level Cross-Phase-Modulation Based on Stored Light Pulses}},\ }\href {https://doi.org/10.1103/PhysRevLett.96.043603} {\bibfield  {journal} {\bibinfo  {journal} {Phys. Rev. Lett.}\ }\textbf {\bibinfo {volume} {96}},\ \bibinfo {pages} {043603} (\bibinfo {year} {2006})}\BibitemShut {NoStop}%
\bibitem [{\citenamefont {Lo}\ \emph {et~al.}(2010)\citenamefont {Lo}, \citenamefont {Su},\ and\ \citenamefont {Chen}}]{Lo2010}%
  \BibitemOpen
  \bibfield  {author} {\bibinfo {author} {\bibfnamefont {H.-Y.}\ \bibnamefont {Lo}}, \bibinfo {author} {\bibfnamefont {P.-C.}\ \bibnamefont {Su}},\ and\ \bibinfo {author} {\bibfnamefont {Y.-F.}\ \bibnamefont {Chen}},\ }\bibfield  {title} {\bibinfo {title} {Low-light-level cross-phase modulation by quantum interference},\ }\href {https://doi.org/10.1103/PhysRevA.81.053829} {\bibfield  {journal} {\bibinfo  {journal} {Phys. Rev. A}\ }\textbf {\bibinfo {volume} {81}},\ \bibinfo {pages} {053829} (\bibinfo {year} {2010})}\BibitemShut {NoStop}%
\bibitem [{\citenamefont {Lo}\ \emph {et~al.}(2011)\citenamefont {Lo}, \citenamefont {Chen}, \citenamefont {Su}, \citenamefont {Chen}, \citenamefont {Chen}, \citenamefont {Chen}, \citenamefont {Yu},\ and\ \citenamefont {Chen}}]{Lo2011}%
  \BibitemOpen
  \bibfield  {author} {\bibinfo {author} {\bibfnamefont {H.-Y.}\ \bibnamefont {Lo}}, \bibinfo {author} {\bibfnamefont {Y.-C.}\ \bibnamefont {Chen}}, \bibinfo {author} {\bibfnamefont {P.-C.}\ \bibnamefont {Su}}, \bibinfo {author} {\bibfnamefont {H.-C.}\ \bibnamefont {Chen}}, \bibinfo {author} {\bibfnamefont {J.-X.}\ \bibnamefont {Chen}}, \bibinfo {author} {\bibfnamefont {Y.-C.}\ \bibnamefont {Chen}}, \bibinfo {author} {\bibfnamefont {I.~A.}\ \bibnamefont {Yu}},\ and\ \bibinfo {author} {\bibfnamefont {Y.-F.}\ \bibnamefont {Chen}},\ }\bibfield  {title} {\bibinfo {title} {Electromagnetically-induced-transparency-based cross-phase-modulation at attojoule levels},\ }\href {https://doi.org/10.1103/PhysRevA.83.041804} {\bibfield  {journal} {\bibinfo  {journal} {Phys. Rev. A}\ }\textbf {\bibinfo {volume} {83}},\ \bibinfo {pages} {041804(R)} (\bibinfo {year} {2011})}\BibitemShut {NoStop}%
\bibitem [{\citenamefont {Firstenberg}\ \emph {et~al.}(2013)\citenamefont {Firstenberg}, \citenamefont {Peyronel}, \citenamefont {Liang}, \citenamefont {Gorshkov}, \citenamefont {Lukin},\ and\ \citenamefont {Vuletić}}]{Firstenberg2013}%
  \BibitemOpen
  \bibfield  {author} {\bibinfo {author} {\bibfnamefont {O.}~\bibnamefont {Firstenberg}}, \bibinfo {author} {\bibfnamefont {T.}~\bibnamefont {Peyronel}}, \bibinfo {author} {\bibfnamefont {Q.-Y.}\ \bibnamefont {Liang}}, \bibinfo {author} {\bibfnamefont {A.~V.}\ \bibnamefont {Gorshkov}}, \bibinfo {author} {\bibfnamefont {M.~D.}\ \bibnamefont {Lukin}},\ and\ \bibinfo {author} {\bibfnamefont {V.}~\bibnamefont {Vuletić}},\ }\bibfield  {title} {\bibinfo {title} {Attractive photons in a quantum nonlinear medium},\ }\href {https://doi.org/10.1038/nature12512} {\bibfield  {journal} {\bibinfo  {journal} {Nature}\ }\textbf {\bibinfo {volume} {502}},\ \bibinfo {pages} {71} (\bibinfo {year} {2013})}\BibitemShut {NoStop}%
\bibitem [{\citenamefont {Feizpour}\ \emph {et~al.}(2015)\citenamefont {Feizpour}, \citenamefont {Hallaji}, \citenamefont {Dmochowski},\ and\ \citenamefont {Steinberg}}]{Feizpour2015}%
  \BibitemOpen
  \bibfield  {author} {\bibinfo {author} {\bibfnamefont {A.}~\bibnamefont {Feizpour}}, \bibinfo {author} {\bibfnamefont {M.}~\bibnamefont {Hallaji}}, \bibinfo {author} {\bibfnamefont {G.}~\bibnamefont {Dmochowski}},\ and\ \bibinfo {author} {\bibfnamefont {A.~M.}\ \bibnamefont {Steinberg}},\ }\bibfield  {title} {\bibinfo {title} {Observation of the nonlinear phase shift due to single post-selected photons},\ }\href {https://doi.org/10.1038/nphys3433} {\bibfield  {journal} {\bibinfo  {journal} {Nat. Phys.}\ }\textbf {\bibinfo {volume} {11}},\ \bibinfo {pages} {905} (\bibinfo {year} {2015})}\BibitemShut {NoStop}%
\bibitem [{\citenamefont {Liu}\ \emph {et~al.}(2016)\citenamefont {Liu}, \citenamefont {Chen}, \citenamefont {Chen}, \citenamefont {Lo}, \citenamefont {Tsai}, \citenamefont {Yu}, \citenamefont {Chen},\ and\ \citenamefont {Chen}}]{Liu2016}%
  \BibitemOpen
  \bibfield  {author} {\bibinfo {author} {\bibfnamefont {Z.-Y.}\ \bibnamefont {Liu}}, \bibinfo {author} {\bibfnamefont {Y.-H.}\ \bibnamefont {Chen}}, \bibinfo {author} {\bibfnamefont {Y.-C.}\ \bibnamefont {Chen}}, \bibinfo {author} {\bibfnamefont {H.-Y.}\ \bibnamefont {Lo}}, \bibinfo {author} {\bibfnamefont {P.-J.}\ \bibnamefont {Tsai}}, \bibinfo {author} {\bibfnamefont {I.~A.}\ \bibnamefont {Yu}}, \bibinfo {author} {\bibfnamefont {Y.-C.}\ \bibnamefont {Chen}},\ and\ \bibinfo {author} {\bibfnamefont {Y.-F.}\ \bibnamefont {Chen}},\ }\bibfield  {title} {\bibinfo {title} {{Large Cross-Phase Modulations at the Few-Photon Level}},\ }\href {https://doi.org/10.1103/PhysRevLett.117.203601} {\bibfield  {journal} {\bibinfo  {journal} {Phys. Rev. Lett.}\ }\textbf {\bibinfo {volume} {117}},\ \bibinfo {pages} {203601} (\bibinfo {year} {2016})}\BibitemShut {NoStop}%
\bibitem [{\citenamefont {Tiarks}\ \emph {et~al.}(2016)\citenamefont {Tiarks}, \citenamefont {Schmidt}, \citenamefont {Rempe},\ and\ \citenamefont {Dürr}}]{Tiarks2016}%
  \BibitemOpen
  \bibfield  {author} {\bibinfo {author} {\bibfnamefont {D.}~\bibnamefont {Tiarks}}, \bibinfo {author} {\bibfnamefont {S.}~\bibnamefont {Schmidt}}, \bibinfo {author} {\bibfnamefont {G.}~\bibnamefont {Rempe}},\ and\ \bibinfo {author} {\bibfnamefont {S.}~\bibnamefont {Dürr}},\ }\bibfield  {title} {\bibinfo {title} {Optical $\pi$ phase shift created with a single-photon pulse},\ }\href {https://doi.org/10.1126/sciadv.1600036} {\bibfield  {journal} {\bibinfo  {journal} {Sci. Adv.}\ }\textbf {\bibinfo {volume} {2}},\ \bibinfo {pages} {e1600036} (\bibinfo {year} {2016})}\BibitemShut {NoStop}%
\bibitem [{\citenamefont {Tiarks}\ \emph {et~al.}(2019)\citenamefont {Tiarks}, \citenamefont {Schmidt-Eberle}, \citenamefont {Stolz}, \citenamefont {Rempe},\ and\ \citenamefont {Dürr}}]{Tiarks2019}%
  \BibitemOpen
  \bibfield  {author} {\bibinfo {author} {\bibfnamefont {D.}~\bibnamefont {Tiarks}}, \bibinfo {author} {\bibfnamefont {S.}~\bibnamefont {Schmidt-Eberle}}, \bibinfo {author} {\bibfnamefont {T.}~\bibnamefont {Stolz}}, \bibinfo {author} {\bibfnamefont {G.}~\bibnamefont {Rempe}},\ and\ \bibinfo {author} {\bibfnamefont {S.}~\bibnamefont {Dürr}},\ }\bibfield  {title} {\bibinfo {title} {{A photon–photon quantum gate based on Rydberg interactions}},\ }\href {https://doi.org/10.1038/s41567-018-0313-7} {\bibfield  {journal} {\bibinfo  {journal} {Nat. Phys.}\ }\textbf {\bibinfo {volume} {15}},\ \bibinfo {pages} {124} (\bibinfo {year} {2019})}\BibitemShut {NoStop}%
\bibitem [{\citenamefont {Matsko}\ \emph {et~al.}(2003)\citenamefont {Matsko}, \citenamefont {Novikova}, \citenamefont {Welch},\ and\ \citenamefont {Zubairy}}]{Matsko2003}%
  \BibitemOpen
  \bibfield  {author} {\bibinfo {author} {\bibfnamefont {A.~B.}\ \bibnamefont {Matsko}}, \bibinfo {author} {\bibfnamefont {I.}~\bibnamefont {Novikova}}, \bibinfo {author} {\bibfnamefont {G.~R.}\ \bibnamefont {Welch}},\ and\ \bibinfo {author} {\bibfnamefont {M.~S.}\ \bibnamefont {Zubairy}},\ }\bibfield  {title} {\bibinfo {title} {{Enhancement of Kerr nonlinearity by multiphoton coherence}},\ }\href {https://doi.org/10.1364/OL.28.000096} {\bibfield  {journal} {\bibinfo  {journal} {Opt. Lett.}\ }\textbf {\bibinfo {volume} {28}},\ \bibinfo {pages} {96} (\bibinfo {year} {2003})}\BibitemShut {NoStop}%
\bibitem [{\citenamefont {Sagona-Stophel}\ \emph {et~al.}(2020)\citenamefont {Sagona-Stophel}, \citenamefont {Shahrokhshahi}, \citenamefont {Jordaan}, \citenamefont {Namazi},\ and\ \citenamefont {Figueroa}}]{Sagona-Stophel2020}%
  \BibitemOpen
  \bibfield  {author} {\bibinfo {author} {\bibfnamefont {S.}~\bibnamefont {Sagona-Stophel}}, \bibinfo {author} {\bibfnamefont {R.}~\bibnamefont {Shahrokhshahi}}, \bibinfo {author} {\bibfnamefont {B.}~\bibnamefont {Jordaan}}, \bibinfo {author} {\bibfnamefont {M.}~\bibnamefont {Namazi}},\ and\ \bibinfo {author} {\bibfnamefont {E.}~\bibnamefont {Figueroa}},\ }\bibfield  {title} {\bibinfo {title} {{Conditional $\pi$-Phase Shift of Single-Photon-Level Pulses at Room Temperature}},\ }\href {https://doi.org/10.1103/PhysRevLett.125.243601} {\bibfield  {journal} {\bibinfo  {journal} {Phys. Rev. Lett.}\ }\textbf {\bibinfo {volume} {125}},\ \bibinfo {pages} {243601} (\bibinfo {year} {2020})}\BibitemShut {NoStop}%
\bibitem [{\citenamefont {Davis}\ \emph {et~al.}(2025)\citenamefont {Davis}, \citenamefont {Burdekin}, \citenamefont {Wasawo}, \citenamefont {Thomas}, \citenamefont {Mosley}, \citenamefont {Nunn},\ and\ \citenamefont {McGarry}}]{Davis2024}%
  \BibitemOpen
  \bibfield  {author} {\bibinfo {author} {\bibfnamefont {W.~O.~C.}\ \bibnamefont {Davis}}, \bibinfo {author} {\bibfnamefont {P.}~\bibnamefont {Burdekin}}, \bibinfo {author} {\bibfnamefont {T.}~\bibnamefont {Wasawo}}, \bibinfo {author} {\bibfnamefont {S.~E.}\ \bibnamefont {Thomas}}, \bibinfo {author} {\bibfnamefont {P.~J.}\ \bibnamefont {Mosley}}, \bibinfo {author} {\bibfnamefont {J.}~\bibnamefont {Nunn}},\ and\ \bibinfo {author} {\bibfnamefont {C.}~\bibnamefont {McGarry}},\ }\bibfield  {title} {\bibinfo {title} {Fast, low-loss, all-optical phase modulation in warm rubidium vapour},\ }\href {https://doi.org/10.1088/2058-9565/ad9fa3} {\bibfield  {journal} {\bibinfo  {journal} {Quantum Sci. Technol.}\ }\textbf {\bibinfo {volume} {10}},\ \bibinfo {pages} {025001} (\bibinfo {year} {2025})}\BibitemShut {NoStop}%
\bibitem [{\citenamefont {Hickman}\ \emph {et~al.}(2015)\citenamefont {Hickman}, \citenamefont {Pittman},\ and\ \citenamefont {Franson}}]{Hickman2015}%
  \BibitemOpen
  \bibfield  {author} {\bibinfo {author} {\bibfnamefont {G.~T.}\ \bibnamefont {Hickman}}, \bibinfo {author} {\bibfnamefont {T.~B.}\ \bibnamefont {Pittman}},\ and\ \bibinfo {author} {\bibfnamefont {J.~D.}\ \bibnamefont {Franson}},\ }\bibfield  {title} {\bibinfo {title} {Low-power cross-phase modulation in a metastable xenon-filled cavity for quantum-information applications},\ }\href {https://doi.org/10.1103/PhysRevA.92.053808} {\bibfield  {journal} {\bibinfo  {journal} {Phys. Rev. A}\ }\textbf {\bibinfo {volume} {92}},\ \bibinfo {pages} {053808} (\bibinfo {year} {2015})}\BibitemShut {NoStop}%
\bibitem [{\citenamefont {Beck}\ \emph {et~al.}(2016)\citenamefont {Beck}, \citenamefont {Hosseini}, \citenamefont {Duan},\ and\ \citenamefont {Vuletić}}]{Beck2016}%
  \BibitemOpen
  \bibfield  {author} {\bibinfo {author} {\bibfnamefont {K.~M.}\ \bibnamefont {Beck}}, \bibinfo {author} {\bibfnamefont {M.}~\bibnamefont {Hosseini}}, \bibinfo {author} {\bibfnamefont {Y.}~\bibnamefont {Duan}},\ and\ \bibinfo {author} {\bibfnamefont {V.}~\bibnamefont {Vuletić}},\ }\bibfield  {title} {\bibinfo {title} {Large conditional single-photon cross-phase modulation},\ }\href {https://doi.org/10.1073/pnas.1524117113} {\bibfield  {journal} {\bibinfo  {journal} {Proc. Natl. Acad. Sci. U.S.A.}\ }\textbf {\bibinfo {volume} {113}},\ \bibinfo {pages} {9740} (\bibinfo {year} {2016})}\BibitemShut {NoStop}%
\bibitem [{\citenamefont {Venkataraman}\ \emph {et~al.}(2013)\citenamefont {Venkataraman}, \citenamefont {Saha},\ and\ \citenamefont {Gaeta}}]{Venkataraman2013}%
  \BibitemOpen
  \bibfield  {author} {\bibinfo {author} {\bibfnamefont {V.}~\bibnamefont {Venkataraman}}, \bibinfo {author} {\bibfnamefont {K.}~\bibnamefont {Saha}},\ and\ \bibinfo {author} {\bibfnamefont {A.~L.}\ \bibnamefont {Gaeta}},\ }\bibfield  {title} {\bibinfo {title} {Phase modulation at the few-photon level for weak-nonlinearity-based quantum computing},\ }\href {https://doi.org/10.1038/nphoton.2012.283} {\bibfield  {journal} {\bibinfo  {journal} {Nat. Photon.}\ }\textbf {\bibinfo {volume} {7}},\ \bibinfo {pages} {138} (\bibinfo {year} {2013})}\BibitemShut {NoStop}%
\bibitem [{\citenamefont {Matsuda}\ \emph {et~al.}(2009)\citenamefont {Matsuda}, \citenamefont {Shimizu}, \citenamefont {Mitsumori}, \citenamefont {Kosaka},\ and\ \citenamefont {Edamatsu}}]{Matsuda2009}%
  \BibitemOpen
  \bibfield  {author} {\bibinfo {author} {\bibfnamefont {N.}~\bibnamefont {Matsuda}}, \bibinfo {author} {\bibfnamefont {R.}~\bibnamefont {Shimizu}}, \bibinfo {author} {\bibfnamefont {Y.}~\bibnamefont {Mitsumori}}, \bibinfo {author} {\bibfnamefont {H.}~\bibnamefont {Kosaka}},\ and\ \bibinfo {author} {\bibfnamefont {K.}~\bibnamefont {Edamatsu}},\ }\bibfield  {title} {\bibinfo {title} {{Observation of optical-fibre Kerr nonlinearity at the single-photon level}},\ }\href {https://doi.org/10.1038/nphoton.2008.292} {\bibfield  {journal} {\bibinfo  {journal} {Nat. Photon.}\ }\textbf {\bibinfo {volume} {3}},\ \bibinfo {pages} {95} (\bibinfo {year} {2009})}\BibitemShut {NoStop}%
\bibitem [{\citenamefont {Kuriakose}\ \emph {et~al.}(2022)\citenamefont {Kuriakose}, \citenamefont {Walker}, \citenamefont {Dowling}, \citenamefont {Kyriienko}, \citenamefont {Shelykh}, \citenamefont {St-Jean}, \citenamefont {Zambon}, \citenamefont {Lemaître}, \citenamefont {Sagnes}, \citenamefont {Legratiet}, \citenamefont {Harouri}, \citenamefont {Ravets}, \citenamefont {Skolnick}, \citenamefont {Amo}, \citenamefont {Bloch},\ and\ \citenamefont {Krizhanovskii}}]{Kuriakose2022}%
  \BibitemOpen
  \bibfield  {author} {\bibinfo {author} {\bibfnamefont {T.}~\bibnamefont {Kuriakose}}, \bibinfo {author} {\bibfnamefont {P.~M.}\ \bibnamefont {Walker}}, \bibinfo {author} {\bibfnamefont {T.}~\bibnamefont {Dowling}}, \bibinfo {author} {\bibfnamefont {O.}~\bibnamefont {Kyriienko}}, \bibinfo {author} {\bibfnamefont {I.~A.}\ \bibnamefont {Shelykh}}, \bibinfo {author} {\bibfnamefont {P.}~\bibnamefont {St-Jean}}, \bibinfo {author} {\bibfnamefont {N.~C.}\ \bibnamefont {Zambon}}, \bibinfo {author} {\bibfnamefont {A.}~\bibnamefont {Lemaître}}, \bibinfo {author} {\bibfnamefont {I.}~\bibnamefont {Sagnes}}, \bibinfo {author} {\bibfnamefont {L.}~\bibnamefont {Legratiet}}, \bibinfo {author} {\bibfnamefont {A.}~\bibnamefont {Harouri}}, \bibinfo {author} {\bibfnamefont {S.}~\bibnamefont {Ravets}}, \bibinfo {author} {\bibfnamefont {M.~S.}\ \bibnamefont {Skolnick}}, \bibinfo {author} {\bibfnamefont {A.}~\bibnamefont {Amo}}, \bibinfo {author} {\bibfnamefont {J.}~\bibnamefont {Bloch}},\ and\ \bibinfo {author} {\bibfnamefont
  {D.~N.}\ \bibnamefont {Krizhanovskii}},\ }\bibfield  {title} {\bibinfo {title} {Few-photon all-optical phase rotation in a quantum-well micropillar cavity},\ }\href {https://doi.org/10.1038/s41566-022-01019-6} {\bibfield  {journal} {\bibinfo  {journal} {Nat. Photon.}\ }\textbf {\bibinfo {volume} {16}},\ \bibinfo {pages} {566} (\bibinfo {year} {2022})}\BibitemShut {NoStop}%
\bibitem [{\citenamefont {Rebi\ifmmode~\acute{c}\else \'{c}\fi{}}\ \emph {et~al.}(2009)\citenamefont {Rebi\ifmmode~\acute{c}\else \'{c}\fi{}}, \citenamefont {Twamley},\ and\ \citenamefont {Milburn}}]{Rebic2009}%
  \BibitemOpen
  \bibfield  {author} {\bibinfo {author} {\bibfnamefont {S.}~\bibnamefont {Rebi\ifmmode~\acute{c}\else \'{c}\fi{}}}, \bibinfo {author} {\bibfnamefont {J.}~\bibnamefont {Twamley}},\ and\ \bibinfo {author} {\bibfnamefont {G.~J.}\ \bibnamefont {Milburn}},\ }\bibfield  {title} {\bibinfo {title} {{Giant Kerr Nonlinearities in Circuit Quantum Electrodynamics}},\ }\href {https://doi.org/10.1103/PhysRevLett.103.150503} {\bibfield  {journal} {\bibinfo  {journal} {Phys. Rev. Lett.}\ }\textbf {\bibinfo {volume} {103}},\ \bibinfo {pages} {150503} (\bibinfo {year} {2009})}\BibitemShut {NoStop}%
\bibitem [{\citenamefont {Hoi}\ \emph {et~al.}(2013)\citenamefont {Hoi}, \citenamefont {Kockum}, \citenamefont {Palomaki}, \citenamefont {Stace}, \citenamefont {Fan}, \citenamefont {Tornberg}, \citenamefont {Sathyamoorthy}, \citenamefont {Johansson}, \citenamefont {Delsing},\ and\ \citenamefont {Wilson}}]{Hoi2013}%
  \BibitemOpen
  \bibfield  {author} {\bibinfo {author} {\bibfnamefont {I.-C.}\ \bibnamefont {Hoi}}, \bibinfo {author} {\bibfnamefont {A.~F.}\ \bibnamefont {Kockum}}, \bibinfo {author} {\bibfnamefont {T.}~\bibnamefont {Palomaki}}, \bibinfo {author} {\bibfnamefont {T.~M.}\ \bibnamefont {Stace}}, \bibinfo {author} {\bibfnamefont {B.}~\bibnamefont {Fan}}, \bibinfo {author} {\bibfnamefont {L.}~\bibnamefont {Tornberg}}, \bibinfo {author} {\bibfnamefont {S.~R.}\ \bibnamefont {Sathyamoorthy}}, \bibinfo {author} {\bibfnamefont {G.}~\bibnamefont {Johansson}}, \bibinfo {author} {\bibfnamefont {P.}~\bibnamefont {Delsing}},\ and\ \bibinfo {author} {\bibfnamefont {C.~M.}\ \bibnamefont {Wilson}},\ }\bibfield  {title} {\bibinfo {title} {{Giant Cross--Kerr Effect for Propagating Microwaves Induced by an Artificial Atom}},\ }\href {https://doi.org/10.1103/PhysRevLett.111.053601} {\bibfield  {journal} {\bibinfo  {journal} {Phys. Rev. Lett.}\ }\textbf {\bibinfo {volume} {111}},\ \bibinfo {pages} {053601} (\bibinfo {year} {2013})}\BibitemShut
  {NoStop}%
\bibitem [{\citenamefont {Shapiro}(2006)}]{Shapiro2006}%
  \BibitemOpen
  \bibfield  {author} {\bibinfo {author} {\bibfnamefont {J.~H.}\ \bibnamefont {Shapiro}},\ }\bibfield  {title} {\bibinfo {title} {{Single-photon Kerr nonlinearities do not help quantum computation}},\ }\href {https://doi.org/10.1103/PhysRevA.73.062305} {\bibfield  {journal} {\bibinfo  {journal} {Phys. Rev. A}\ }\textbf {\bibinfo {volume} {73}},\ \bibinfo {pages} {062305} (\bibinfo {year} {2006})}\BibitemShut {NoStop}%
\bibitem [{\citenamefont {Gea-Banacloche}(2010)}]{Gea-Banacloche2010}%
  \BibitemOpen
  \bibfield  {author} {\bibinfo {author} {\bibfnamefont {J.}~\bibnamefont {Gea-Banacloche}},\ }\bibfield  {title} {\bibinfo {title} {{Impossibility of large phase shifts via the giant Kerr effect with single-photon wave packets}},\ }\href {https://doi.org/10.1103/PhysRevA.81.043823} {\bibfield  {journal} {\bibinfo  {journal} {Phys. Rev. A}\ }\textbf {\bibinfo {volume} {81}},\ \bibinfo {pages} {043823} (\bibinfo {year} {2010})}\BibitemShut {NoStop}%
\bibitem [{\citenamefont {Dove}\ \emph {et~al.}(2014)\citenamefont {Dove}, \citenamefont {Chudzicki},\ and\ \citenamefont {Shapiro}}]{Dove2014}%
  \BibitemOpen
  \bibfield  {author} {\bibinfo {author} {\bibfnamefont {J.}~\bibnamefont {Dove}}, \bibinfo {author} {\bibfnamefont {C.}~\bibnamefont {Chudzicki}},\ and\ \bibinfo {author} {\bibfnamefont {J.~H.}\ \bibnamefont {Shapiro}},\ }\bibfield  {title} {\bibinfo {title} {Phase-noise limitations on single-photon cross-phase modulation with differing group velocities},\ }\href {https://doi.org/10.1103/PhysRevA.90.062314} {\bibfield  {journal} {\bibinfo  {journal} {Phys. Rev. A}\ }\textbf {\bibinfo {volume} {90}},\ \bibinfo {pages} {062314} (\bibinfo {year} {2014})}\BibitemShut {NoStop}%
\bibitem [{\citenamefont {Duan}\ and\ \citenamefont {Kimble}(2004)}]{Duan2004}%
  \BibitemOpen
  \bibfield  {author} {\bibinfo {author} {\bibfnamefont {L.-M.}\ \bibnamefont {Duan}}\ and\ \bibinfo {author} {\bibfnamefont {H.~J.}\ \bibnamefont {Kimble}},\ }\bibfield  {title} {\bibinfo {title} {{Scalable Photonic Quantum Computation through Cavity-Assisted Interactions}},\ }\href {https://doi.org/10.1103/PhysRevLett.92.127902} {\bibfield  {journal} {\bibinfo  {journal} {Phys. Rev. Lett.}\ }\textbf {\bibinfo {volume} {92}},\ \bibinfo {pages} {127902} (\bibinfo {year} {2004})}\BibitemShut {NoStop}%
\bibitem [{\citenamefont {Koshino}\ \emph {et~al.}(2010)\citenamefont {Koshino}, \citenamefont {Ishizaka},\ and\ \citenamefont {Nakamura}}]{Koshino2010}%
  \BibitemOpen
  \bibfield  {author} {\bibinfo {author} {\bibfnamefont {K.}~\bibnamefont {Koshino}}, \bibinfo {author} {\bibfnamefont {S.}~\bibnamefont {Ishizaka}},\ and\ \bibinfo {author} {\bibfnamefont {Y.}~\bibnamefont {Nakamura}},\ }\bibfield  {title} {\bibinfo {title} {{Deterministic photon-photon $\sqrt{\text{SWAP}}$ gate using a $\ensuremath{\Lambda}$ system}},\ }\href {https://doi.org/10.1103/PhysRevA.82.010301} {\bibfield  {journal} {\bibinfo  {journal} {Phys. Rev. A}\ }\textbf {\bibinfo {volume} {82}},\ \bibinfo {pages} {010301(R)} (\bibinfo {year} {2010})}\BibitemShut {NoStop}%
\bibitem [{\citenamefont {Johne}\ and\ \citenamefont {Fiore}(2012)}]{Johne2012}%
  \BibitemOpen
  \bibfield  {author} {\bibinfo {author} {\bibfnamefont {R.}~\bibnamefont {Johne}}\ and\ \bibinfo {author} {\bibfnamefont {A.}~\bibnamefont {Fiore}},\ }\bibfield  {title} {\bibinfo {title} {Proposal for a two-qubit quantum phase gate for quantum photonic integrated circuits},\ }\href {https://doi.org/10.1103/PhysRevA.86.063815} {\bibfield  {journal} {\bibinfo  {journal} {Phys. Rev. A}\ }\textbf {\bibinfo {volume} {86}},\ \bibinfo {pages} {063815} (\bibinfo {year} {2012})}\BibitemShut {NoStop}%
\bibitem [{\citenamefont {Chudzicki}\ \emph {et~al.}(2013)\citenamefont {Chudzicki}, \citenamefont {Chuang},\ and\ \citenamefont {Shapiro}}]{Chudzicki2013}%
  \BibitemOpen
  \bibfield  {author} {\bibinfo {author} {\bibfnamefont {C.}~\bibnamefont {Chudzicki}}, \bibinfo {author} {\bibfnamefont {I.~L.}\ \bibnamefont {Chuang}},\ and\ \bibinfo {author} {\bibfnamefont {J.~H.}\ \bibnamefont {Shapiro}},\ }\bibfield  {title} {\bibinfo {title} {Deterministic and cascadable conditional phase gate for photonic qubits},\ }\href {https://doi.org/10.1103/PhysRevA.87.042325} {\bibfield  {journal} {\bibinfo  {journal} {Phys. Rev. A}\ }\textbf {\bibinfo {volume} {87}},\ \bibinfo {pages} {042325} (\bibinfo {year} {2013})}\BibitemShut {NoStop}%
\bibitem [{\citenamefont {Krastanov}\ \emph {et~al.}(2022)\citenamefont {Krastanov}, \citenamefont {Jacobs}, \citenamefont {Gilbert}, \citenamefont {Englund},\ and\ \citenamefont {Heuck}}]{Krastanov2022}%
  \BibitemOpen
  \bibfield  {author} {\bibinfo {author} {\bibfnamefont {S.}~\bibnamefont {Krastanov}}, \bibinfo {author} {\bibfnamefont {K.}~\bibnamefont {Jacobs}}, \bibinfo {author} {\bibfnamefont {G.}~\bibnamefont {Gilbert}}, \bibinfo {author} {\bibfnamefont {D.~R.}\ \bibnamefont {Englund}},\ and\ \bibinfo {author} {\bibfnamefont {M.}~\bibnamefont {Heuck}},\ }\bibfield  {title} {\bibinfo {title} {Controlled-phase gate by dynamic coupling of photons to a two-level emitter},\ }\href {https://doi.org/10.1038/s41534-022-00604-5} {\bibfield  {journal} {\bibinfo  {journal} {npj Quantum Inf.}\ }\textbf {\bibinfo {volume} {8}},\ \bibinfo {pages} {103} (\bibinfo {year} {2022})}\BibitemShut {NoStop}%
\bibitem [{\citenamefont {Tian}\ \emph {et~al.}(2026)\citenamefont {Tian}, \citenamefont {Gu},\ and\ \citenamefont {Chen}}]{Tian2024}%
  \BibitemOpen
  \bibfield  {author} {\bibinfo {author} {\bibfnamefont {Z.}~\bibnamefont {Tian}}, \bibinfo {author} {\bibfnamefont {Y.}~\bibnamefont {Gu}},\ and\ \bibinfo {author} {\bibfnamefont {X.-W.}\ \bibnamefont {Chen}},\ }\bibfield  {title} {\bibinfo {title} {Deterministic photonic controlled-$\ensuremath{\pi}$-phase gate enabled by passive time-reversal-symmetric photon transport},\ }\href {https://doi.org/10.1103/49mp-bx59} {\bibfield  {journal} {\bibinfo  {journal} {Phys. Rev. Appl.}\ }\textbf {\bibinfo {volume} {26}},\ \bibinfo {pages} {024069} (\bibinfo {year} {2026})}\BibitemShut {NoStop}%
\bibitem [{\citenamefont {Tang}\ \emph {et~al.}(2025)\citenamefont {Tang}, \citenamefont {Zhao}, \citenamefont {Yang}, \citenamefont {Tian}, \citenamefont {Tang}, \citenamefont {Wang}, \citenamefont {Lamata},\ and\ \citenamefont {Peng}}]{Tang2024}%
  \BibitemOpen
  \bibfield  {author} {\bibinfo {author} {\bibfnamefont {J.-C.}\ \bibnamefont {Tang}}, \bibinfo {author} {\bibfnamefont {J.}~\bibnamefont {Zhao}}, \bibinfo {author} {\bibfnamefont {H.}~\bibnamefont {Yang}}, \bibinfo {author} {\bibfnamefont {J.}~\bibnamefont {Tian}}, \bibinfo {author} {\bibfnamefont {P.}~\bibnamefont {Tang}}, \bibinfo {author} {\bibfnamefont {S.-P.}\ \bibnamefont {Wang}}, \bibinfo {author} {\bibfnamefont {L.}~\bibnamefont {Lamata}},\ and\ \bibinfo {author} {\bibfnamefont {J.}~\bibnamefont {Peng}},\ }\bibfield  {title} {\bibinfo {title} {{Deterministic two-photon controlled-$Z$ gate with the two-photon quantum Rabi model}},\ }\href {https://doi.org/10.1103/PhysRevA.111.052601} {\bibfield  {journal} {\bibinfo  {journal} {Phys. Rev. A}\ }\textbf {\bibinfo {volume} {111}},\ \bibinfo {pages} {052601} (\bibinfo {year} {2025})}\BibitemShut {NoStop}%
\bibitem [{\citenamefont {Turchette}\ \emph {et~al.}(1995)\citenamefont {Turchette}, \citenamefont {Hood}, \citenamefont {Lange}, \citenamefont {Mabuchi},\ and\ \citenamefont {Kimble}}]{Turchette1995}%
  \BibitemOpen
  \bibfield  {author} {\bibinfo {author} {\bibfnamefont {Q.~A.}\ \bibnamefont {Turchette}}, \bibinfo {author} {\bibfnamefont {C.~J.}\ \bibnamefont {Hood}}, \bibinfo {author} {\bibfnamefont {W.}~\bibnamefont {Lange}}, \bibinfo {author} {\bibfnamefont {H.}~\bibnamefont {Mabuchi}},\ and\ \bibinfo {author} {\bibfnamefont {H.~J.}\ \bibnamefont {Kimble}},\ }\bibfield  {title} {\bibinfo {title} {{Measurement of Conditional Phase Shifts for Quantum Logic}},\ }\href {https://doi.org/10.1103/PhysRevLett.75.4710} {\bibfield  {journal} {\bibinfo  {journal} {Phys. Rev. Lett.}\ }\textbf {\bibinfo {volume} {75}},\ \bibinfo {pages} {4710} (\bibinfo {year} {1995})}\BibitemShut {NoStop}%
\bibitem [{\citenamefont {Fushman}\ \emph {et~al.}(2008)\citenamefont {Fushman}, \citenamefont {Englund}, \citenamefont {Faraon}, \citenamefont {Stoltz}, \citenamefont {Petroff},\ and\ \citenamefont {Vučković}}]{Fushman2008}%
  \BibitemOpen
  \bibfield  {author} {\bibinfo {author} {\bibfnamefont {I.}~\bibnamefont {Fushman}}, \bibinfo {author} {\bibfnamefont {D.}~\bibnamefont {Englund}}, \bibinfo {author} {\bibfnamefont {A.}~\bibnamefont {Faraon}}, \bibinfo {author} {\bibfnamefont {N.}~\bibnamefont {Stoltz}}, \bibinfo {author} {\bibfnamefont {P.}~\bibnamefont {Petroff}},\ and\ \bibinfo {author} {\bibfnamefont {J.}~\bibnamefont {Vučković}},\ }\bibfield  {title} {\bibinfo {title} {{Controlled Phase Shifts with a Single Quantum Dot}},\ }\href {https://doi.org/10.1126/science.1154643} {\bibfield  {journal} {\bibinfo  {journal} {Science}\ }\textbf {\bibinfo {volume} {320}},\ \bibinfo {pages} {769} (\bibinfo {year} {2008})}\BibitemShut {NoStop}%
\bibitem [{\citenamefont {Volz}\ \emph {et~al.}(2014)\citenamefont {Volz}, \citenamefont {Scheucher}, \citenamefont {Junge},\ and\ \citenamefont {Rauschenbeutel}}]{Volz2014}%
  \BibitemOpen
  \bibfield  {author} {\bibinfo {author} {\bibfnamefont {J.}~\bibnamefont {Volz}}, \bibinfo {author} {\bibfnamefont {M.}~\bibnamefont {Scheucher}}, \bibinfo {author} {\bibfnamefont {C.}~\bibnamefont {Junge}},\ and\ \bibinfo {author} {\bibfnamefont {A.}~\bibnamefont {Rauschenbeutel}},\ }\bibfield  {title} {\bibinfo {title} {Nonlinear $\pi$ phase shift for single fibre-guided photons interacting with a single resonator-enhanced atom},\ }\href {https://doi.org/10.1038/nphoton.2014.253} {\bibfield  {journal} {\bibinfo  {journal} {Nat. Photon.}\ }\textbf {\bibinfo {volume} {8}},\ \bibinfo {pages} {965} (\bibinfo {year} {2014})}\BibitemShut {NoStop}%
\bibitem [{\citenamefont {Hacker}\ \emph {et~al.}(2016)\citenamefont {Hacker}, \citenamefont {Welte}, \citenamefont {Rempe},\ and\ \citenamefont {Ritter}}]{Hacker2016}%
  \BibitemOpen
  \bibfield  {author} {\bibinfo {author} {\bibfnamefont {B.}~\bibnamefont {Hacker}}, \bibinfo {author} {\bibfnamefont {S.}~\bibnamefont {Welte}}, \bibinfo {author} {\bibfnamefont {G.}~\bibnamefont {Rempe}},\ and\ \bibinfo {author} {\bibfnamefont {S.}~\bibnamefont {Ritter}},\ }\bibfield  {title} {\bibinfo {title} {A photon–photon quantum gate based on a single atom in an optical resonator},\ }\href {https://doi.org/10.1038/nature18592} {\bibfield  {journal} {\bibinfo  {journal} {Nature}\ }\textbf {\bibinfo {volume} {536}},\ \bibinfo {pages} {193} (\bibinfo {year} {2016})}\BibitemShut {NoStop}%
\bibitem [{\citenamefont {Ralph}\ \emph {et~al.}(2015)\citenamefont {Ralph}, \citenamefont {S\"ollner}, \citenamefont {Mahmoodian}, \citenamefont {White},\ and\ \citenamefont {Lodahl}}]{Ralph2015}%
  \BibitemOpen
  \bibfield  {author} {\bibinfo {author} {\bibfnamefont {T.~C.}\ \bibnamefont {Ralph}}, \bibinfo {author} {\bibfnamefont {I.}~\bibnamefont {S\"ollner}}, \bibinfo {author} {\bibfnamefont {S.}~\bibnamefont {Mahmoodian}}, \bibinfo {author} {\bibfnamefont {A.~G.}\ \bibnamefont {White}},\ and\ \bibinfo {author} {\bibfnamefont {P.}~\bibnamefont {Lodahl}},\ }\bibfield  {title} {\bibinfo {title} {{Photon Sorting, Efficient Bell Measurements, and a Deterministic Controlled-$Z$ Gate Using a Passive Two-Level Nonlinearity}},\ }\href {https://doi.org/10.1103/PhysRevLett.114.173603} {\bibfield  {journal} {\bibinfo  {journal} {Phys. Rev. Lett.}\ }\textbf {\bibinfo {volume} {114}},\ \bibinfo {pages} {173603} (\bibinfo {year} {2015})}\BibitemShut {NoStop}%
\bibitem [{\citenamefont {Schrinski}\ \emph {et~al.}(2022)\citenamefont {Schrinski}, \citenamefont {Lamaison},\ and\ \citenamefont {S\o{}rensen}}]{Schrinski2022}%
  \BibitemOpen
  \bibfield  {author} {\bibinfo {author} {\bibfnamefont {B.}~\bibnamefont {Schrinski}}, \bibinfo {author} {\bibfnamefont {M.}~\bibnamefont {Lamaison}},\ and\ \bibinfo {author} {\bibfnamefont {A.~S.}\ \bibnamefont {S\o{}rensen}},\ }\bibfield  {title} {\bibinfo {title} {{Passive Quantum Phase Gate for Photons Based on Three Level Emitters}},\ }\href {https://doi.org/10.1103/PhysRevLett.129.130502} {\bibfield  {journal} {\bibinfo  {journal} {Phys. Rev. Lett.}\ }\textbf {\bibinfo {volume} {129}},\ \bibinfo {pages} {130502} (\bibinfo {year} {2022})}\BibitemShut {NoStop}%
\bibitem [{\citenamefont {Pettersson}\ \emph {et~al.}(2026)\citenamefont {Pettersson}, \citenamefont {Christiansen}, \citenamefont {Mølmer},\ and\ \citenamefont {Sørensen}}]{Pettersson2026}%
  \BibitemOpen
  \bibfield  {author} {\bibinfo {author} {\bibfnamefont {L.~A.}\ \bibnamefont {Pettersson}}, \bibinfo {author} {\bibfnamefont {V.~R.}\ \bibnamefont {Christiansen}}, \bibinfo {author} {\bibfnamefont {K.}~\bibnamefont {Mølmer}},\ and\ \bibinfo {author} {\bibfnamefont {A.~S.}\ \bibnamefont {Sørensen}},\ }\href@noop {} {\bibinfo {title} {High fidelity photon-photon gates by scattering off a two-level quantum emitter}} (\bibinfo {year} {2026}),\ \Eprint {https://arxiv.org/abs/2603.10805} {arXiv:2603.10805 [quant-ph]} \BibitemShut {NoStop}%
\bibitem [{\citenamefont {Nysteen}\ \emph {et~al.}(2017)\citenamefont {Nysteen}, \citenamefont {McCutcheon}, \citenamefont {Heuck}, \citenamefont {M\o{}rk},\ and\ \citenamefont {Englund}}]{Nysteen2017}%
  \BibitemOpen
  \bibfield  {author} {\bibinfo {author} {\bibfnamefont {A.}~\bibnamefont {Nysteen}}, \bibinfo {author} {\bibfnamefont {D.~P.~S.}\ \bibnamefont {McCutcheon}}, \bibinfo {author} {\bibfnamefont {M.}~\bibnamefont {Heuck}}, \bibinfo {author} {\bibfnamefont {J.}~\bibnamefont {M\o{}rk}},\ and\ \bibinfo {author} {\bibfnamefont {D.~R.}\ \bibnamefont {Englund}},\ }\bibfield  {title} {\bibinfo {title} {{Limitations of two-level emitters as nonlinearities in two-photon controlled-PHASE gates}},\ }\href {https://doi.org/10.1103/PhysRevA.95.062304} {\bibfield  {journal} {\bibinfo  {journal} {Phys. Rev. A}\ }\textbf {\bibinfo {volume} {95}},\ \bibinfo {pages} {062304} (\bibinfo {year} {2017})}\BibitemShut {NoStop}%
\bibitem [{\citenamefont {Xu}\ \emph {et~al.}(2013)\citenamefont {Xu}, \citenamefont {Rephaeli},\ and\ \citenamefont {Fan}}]{Xu2013}%
  \BibitemOpen
  \bibfield  {author} {\bibinfo {author} {\bibfnamefont {S.}~\bibnamefont {Xu}}, \bibinfo {author} {\bibfnamefont {E.}~\bibnamefont {Rephaeli}},\ and\ \bibinfo {author} {\bibfnamefont {S.}~\bibnamefont {Fan}},\ }\bibfield  {title} {\bibinfo {title} {{Analytic Properties of Two-Photon Scattering Matrix in Integrated Quantum Systems Determined by the Cluster Decomposition Principle}},\ }\href {https://doi.org/10.1103/PhysRevLett.111.223602} {\bibfield  {journal} {\bibinfo  {journal} {Phys. Rev. Lett.}\ }\textbf {\bibinfo {volume} {111}},\ \bibinfo {pages} {223602} (\bibinfo {year} {2013})}\BibitemShut {NoStop}%
\bibitem [{\citenamefont {Brod}\ and\ \citenamefont {Combes}(2016)}]{Brod2016_1}%
  \BibitemOpen
  \bibfield  {author} {\bibinfo {author} {\bibfnamefont {D.~J.}\ \bibnamefont {Brod}}\ and\ \bibinfo {author} {\bibfnamefont {J.}~\bibnamefont {Combes}},\ }\bibfield  {title} {\bibinfo {title} {{Passive CPHASE Gate via Cross-Kerr Nonlinearities}},\ }\href {https://doi.org/10.1103/PhysRevLett.117.080502} {\bibfield  {journal} {\bibinfo  {journal} {Phys. Rev. Lett.}\ }\textbf {\bibinfo {volume} {117}},\ \bibinfo {pages} {080502} (\bibinfo {year} {2016})}\BibitemShut {NoStop}%
\bibitem [{\citenamefont {Brod}\ \emph {et~al.}(2016)\citenamefont {Brod}, \citenamefont {Combes},\ and\ \citenamefont {Gea-Banacloche}}]{Brod2016_2}%
  \BibitemOpen
  \bibfield  {author} {\bibinfo {author} {\bibfnamefont {D.~J.}\ \bibnamefont {Brod}}, \bibinfo {author} {\bibfnamefont {J.}~\bibnamefont {Combes}},\ and\ \bibinfo {author} {\bibfnamefont {J.}~\bibnamefont {Gea-Banacloche}},\ }\bibfield  {title} {\bibinfo {title} {{Two photons co- and counterpropagating through $N$ cross-Kerr sites}},\ }\href {https://doi.org/10.1103/PhysRevA.94.023833} {\bibfield  {journal} {\bibinfo  {journal} {Phys. Rev. A}\ }\textbf {\bibinfo {volume} {94}},\ \bibinfo {pages} {023833} (\bibinfo {year} {2016})}\BibitemShut {NoStop}%
\bibitem [{\citenamefont {Konyk}\ and\ \citenamefont {Gea-Banacloche}(2019)}]{Konyk2019}%
  \BibitemOpen
  \bibfield  {author} {\bibinfo {author} {\bibfnamefont {W.}~\bibnamefont {Konyk}}\ and\ \bibinfo {author} {\bibfnamefont {J.}~\bibnamefont {Gea-Banacloche}},\ }\bibfield  {title} {\bibinfo {title} {{Passive, deterministic photonic conditional-PHASE gate via two-level systems}},\ }\href {https://doi.org/10.1103/PhysRevA.99.010301} {\bibfield  {journal} {\bibinfo  {journal} {Phys. Rev. A}\ }\textbf {\bibinfo {volume} {99}},\ \bibinfo {pages} {010301(R)} (\bibinfo {year} {2019})}\BibitemShut {NoStop}%
\bibitem [{\citenamefont {Levy-Yeyati}\ \emph {et~al.}(2025)\citenamefont {Levy-Yeyati}, \citenamefont {Vega}, \citenamefont {Ramos},\ and\ \citenamefont {Gonz\'alez-Tudela}}]{Levy-Yeyati2024}%
  \BibitemOpen
  \bibfield  {author} {\bibinfo {author} {\bibfnamefont {T.}~\bibnamefont {Levy-Yeyati}}, \bibinfo {author} {\bibfnamefont {C.}~\bibnamefont {Vega}}, \bibinfo {author} {\bibfnamefont {T.}~\bibnamefont {Ramos}},\ and\ \bibinfo {author} {\bibfnamefont {A.}~\bibnamefont {Gonz\'alez-Tudela}},\ }\bibfield  {title} {\bibinfo {title} {{Passive Photonic CZ Gate with Two-Level Emitters in Chiral Multimode Waveguide QED}},\ }\href {https://doi.org/10.1103/PRXQuantum.6.010342} {\bibfield  {journal} {\bibinfo  {journal} {PRX Quantum}\ }\textbf {\bibinfo {volume} {6}},\ \bibinfo {pages} {010342} (\bibinfo {year} {2025})}\BibitemShut {NoStop}%
\bibitem [{\citenamefont {Heuck}\ \emph {et~al.}(2020{\natexlab{a}})\citenamefont {Heuck}, \citenamefont {Jacobs},\ and\ \citenamefont {Englund}}]{Heuck2020}%
  \BibitemOpen
  \bibfield  {author} {\bibinfo {author} {\bibfnamefont {M.}~\bibnamefont {Heuck}}, \bibinfo {author} {\bibfnamefont {K.}~\bibnamefont {Jacobs}},\ and\ \bibinfo {author} {\bibfnamefont {D.~R.}\ \bibnamefont {Englund}},\ }\bibfield  {title} {\bibinfo {title} {Photon-photon interactions in dynamically coupled cavities},\ }\href {https://doi.org/10.1103/PhysRevA.101.042322} {\bibfield  {journal} {\bibinfo  {journal} {Phys. Rev. A}\ }\textbf {\bibinfo {volume} {101}},\ \bibinfo {pages} {042322} (\bibinfo {year} {2020}{\natexlab{a}})}\BibitemShut {NoStop}%
\bibitem [{\citenamefont {Heuck}\ \emph {et~al.}(2020{\natexlab{b}})\citenamefont {Heuck}, \citenamefont {Jacobs},\ and\ \citenamefont {Englund}}]{Heuck2020_2}%
  \BibitemOpen
  \bibfield  {author} {\bibinfo {author} {\bibfnamefont {M.}~\bibnamefont {Heuck}}, \bibinfo {author} {\bibfnamefont {K.}~\bibnamefont {Jacobs}},\ and\ \bibinfo {author} {\bibfnamefont {D.~R.}\ \bibnamefont {Englund}},\ }\bibfield  {title} {\bibinfo {title} {{Controlled-Phase Gate Using Dynamically Coupled Cavities and Optical Nonlinearities}},\ }\href {https://doi.org/10.1103/PhysRevLett.124.160501} {\bibfield  {journal} {\bibinfo  {journal} {Phys. Rev. Lett.}\ }\textbf {\bibinfo {volume} {124}},\ \bibinfo {pages} {160501} (\bibinfo {year} {2020}{\natexlab{b}})}\BibitemShut {NoStop}%
\bibitem [{\citenamefont {Hassan}\ and\ \citenamefont {Gea-Banacloche}(2023)}]{Hassan2023}%
  \BibitemOpen
  \bibfield  {author} {\bibinfo {author} {\bibfnamefont {A.}~\bibnamefont {Hassan}}\ and\ \bibinfo {author} {\bibfnamefont {J.}~\bibnamefont {Gea-Banacloche}},\ }\bibfield  {title} {\bibinfo {title} {Conditional phase gate between two photons through control of the interaction time with a single atom in a cavity},\ }\href {https://doi.org/10.1103/PhysRevA.108.052601} {\bibfield  {journal} {\bibinfo  {journal} {Phys. Rev. A}\ }\textbf {\bibinfo {volume} {108}},\ \bibinfo {pages} {052601} (\bibinfo {year} {2023})}\BibitemShut {NoStop}%
\bibitem [{\citenamefont {Chen}\ \emph {et~al.}(2021)\citenamefont {Chen}, \citenamefont {Zhou}, \citenamefont {Shen}, \citenamefont {Ku},\ and\ \citenamefont {Steel}}]{Chen2021}%
  \BibitemOpen
  \bibfield  {author} {\bibinfo {author} {\bibfnamefont {Z.}~\bibnamefont {Chen}}, \bibinfo {author} {\bibfnamefont {Y.}~\bibnamefont {Zhou}}, \bibinfo {author} {\bibfnamefont {J.-T.}\ \bibnamefont {Shen}}, \bibinfo {author} {\bibfnamefont {P.-C.}\ \bibnamefont {Ku}},\ and\ \bibinfo {author} {\bibfnamefont {D.}~\bibnamefont {Steel}},\ }\bibfield  {title} {\bibinfo {title} {Two-photon controlled-phase gates enabled by photonic dimers},\ }\href {https://doi.org/10.1103/PhysRevA.103.052610} {\bibfield  {journal} {\bibinfo  {journal} {Phys. Rev. A}\ }\textbf {\bibinfo {volume} {103}},\ \bibinfo {pages} {052610} (\bibinfo {year} {2021})}\BibitemShut {NoStop}%
\bibitem [{\citenamefont {Gardiner}\ and\ \citenamefont {Collett}(1985)}]{Gardiner1985}%
  \BibitemOpen
  \bibfield  {author} {\bibinfo {author} {\bibfnamefont {C.~W.}\ \bibnamefont {Gardiner}}\ and\ \bibinfo {author} {\bibfnamefont {M.~J.}\ \bibnamefont {Collett}},\ }\bibfield  {title} {\bibinfo {title} {{Input and output in damped quantum systems: Quantum stochastic differential equations and the master equation}},\ }\href {https://doi.org/10.1103/PhysRevA.31.3761} {\bibfield  {journal} {\bibinfo  {journal} {Phys. Rev. A}\ }\textbf {\bibinfo {volume} {31}},\ \bibinfo {pages} {3761} (\bibinfo {year} {1985})}\BibitemShut {NoStop}%
\bibitem [{\citenamefont {Rephaeli}\ and\ \citenamefont {Fan}(2012)}]{Rephaeli2012}%
  \BibitemOpen
  \bibfield  {author} {\bibinfo {author} {\bibfnamefont {E.}~\bibnamefont {Rephaeli}}\ and\ \bibinfo {author} {\bibfnamefont {S.}~\bibnamefont {Fan}},\ }\bibfield  {title} {\bibinfo {title} {{Few-Photon Single-Atom Cavity QED With Input-Output Formalism in Fock Space}},\ }\href {https://doi.org/10.1109/JSTQE.2012.2196261} {\bibfield  {journal} {\bibinfo  {journal} {IEEE J. Sel. Top. Quantum Electron.}\ }\textbf {\bibinfo {volume} {18}},\ \bibinfo {pages} {1754} (\bibinfo {year} {2012})}\BibitemShut {NoStop}%
\bibitem [{\citenamefont {Gough}\ and\ \citenamefont {James}(2009)}]{Gough2009}%
  \BibitemOpen
  \bibfield  {author} {\bibinfo {author} {\bibfnamefont {J.}~\bibnamefont {Gough}}\ and\ \bibinfo {author} {\bibfnamefont {M.~R.}\ \bibnamefont {James}},\ }\bibfield  {title} {\bibinfo {title} {{The Series Product and Its Application to Quantum Feedforward and Feedback Networks}},\ }\href {https://doi.org/10.1109/TAC.2009.2031205} {\bibfield  {journal} {\bibinfo  {journal} {IEEE Trans. Automat. Contr.}\ }\textbf {\bibinfo {volume} {54}},\ \bibinfo {pages} {2530} (\bibinfo {year} {2009})}\BibitemShut {NoStop}%
\bibitem [{\citenamefont {Combes}\ \emph {et~al.}(2017)\citenamefont {Combes}, \citenamefont {Kerckhoff},\ and\ \citenamefont {Sarovar}}]{Combes2017}%
  \BibitemOpen
  \bibfield  {author} {\bibinfo {author} {\bibfnamefont {J.}~\bibnamefont {Combes}}, \bibinfo {author} {\bibfnamefont {J.}~\bibnamefont {Kerckhoff}},\ and\ \bibinfo {author} {\bibfnamefont {M.}~\bibnamefont {Sarovar}},\ }\bibfield  {title} {\bibinfo {title} {{The SLH framework for modeling quantum input-output networks}},\ }\href {https://doi.org/10.1080/23746149.2017.1343097} {\bibfield  {journal} {\bibinfo  {journal} {Adv. Phys.: X}\ }\textbf {\bibinfo {volume} {2}},\ \bibinfo {pages} {784} (\bibinfo {year} {2017})}\BibitemShut {NoStop}%
\bibitem [{\citenamefont {Hong}\ \emph {et~al.}(1987)\citenamefont {Hong}, \citenamefont {Ou},\ and\ \citenamefont {Mandel}}]{Hong1987}%
  \BibitemOpen
  \bibfield  {author} {\bibinfo {author} {\bibfnamefont {C.~K.}\ \bibnamefont {Hong}}, \bibinfo {author} {\bibfnamefont {Z.~Y.}\ \bibnamefont {Ou}},\ and\ \bibinfo {author} {\bibfnamefont {L.}~\bibnamefont {Mandel}},\ }\bibfield  {title} {\bibinfo {title} {Measurement of subpicosecond time intervals between two photons by interference},\ }\href {https://doi.org/10.1103/PhysRevLett.59.2044} {\bibfield  {journal} {\bibinfo  {journal} {Phys. Rev. Lett.}\ }\textbf {\bibinfo {volume} {59}},\ \bibinfo {pages} {2044} (\bibinfo {year} {1987})}\BibitemShut {NoStop}%
\bibitem [{\citenamefont {Duda}\ \emph {et~al.}(2024)\citenamefont {Duda}, \citenamefont {Brunswick}, \citenamefont {Wilson},\ and\ \citenamefont {Kok}}]{Duda2024}%
  \BibitemOpen
  \bibfield  {author} {\bibinfo {author} {\bibfnamefont {M.}~\bibnamefont {Duda}}, \bibinfo {author} {\bibfnamefont {L.}~\bibnamefont {Brunswick}}, \bibinfo {author} {\bibfnamefont {L.~R.}\ \bibnamefont {Wilson}},\ and\ \bibinfo {author} {\bibfnamefont {P.}~\bibnamefont {Kok}},\ }\bibfield  {title} {\bibinfo {title} {Efficient, high-fidelity single-photon switch based on waveguide-coupled cavities},\ }\href {https://doi.org/10.1103/PhysRevA.110.042615} {\bibfield  {journal} {\bibinfo  {journal} {Phys. Rev. A}\ }\textbf {\bibinfo {volume} {110}},\ \bibinfo {pages} {042615} (\bibinfo {year} {2024})}\BibitemShut {NoStop}%
\bibitem [{\citenamefont {Fan}\ \emph {et~al.}(2010)\citenamefont {Fan}, \citenamefont {\c{S}. E.~Kocaba\c{s}},\ and\ \citenamefont {Shen}}]{Fan2010}%
  \BibitemOpen
  \bibfield  {author} {\bibinfo {author} {\bibfnamefont {S.}~\bibnamefont {Fan}}, \bibinfo {author} {\bibnamefont {\c{S}. E.~Kocaba\c{s}}},\ and\ \bibinfo {author} {\bibfnamefont {J.-T.}\ \bibnamefont {Shen}},\ }\bibfield  {title} {\bibinfo {title} {Input-output formalism for few-photon transport in one-dimensional nanophotonic waveguides coupled to a qubit},\ }\href {https://doi.org/10.1103/PhysRevA.82.063821} {\bibfield  {journal} {\bibinfo  {journal} {Phys. Rev. A}\ }\textbf {\bibinfo {volume} {82}},\ \bibinfo {pages} {063821} (\bibinfo {year} {2010})}\BibitemShut {NoStop}%
\bibitem [{\citenamefont {Rephaeli}\ and\ \citenamefont {Fan}(2013)}]{Rephaeli2013}%
  \BibitemOpen
  \bibfield  {author} {\bibinfo {author} {\bibfnamefont {E.}~\bibnamefont {Rephaeli}}\ and\ \bibinfo {author} {\bibfnamefont {S.}~\bibnamefont {Fan}},\ }\bibfield  {title} {\bibinfo {title} {Dissipation in few-photon waveguide transport [\uppercase{I}nvited]},\ }\href {https://doi.org/10.1364/PRJ.1.000110} {\bibfield  {journal} {\bibinfo  {journal} {Photon. Res.}\ }\textbf {\bibinfo {volume} {1}},\ \bibinfo {pages} {110} (\bibinfo {year} {2013})}\BibitemShut {NoStop}%
\end{thebibliography}

%

\end{document}